\documentclass{elsarticle}

\usepackage{hyperref}
\journal{Elsevier}

\usepackage{amsmath}
\usepackage{amsthm}
\usepackage{amssymb}
\usepackage{eucal}
\usepackage{mathrsfs}
\usepackage{subfigure}
\usepackage{here}
\usepackage{multirow}
\usepackage{float}
\usepackage{fix-cm}
\usepackage{bm}

\usepackage[dvipdfmx]{color}
\biboptions{sort&compress}
\usepackage{ulem}

\def\dG{\ \text{d}\hspace{-0.3mm}\Gamma}

\def \div{\mbox{\rm div}}
\def\x{\bm{x}}

\newcounter{num}
\begin{document}

\begin{frontmatter}

\title{Optimal design of compliant displacement magnification mechanisms using stress-constrained topology optimization based on effective energy}

\author[osakaaddress,tokyoengaddress]{K. Miyajima}
\author[tokyoengaddress,tokyoinovaddress]{Y. Noguchi}
\author[tokyoengaddress,tokyoinovaddress]{T. Yamada\corref{mycorrespondingauthor}}
\ead{t.yamada@mech.t.u-tokyo.ac.jp}

\cortext[mycorrespondingauthor]{Corresponding author}
\address[osakaaddress]{Osaka Research Institute of Industrial Science and Technology, 7-1, Ayumino-2, Izumi-city, Osaka, 594-1157, Japan}
\address[tokyoengaddress]{Department of Mechanical Engineering, Graduate School of Engineering, The University of Tokyo, Yayoi 2-11-16, Bunkyo-ku, Tokyo 113-8656, Japan.}
\address[tokyoinovaddress]{Department of Strategic Studies, Institute of Engineering Innovation, School of Engineering, The University of Tokyo, Yayoi 2-11-16, Bunkyo-ku, Tokyo 113-8656, Japan.}

\begin{abstract}
In this paper, stress-constrained topology optimization is applied to the design of compliant displacement magnification mechanisms. By formulating the objective function based on the concept of effective energy, it is not necessary to place artificial spring components at the boundaries of the output and input ports as in previous methods. This makes it possible to design mechanisms that do not receive a reaction force at the output port, such as sensors. Furthermore, by imposing a constraint on the maximum stress evaluated in terms of the p-norm of the von Mises equivalent stress, problems such as stress concentration can be avoided. Several numerical examples of displacement magnification mechanisms are provided to demonstrate the effectiveness of the proposed method.
\end{abstract}


\end{frontmatter}

\section{Introduction}
\label{sec:intro}

A compliant mechanism transfers or transforms motion, force or energy through the deflection of flexible members \cite{howell2001}.
Unlike rigid-body mechanisms, compliant mechanisms gain some of their mobility from the deflection of flexible members rather than from movable joints alone.
Consequently, compliant mechanisms require less assembly than rigid-body mechanisms and can be applied to electronic devices, such as microelectromechanical systems (MEMS) and sensors \cite{kota2001design}.
In particular, displacement magnification mechanisms are used to increase the sensitivity of sensors that detect displacement.
To design high-performance compliant mechanisms, design methods for compliant mechanisms using topology optimization \cite{BENDSOE1988197} have been studied.
Sigmund \cite{sigmund1997design} maximized the mechanical advantage, which is the ratio of the input force to the output force, as the objective function, 
while Kota et al. \cite{kota2001design} maximized the geometric advantage, which is the ratio of the displacement at the input port to the displacement at the output port, as the objective function.
Frecker et al. \cite{frecker1997topological} formulated the flexibility of compliant mechanisms using the concept of mutual energy.
In these studies, artificial spring components at the input and output ports provided stiffness against both the input force and the reaction force at the output port in the optimal structure.
These formulations can be widely applied to the design of compliant devices \cite{pedersen2001topology, bruns2001topology, zhu2014topology};
however, they cannot be applied to mechanisms such as sensors, where no reaction force is applied at the output port.
Therefore, the aforementioned studies are not applicable to the displacement magnification mechanisms.
In other work, Yamada et al. \cite{YAMADA201717-00453} formulated the objective function based on the concept of effective energy, which enables the design of compliant devices without placing spring components at the output and input ports.
This makes it possible to design displacement magnification mechanisms based on topology optimization.
However, the mechanisms obtained by this method have several hinges and are difficult to fabricate in practice.

The problem of hinges in structures obtained by topology optimization has been discussed extensively, and various methods have been proposed to avoid hinges.
Representative methods involve increasing the stiffness of the structure \cite{rahmatalla2005sparse, zhu2012new, zhu2014multi}, using filters \cite{sigmund2007morphology, poulsen2003new, bendsoe2003topology, bendsoe1993topology}, using subsequent local shape optimization \cite{christiansen2014topology, nguyen2020efficient, stankiewicz2021coupled}, and imposing stress constraints \cite{yang1996stress, duysinx1998topology, duysinx1998new}.
%Several operations on the structure obtained by topology optimization may promote localization unnecessarily.
Subsequent local shape optimization can modify structures with stress concentrations by applying shape optimization to the topology-optimized structure. 
However, it has a higher probability of falling into a local solution.
Furthermore, compared to the algorithm of methods imposing stress constraints, the algorithm of methods using filters or local shape optimization is complex. 
For these reasons, we focus on the method of imposing stress constraints to avoid hinges and reduce stress concentration in this paper.
Methods of imposing stress constraints on topology optimization have been widely studied; however, they possess significant challenges, which are discussed below \cite{bendsoe2003topology, le2010stress}.
The first is the singularity problem \cite{Cheng1997}, while the second is the implementation of the constraint related to the local quantity \cite{yang1996stress}.
The singularity problem is a phenomenon in which stress values exhibit a singular behavior when beams or other components are lost in the process of structural optimization.
To overcome this problem, several approaches for relaxing the stress constraints have been applied \cite{Cheng1997, bruggi2008alternative, le2010stress, holmberg2013stress, xu2021stress}.
In addition, several different formulations of stress constraints have been proposed, such as local stress constraints \cite{duysinx1998topology, amstutz2010topological}, global stress constraints \cite{yang1996stress, duysinx1998new, ogawa2022topology}, and regional stress constraints \cite{le2010stress, holmberg2013stress}.
Local stress constraints control the local stress behavior with a large number of constraints, whereas global stress constraints control the approximate maximum stress with only one constraint, such as the p-norm and the Kreisselmeier--Steinhauser function.
Stress constraint methods based on the p-norm have been particularly well studied \cite{duysinx1998new, allaire2008minimum, amstutz2010topological}.

Several methods have been proposed to impose stress constraints on the design of compliant mechanisms: those imposing local stress constraints \cite{da2019topology, emmendoerfer2020stress}, those imposing regional stress constraints \cite{conlan2019stress}, and those imposing global stress constraints \cite{lopes2016topology, otomori2011level}.
In addition, Pereira et al. \cite{de2018influence} compared the effect of local and global stress constraints and filters on the creation of hinge-free structures.
However, the above-mentioned methods are all based on placing artificial spring elements at the output port, and thus do not allow for the design of displacement magnification mechanisms.

In this paper, we propose a method for designing optimal compliant displacement magnification mechanisms using stress-constrained topology optimization based on the concept of effective energy.
We refer to the global stress constraints and relaxation method formulated by Holmberg et al. \cite{holmberg2013stress} and modify it to fit the formulation of topology optimization based on the level set method.

The remainder of this paper is organized as follows.
Section \ref{sec:Top} describes the topology optimization method based on the level set method, while Section \ref{sec:formulation} presents the formulation for optimal design of compliant mechanisms based on the concept of effective energy and stress constraint.
Section \ref{sec:impli} presents the numerical implementation of the proposed method, while Section \ref{sec:example} provides numerical examples, where a benchmark model is used to verify the effectiveness of the proposed method.
Section \ref{sec:conclusion} presents the conclusions.
Furthermore,  \ref{Lbeam} presents several numerical examples that minimize the p-norm of the von Mises stress.
\ref{comp_inv} presents comparison of displacement between proposed method and conventional topology optimization results.
Lastly, \ref{proto} presents prototypes of the displacement magnification mechanisms printed by a 3D printer.

\section{Topology optimization}
\label{sec:Top}

\subsection{Concept of topology optimization}

Topology optimization is a type of structural optimization method.
Structural optimization is used to obtain a structure $\Omega$ that minimizes or maximizes an objective function.
The objective function often includes physical properties, such as stiffness \cite{BENDSOE1988197, YAMADA20102876, wang2018level, noda2021extended}, thermal properties \cite{haslinger2002optimization, yamada2011level, JING201561, wu2019multi, tang2019topology, miki2021topology, NODA2021}, and acoustic properties \cite{sigmund2003systematic, hu2020topology, noguchi2021topology}.
Therefore, the optimal structure is obtained on the assumption that the objective function satisfies the governing equations that describe the physical phenomena.
The governing equations are treated as constraints in the optimization problem, and the basic structural optimization problem can be formulated as follows:
	\begin{eqnarray}
	\begin{split}
		\underset{\Omega}{\text{inf}} \qquad &F(u,\Omega) = \int_{\Omega}f(u)\mathrm{d\Omega}\\
		\text{subject to} \qquad &\text{governing equation system},
	\label{strct_optim}
	\end{split}
	\end{eqnarray}
where $u$ is the state variable obtained as the solution of the governing equation, and $f(u)$ is the objective function.

Next, we consider the application of topology optimization to the structural optimization problem (\ref{strct_optim}).
We introduce a domain $\Omega_D\subset\mathbb{R}^n (n = 2 \; \text{or} \; 3)$ where the structure can be placed.
Here the domain $\Omega_D$ is called a fixed design domain because it does not change during the optimization process.
The fixed design domain consists of a domain filled with the structure (hereafter referred to as the material domain) and a domain not filled with the structure (hereafter referred to as the void domain), and these domains are expressed by the characteristic function $\chi$ defined as follows:
	\begin{eqnarray}
	\chi(\bm{x}) := \left\{ 
	\begin{array}{ll}
	1 \qquad \mathrm{for} \quad\bm{x}\in \Omega \\
	0 \qquad \mathrm{for} \quad\bm{x}\in\Omega_D\backslash\Omega , \\
	\end{array} 
	\right.
	\end{eqnarray}
where the boundary between the material and void domains is included in the material domain.
Using the characteristic function $\chi$, the topology optimization problem can be formulated as follows:
	\begin{eqnarray}
	\begin{split}
		\underset{\chi}{\text{inf}} \qquad &F(u,\Omega) = \int_{\Omega_D}f(u)\chi\mathrm{d\Omega}\\
		\text{subject to} \qquad &\text{governing equation system} .
	\label{top_optim}
	\end{split}
	\end{eqnarray}
In topology optimization, the optimization problem (\ref{strct_optim}) is replaced with a material distribution problem, which allows topological changes, such as an increase or decrease in the number of holes, during the optimization procedure.

However, topology optimization problems are commonly ill-posed \cite{allaire2001shape}; therefore, the space of admissible design should incorporate relaxation or regularization techniques to make the problem well-posed.
A typical method based on relaxation of the space of admissible design is the homogenization method \cite{BENDSOE1988197}.
In this paper, we use a level set-based topology optimization method \cite{YAMADA20102876} to transform an ill-posed problem into a well-posed problem.
In this method, the boundary surface of the material domain is represented by the isosurface of a scalar function called the level set function, and changes in the level set function represent changes in the shape of the material domain.
The topology optimization problem is regularized by ensuring proper smoothness of the level set function.
This method is described in Section \ref{sec:lev}.

\subsection{Level set-based topology optimization}
\label{sec:lev}
In the level set-based method, the scalar function $\phi(\bm{x})$, called the level set function, illustrated in the following equation is introduced to represent the shape:
	\begin{eqnarray}
	\left\{ 
	\begin{array}{lll}
	-1 \leq \phi(\bm{x}) <0 \qquad \mathrm{for} \quad \bm{x} \in \Omega_D\backslash\Omega \\
	\phi(\bm{x})=0\qquad \mathrm{for} \quad \bm{x} \in \partial\Omega \\
	1 \geq \phi(\bm{x}) >0 \qquad \mathrm{for} \quad \bm{x} \in \Omega \\
	\end{array} ,
	\right.
	\end{eqnarray}
where $\partial \Omega$ denotes the boundaries between the material and void domains.
We redefine the characteristic function using the level set function as follows:
	\begin{eqnarray}
	\chi_\phi(\bm{x}) := \left\{ 
	\begin{array}{ll}
	1 \qquad \mathrm{for} \quad\phi(\bm{x}) \geq 0 \\
	0 \qquad \mathrm{for} \quad\phi(\bm{x})< 0  \\
	\end{array} 
	\right . .
	\label{eq:chi}
	\end{eqnarray}
In the level set-based method, the topology optimization problem is formulated using the characteristic function defined in (\ref{eq:chi}).

Next, we describe how to update the level set function.
Assuming that the level set function is a function of the fictitious time $t$, it is updated by the reaction-diffusion equation as follows:
\begin{eqnarray}
\begin{split}
\frac{\partial \phi(\bm{x},t)}{\partial t} = -K\{-\tilde{C} d_t F - \tau \nabla^2 \phi(\bm{x},t)  \} ,\\
\tilde{C} := \frac{C \int_{\Omega_D}\mathrm{d\Omega} }{\int_{\Omega_D} | d_t F | \mathrm{d\Omega} } ,
\end{split}
\label{eq:RDE}
\end{eqnarray}
where $K\in \mathbb{R}_+$ is the proportionality coefficient, $C\in \mathbb{R}_+$ is the normalization coefficient, $\tau \in \mathbb{R}_+$ is the  regularization coefficient, and $d_t F$ is the topological derivative \cite{amstutz2006new}.
In this paper, we set $K=1.0$ and $C=0.8$.
The topological derivative is defined as follows:
\begin{eqnarray}
d_t F = \underset{\epsilon \to 0}{\text{lim}} \dfrac{(F+\delta F) - F}{meas(\Omega \backslash \Omega_\epsilon) - meas(\Omega)} ,
\label{eq:dtF}
\end{eqnarray}
where $\Omega_\epsilon$ is a small hole of radius $\epsilon$ in the material domain, and $\delta F$ is the change in the objective function $F$ due to the opening of the small hole $\Omega_\epsilon$.

\section{Formulation for topology optimization of compliant mechanisms}
\label{sec:formulation}

\begin{figure*}[htb]
	\begin{center}
		\includegraphics[height=7.5cm]{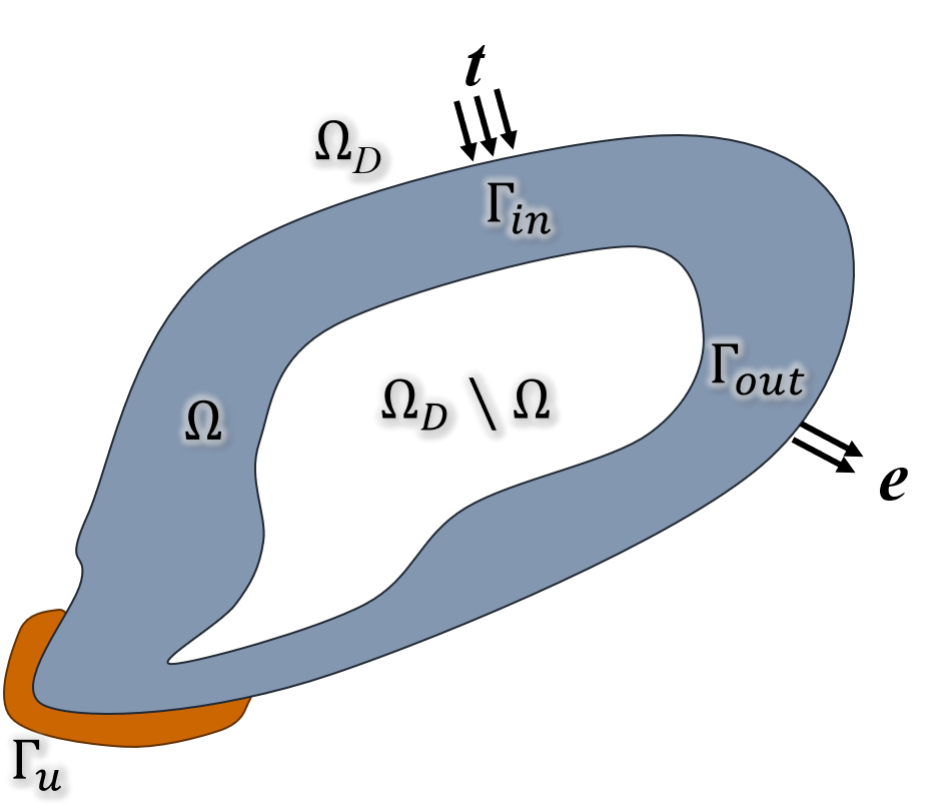}
		\caption{Problem setup for the optimal design of a compliant mechanism}
		\label{fig:concept}
	\end{center}
\end{figure*}

Figure \ref{fig:concept} presents an overview of the problem setup for the optimal design of a compliant mechanism. 
The compliant mechanism is represented by a material domain $\Omega$ filled with linear elastic material and whose displacement is equal to zero at the boundary $\Gamma_{u}$. 
The design requirements of the compliant mechanism are that the traction $\bm{t}$ input to the boundary $\Gamma_{in}$ should be output as a displacement in the direction represented by vector $\bm{e}$ at boundary $\Gamma_{out}$. 

\subsection{Standard formulation}

In the standard formulation \cite{sigmund1997design}, spring components are set at the input port $\Gamma_{in}$ and output port $\Gamma_{out}$, and the objective function $F_s$ is set to maximize the output displacement in the desired direction $\bm{e}$ as follows:
	\begin{eqnarray}
	\begin{split}
		\underset{\Omega}{\text{maximize}} \qquad J_s :&= \int_{\Gamma_{out}} \bm{e}\cdot \bm{u} \mathrm{d\Gamma}\\
		\text{subject to} \qquad G :&= - \div \left\lbrace  \frac{1}{2} \bm{D}(\bm{\nabla u}   + (\bm{\nabla u})^T ) \right\rbrace  = 0
		\qquad \mathrm{in} \hspace{2mm} \Omega\\
		\bm{u} &= 0 \hspace{49mm}\mathrm{on} \hspace{2mm} \Gamma_u\\
		\bm{\sigma_{n}} &= -k_{out} \bm{u} \hspace{38mm}\mathrm{on} \hspace{2mm} \Gamma_{out}\\
		\bm{\sigma_{n}} &= \bm{t} - k_{in}\bm{u} \hspace{36mm}\mathrm{on} \hspace{2mm} \Gamma_{in}\\
		G_{V} :&= \dfrac{\int_{\Omega} \mathrm{d\Omega}}{\int_{\Omega_D} \mathrm{d\Omega}} -V_{\mathrm{max}} \leq 0,
	\end{split}
	\end{eqnarray}
where $\bm{u}$ is the displacement, $\bm{\sigma_n}$ is the traction force, $\bm{D}$ is the elastic tensor, $k_{in}$ and $k_{out}$ are the spring constants on $\Gamma_{in}$ and $\Gamma_{out}$, respectively, and $V_{max}$ is the upper limit of the volume constraint $G_V$.
In this standard formulation, the maximum stress in the compliant mechanism can be controlled indirectly by the spring component at the input port. 
The spring component at the output port provides implicit stiffness against the reaction force from the workpiece.
It is possible to control the characteristics of the obtained structure by adjusting the spring constants of the spring components at the input and output ports.
This formulation is widely used because of its simplicity and stability of calculation; however, it does not completely consider the problem of hinge generation \cite{sigmund1997design}.
Therefore, a number of studies have been conducted to overcome the problem of hinging in this formulation \cite{de2015stress, da2019topology}, and  compliant mechanisms without hinges have been successfully obtained.

\subsection{Formulation based on effective energy}

In another approach, a formulation based on the concept of effective energy, has been proposed \cite{YAMADA201717-00453}. 
In this formulation, the compliant mechanism is considered an energy transformation device to transfer the input force into the desired deformation and maximize the effective energy, which is the energy conversion efficiency, as follows:
	\begin{eqnarray}
	\begin{split}
		\underset{\Omega}{\text{maximize}} \qquad J_p :=& 	\frac{U_0(\bm{u})}	{E_0(\bm{u})}
		\\
		\text{subject to} \qquad G =& 0\hspace{46mm} \mathrm{in} \hspace{2mm} \Omega\\
		\bm{u} =& 0 \hspace{46mm}\mathrm{on} \hspace{2mm} \Gamma_u\\
		\bm{\sigma_{n}} =& \bm{t} \hspace{46mm}\mathrm{on} \hspace{2mm} \Gamma_{in}\\
		G_{V} \leq& 0 ,
	\end{split}
	\end{eqnarray}
where $U_0$ and $E_0$ are the average displacement at the output port and the input energy, respectively, defined as follows:
	\begin{eqnarray}
		U_0(\bm{u}) :=& \frac{\int _{\Gamma _{out}} \bm{e} \cdot \bm{u} \, \mathrm{d\Gamma}}{ \int _{\Gamma _{out}}  \, \mathrm{d\Gamma}} ,
		\\
		E_0(\bm{u}) :=& \frac{1}{2} \int _{\Gamma _{in}} \bm{t} \cdot \bm{u} \, \mathrm{d\Gamma} .
	\end{eqnarray}
In this formulation, stiffness against the input force is implicitly provided by minimizing the input energy.
This formulation can produce displacement magnification mechanisms as described in \cite{YAMADA201717-00453}; however, it cannot avoid hinges.
Furthermore, unlike the standard formulation, it cannot control the characteristics of the obtained structure  because only maximizes the effective energy.

\subsection{Proposed formulation}
\label{subsec:prop}

Extending the formulation proposed by Yamada et al. \cite{YAMADA201717-00453}, we propose a formulation with a stress constraint that can produce displacement magnification mechanisms, and control the qualitative stiffness of the structure as follows:
	\begin{eqnarray}
	\begin{split}
		\underset{\Omega}{\text{maximize}} \qquad J(\bm{u}) :=\frac{W(\bm{u})+\alpha}	{E(\bm{u})+\beta}\hspace{27mm}
		\label{eq:F}
	\end{split}\\
	\text{subject to} \qquad G =0\hspace{30mm} \mathrm{in} \hspace{2mm} \Omega \hspace{4mm}\label{eq:G}\\
		\bm{u} =0 \hspace{30mm}\mathrm{on} \hspace{2mm} \Gamma_u \hspace{2mm}\label{eq:Gamma_u} \\
		\bm{\sigma_{n}} = \bm{t} \hspace{30mm}\mathrm{on} \hspace{2mm} \Gamma_{in} \hspace{1mm}\label{eq:Gamma_in}\\
		G_{V} \leq0 \hspace{42mm}\\
		G_{\sigma} \leq 0\label{eq:G_sigma} ,\hspace{42mm}
	\end{eqnarray}	
	\begin{eqnarray}
	\begin{split}			
		W(\bm{u}) :=\frac{\int _{\Gamma _{out}} \bm{e} \cdot \bm{u} \, \mathrm{d\Gamma}}{ \bar{W}} ,\hspace{20mm}
		\label{eq:W}
	\end{split}\\
		\bar{W} :=\int _{\Gamma _{out}} |\bm{e} \cdot \bm{U} _{out}| \, \mathrm{d\Gamma} , \hspace{20mm}\\
	\begin{split}
		E(\bm{u}) :=\frac{\int _{\Gamma _{in}} \bm{t} \cdot \bm{u} \, \mathrm{d\Gamma}}{\bar{E} } , \hspace{23mm}
		\label{eq:E}
	\end{split}\\
		\bar{E} := \int _{\Gamma _{in}} |\bm{t} \cdot \bm{U} _{in}| \, \mathrm{d\Gamma} ,\hspace{25mm}
	\end{eqnarray}
where $\bm{U}_{out}$ and $\bm{U}_{in}$ are the representative displacement for normalizing displacement on $\Gamma_{out}$ and $\Gamma_{in}$, respectively, $\alpha$ and $\beta$ are parameters to adjust the balance between minimizing the denominator and maximizing the numerator of the objective function; and $G_{\sigma}$ is the stress constraint function.

First, we describe parameters $\alpha$ and $\beta$ in (\ref{eq:F}).
By increasing parameter $\alpha$, changes in the numerator $W$ will have less effect on the objective function $J$, and the optimization procedure will minimize the denominator $E$, thereby increasing the stiffness of the structure against the input force $\bm{t}$.
By increasing parameter $\beta$, changes in the denominator $E$ will have less effect on the objective function $J$, and the optimization procedure will maximize the numerator $W$, thereby causing the obtained structure to allow a larger displacement.
Therefore, the characteristics of the structure, namely, its stiffness and allowance of a larger displacement, can be controlled by adjusting parameters $\alpha$ and $\beta$.

Next, we describe the stress constraint.
In this paper, we use a method that approximates the p-norm of the relaxed von Mises stress $\sigma_{vm}$ as the maximum stress of the structure \cite{duysinx1998new} \cite{holmberg2013stress}.
The stress constraint function is formulated as follows:
	\begin{equation}
		G_{\sigma} :=\sigma_{pn} - 1 ,
		\label{eq:dif_Gs}
	\end{equation}	
	\begin{eqnarray}
			\sigma_{pn} := 
			\left(\int_\Omega (\frac{\sigma_{vm}}{\Phi(\bm{x}) \sigma_{max}})^{p}  \mathrm{d\Omega} \right)^{\frac{1}{p}} ,
			\label{eq:pn}
	\end{eqnarray}

	\begin{eqnarray}
	\sigma_{vm} = \sqrt{ (\bm{BD\nabla u})^T \bm{D\nabla u} } ,\\
	\bm{B} = 3\bm{ \Roman{num}} - ( \bm{I} \otimes \bm{I}) ,
	\end{eqnarray}
where $\Phi(\bm{x})$ is the relaxation function to avoid the singularity problem, $\sigma_{max}$ is the upper limit of the stress constraint, $p$ is the parameter, and $\bm{I}$ and $\bm{\Roman{num}}$ are unit tensors of the second and fourth order, respectively.
By normalizing the terms in the p-norm by $\sigma_{max}$, constraint (\ref{eq:G_sigma}) becomes equivalent to the inequality constraint stating that the value of the p-norm should be less than $\sigma_{max}$.

\ref{Lbeam} provides several numerical examples that minimize the p-norm of the von Mises stress defined in (\ref{eq:pn}).

\section{Numerical implementation}
\label{sec:impli}

\subsection{Optimization algorithm}
The optimization algorithm is as follows.
\begin{itemize}
	\item[\bf{Step 1}] The initial level set function is set.
	\item[\bf{Step 2}] The displacement fields $\bm{u}$ defined in (\ref{eq:G}) are solved using the finite element method (FEM).
	\item[\bf{Step 3}] The objective function $J$ formulated using (\ref{eq:F}) and the constraint function $G_{\sigma}$ formulated using (\ref{eq:dif_Gs}) are calculated.
	\item[\bf{Step 4}] If the objective function converges, the optimization procedure is terminated; otherwise, the adjoint field $\bm{v}$ defined in (\ref{eq:v}) is solved using FEM, and the topological derivatives with respect to the objective function are calculated using (\ref{eq:dTL}).
	\item[\bf{Step 5}]The level set function is updated using the time evolution equation given by (\ref{eq:RDE}); then, the optimization procedure returns to step 2.
\end{itemize}

We use FreeFEM++ \cite{MR3043640} as a FEM solver.
In \ref{subsec:euler}, we explain about approximation of displacement field.
Next, in \ref{subsec:relax_stress}, we explain the relaxation function for stress constraint.
Finally, in \ref{subsec:sens}, sensitivity analysis is explained.

\subsection{Approximate solution of displacement field based on Eulerian coordinate system}
\label{subsec:euler}

The fixed design domain is represented by the Eulerian coordinate system; therefore, it requires the generation of finite elements for each iteration of the optimization procedure. 
To reduce the computational cost, an approximate solution method using the ersatz material approach \cite{allaire2004structural} is applied.
Specifically, the void domain is assumed to be a structural material with a relatively small elastic tensor, and the material properties are assumed to be smoothly distributed in the neighborhood of the interface.
The governing equations for the displacement field in the FEM are extended to the fixed design domain $\Omega_D$ using the extended elastic tensor $\tilde{\bm{D}}$ and an approximate Heaviside function $h(\phi)$ as follows:
	\begin{eqnarray}
	\begin{split}
	\tilde{\bm{D}} :&= h(\phi) \bm{D} ,\\
	h(\phi) :&= \left\{ 
		\begin{array}{lll}
		d \qquad \mathrm{for} \quad \phi < -w  \\
		\left\{\frac{1}{2}+\frac{\phi}{w} \left[ \frac{15}{16} - \frac{\phi^2}{w^2}(\frac{5}{8}-\frac{3}{16}\frac{\phi^2}{w^2}) \right]\right\}(1-d) + d \qquad \mathrm{for} \quad -w \leq \phi \leq w \\
		1  \qquad \mathrm{for} \quad w<\phi \\
		\end{array} 
		,\right.
	\end{split}
	\end{eqnarray}
where $\bm{D}$ is the elastic tensor, $w$ is the transition width of the Heaviside function, and $d$ is a sufficiently small positive number.
In this paper, we set $w=0.9$ and $d=0.01$.

\subsection{Relaxation function for stress constraint}
\label{subsec:relax_stress}

To avoid the singularity problem mentioned in Section \ref{sec:intro} \cite{Cheng1997}, in which the stress exhibits singular behavior when the material domain changes to the void domain, the stress is evaluated using the relaxed stress.
We redefine the stress constraint as follows:
	\begin{equation}
		\tilde{G}_{\sigma}(\phi) :=\tilde{\sigma}_{pn}(\phi) - 1 ,
		\label{eq:redif_Gs}
	\end{equation}	
	\begin{eqnarray}
			\tilde{\sigma}_{pn}(\phi) := 
			\left(\int_{\Omega_D} (\frac{\tilde{\sigma}_{vm}(\phi)}{\Phi(\bm{x}) \sigma_{max}})^{p}  \mathrm{d\Omega} \right)^{\frac{1}{p}} ,
			\label{eq:redef_pn}
	\end{eqnarray}
In the same way as the displacement field, the stress field can be expressed using an approximate Heaviside function as follows:
	\begin{eqnarray}
	\begin{split}
	\tilde{\sigma}_{vm}(\phi) :=  \sqrt{(\bm{B\tilde{D}\nabla u})^T \bm{\tilde{D}\nabla u}}
						 = h(\phi) \sigma_{vm}.
	\end{split}
	\end{eqnarray}
In this paper, we set the relaxation function $\Phi$ in (\ref{eq:redef_pn}) to the following:
	\begin{equation}
	\Phi(\bm{x}) := h(\phi)^{\frac{1}{2}} .
	\end{equation}
This relaxation function is based on the qp approach \cite{le2010stress}\cite{bruggi2008alternative}, and the exponent is proposed by Holmberg et al. \cite{holmberg2013stress}. The relaxation function is modified to fit the formulation of level set-based topology optimization.

\subsection{Sensitivity analysis}
\label{subsec:sens}

Here, we describe a procedure for obtaining topological derivatives for updating the level set functions.
For stability of calculation, we change the formulation from maximization to minimization as follows:
\begin{eqnarray}
\begin{split}
  \underset{\phi}{\text{minimize}} \qquad \qquad &-J(\bm{u})\\
		\text{subject to} \qquad G_\phi :=& - \div \left\lbrace  \frac{1}{2} \tilde{\bm{D}}(\bm{\nabla u}   + (\bm{\nabla u})^T ) \right\rbrace  = 0
		\qquad \mathrm{in} \hspace{2mm} \Omega_D\\
 		\bm{u} =& 0 \hspace{46mm}\mathrm{on} \hspace{2mm} \Gamma_u\\
 		\bm{\sigma_{n}} =& \bm{t} \hspace{46mm}\mathrm{on} \hspace{2mm} \Gamma_{in}\\
 		G_{V\phi}:=& \dfrac{\int_{\Omega_D} h(\phi) \mathrm{d\Omega}}{\int_{\Omega_D} \mathrm{d\Omega}} -V_{\mathrm{max}} \leq 0\\
 		\tilde{G}_{\sigma}  \leq& 0	.
\end{split}
\end{eqnarray}
The Lagrangean is defined as follows:
\begin{eqnarray}
\begin{split}
	L(\phi, \bm{\bm{u}}, \bm{v})
	:=&
	-J(\bm{u}) + \int_{\Omega} (\bm{\nabla} \bm{v})^T \tilde{\bm{D}} \bm{\nabla} \bm{u} \mathrm{d\Omega} 
	- \int_{\Gamma} \bm{v} \cdot (\bm{n}\cdot \tilde{\bm{D}}\bm{\nabla}\bm{u}) \mathrm{d\Gamma}
	+\int _{\Gamma_{u}}	\bm{v}  \cdot \bm{u}\, \mathrm{d\Gamma}	\\
	&+\int _{\Gamma_{in}}\bm{v} \cdot	(\bm{n} \cdot (\tilde{\bm{D}} \bm{\nabla} \bm{u})-\bm{t})\, \mathrm{d\Gamma}
	+\int _{\Gamma \backslash (\Gamma_{in} \cup \Gamma_{u})}\bm{v} \cdot	(\bm{n} \cdot (\tilde{\bm{D}}\bm{\nabla} \bm{u}))\, \mathrm{d\Gamma}\\
	&+ \mu \tilde{G}_{\sigma}
	+ \lambda G_{V} ,\\	
	\end{split}
\end{eqnarray}
where $\bm{v}, \mu$ and $\lambda$ are the Lagrange multipliers.
In this paper, $\mu$ is treated as a constant value.
We choose $\bm{v}$ as the adjoint state which is the solution of the following equation:
%The adjoint equation is defined for adjoint variable $\bm{v}$ as follows:
\begin{eqnarray}
\begin{split}
&\int_{\Omega \backslash \Omega_\epsilon} (\bm{\nabla} \bm{v})^T \tilde{\bm{D}}\bm{\nabla} \bm{\psi} (\bm{x}) \mathrm{d\Omega}
=
\frac{J(\bm{u})}{W(\bm{u})\bar{W}}\int_{\Gamma _{out}} \bm{e} \cdot \bm{\psi} (\bm{x}) \mathrm{d\Gamma}
-\frac{J(\bm{u})}{E(\bm{u})\bar{E}}\int_{\Gamma _{in}} \bm{t} \cdot \bm{\psi} (\bm{x}) \mathrm{d\Gamma}\\
&-	\mu 
	\left(\int_{\Omega} (\frac{\tilde{\sigma}_{vm}(\phi) }{\Phi(\bm{x}) \sigma_{max}})^{p}  \mathrm{d\Omega}\right)^{\frac{1}{p} - 1}
	\int_{\Omega\backslash\Omega_\epsilon} (\frac{\tilde{\sigma}_{vm}(\phi ) }{\Phi(\bm{x}) \sigma_{max}})^{p-1}  (\frac{h(\phi) (\bm{D}\nabla \bm{u} )^T  \bm{B} (\bm{D} \bm{\nabla} \bm{\psi} (\bm{x}) ) }	{ \sigma_{max}\sqrt{(\bm{D}\nabla \bm{u} )^T  \bm{B} (\bm{D} \bm{\nabla \bm{u}} )} })	\mathrm{d\Omega} ,
	\label{eq:v}
\end{split}
\end{eqnarray} 
where $\bm{\psi} (\bm{x})$ is a test function.
In this problem, the topological derivative is given as the following equation using the tensor $\bm{A}$ defined by Otomori et al. \cite{otomori2015matlab}:
\begin{eqnarray}
d_t L =& (\bm{\nabla}\bm{v})^T h(\phi) \bm{A\nabla}\bm{u} +\lambda 
	+\mu 
	\tilde{\tilde{dG_{\sigma}}} ,
	\label{eq:dTL}
\end{eqnarray}
\begin{eqnarray}
	A_{ikjl} :=& \frac{3(1-\nu)}{2(1+\nu)(7-5\nu)}\left[ \frac{-(1-14\nu+15\nu^2)E}{(1-2\nu)^2}\delta_{ij}\delta_{kl} + 5E(\delta_{ik}\delta_{il}+\delta_{il}\delta_{jk}) \right] , \label{eq:A}
\end{eqnarray}
where $\bm{v}$ is the adjoint variable, $\lambda$ and $\mu$ are the Lagrange multipliers, and $\tilde{\tilde{dG_{\sigma}}}$ is the term for the stress constraint.

The stress is distributed locally, and the term representing the von Mises stress in the topological derivative is also local, which leads to unstable calculation.
Therefore, the term for the stress constraint in the sensitivity is updated using the sensitivity from the previous iteration.
The sensitivity in the $n$th iteration is updated using the sensitivity in the ($n-1$)th iteration as follows:
\begin{eqnarray}
d\tilde{G}_{\sigma}^{(n)} = \mu \frac{1}{p}	\left(\int_{\Omega} (\frac{\tilde{\sigma}_{vm}(\bm{u}^{(n)}) }{\Phi \sigma_{max}})^{p}  \mathrm{d\Omega}\right)^{\frac{1}{p} - 1}  
	(\frac{\tilde{\sigma}_{vm}(\bm{u}^{(n)} )}{\Phi \sigma_{max}})^{p} ,\\
\left\{ 
	\begin{array}{lll}
\tilde{\tilde{dG_{\sigma}}}^{(0)} = d\tilde{G}_{\sigma}^{(0)} \qquad \qquad \qquad \qquad \qquad \qquad \mathrm{for} \quad n=0\\
\tilde{\tilde{dG_{\sigma}}}^{(n)} = (1-w_p)d\tilde{G}_{\sigma}^{(n)}+ w_p \tilde{\tilde{dG_{\sigma}}}^{(n-1)} \qquad \mathrm{for} \quad n>0
	\end{array} ,
	\right.	
\end{eqnarray}
where $\tilde{\tilde{dG_{\sigma}}}^{(n)}$ is the sensitivity used to update the level set function for the $n$th iteration, $\bm{u}^{(n)}$ is the displacement used to calculate the sensitivity for the $n$th iteration, and $w_p$ is the ratio of using the previous sensitivity to the present sensitivity.
In this paper, $w_p$ is set to 0.9.

\section{Numerical examples}
\label{sec:example}
In this section, several numerical examples are presented to demonstrate the utility and validity of the proposed method.
First, we present numerical examples of the displacement inverter used as a benchmark in previous studies \cite{sigmund1997design, YAMADA201717-00453}.
Furthermore, we present numerical examples of displacement magnification mechanisms using the proposed method.

\subsection{Displacement inverter}
\label{subsec:inv}

\begin{figure*}[htbp]
	\begin{center}
		\includegraphics[height=5.5cm]{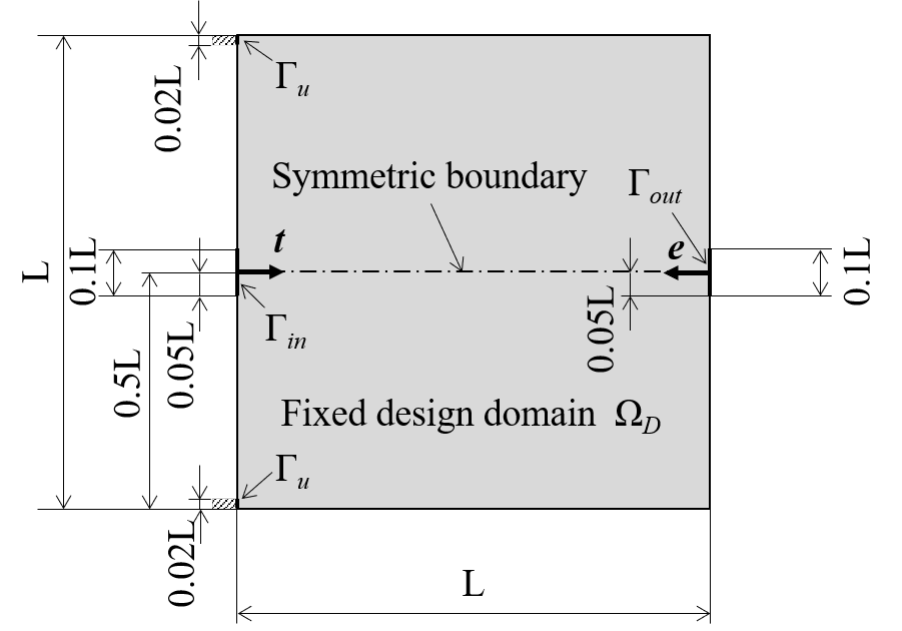}
		\caption{Design settings for displacement inverter}
		\label{fig:inv_cond}
	\end{center}
\end{figure*}

\begin{table}[htbp]
 \caption{Parameters for the design of displacement inverters}
 \label{tab:inv_param}
 \centering
 \begin{tabular}{cccccccc}
 \hline
$K$ & $C$ & $\tau$ & $p$ & $d$ & $w_p$& $V_{max}$&$\sigma_{max}$[Pa]\\
\hline
1.0 &  0.8 & 5.0$\times 10^{-5}$ & 2.0 & 0.01 & 0.9 &0.3 & $2.0\times 10^7$\\ 
 \hline
 \end{tabular}
\end{table}

Figure \ref{fig:inv_cond} illustrates the fixed design domain and the boundary conditions of the displacement inverter problem.
Parameters for optimization are listed in Table \ref{tab:inv_param}.
The fixed design domain is L $\times$ L domain that has a symmetric boundary at the center between the top and bottom parts.
We set the bottom-left and top-left boundaries of width 0.02L as the fixed boundary $\Gamma_{u}$, the center-left boundary of width 0.1L as the input port $\Gamma_{in}$, and the center-right boundary of width 0.1L as the output port $\Gamma_{out}$.
In this section, L is set to 1.0 m.
The input load vector $\bm{t}$ is a right--direction vector of size $1.0\times 10^7$ Pa, while the output vector $\bm{e}$ is a left--direction vector of size $1.0$.
In these examples, the isotropic linear elastic material has a Young's modulus set to $210$ GPa and a Poisson ratio set to 0.3.
The upper limit of the volume constraint $V_{max}$ is set to 0.3, while the upper limit of the von Mises stress $\sigma_{max}$ is set to $2.0\times 10^7$ Pa.
The parameter $p$ for the p-norm is set to 2.0, because preliminary numerical experiments showed that the objective function oscillates for large values of $p$. 
In addition, the regularization parameter $\tau$ is set to $5.0\times 10^{-5}$.
The fixed design domain is discretized using a structural mesh and three-node triangular plane stress elements whose length is $2.5 \times 10^{-3}$ L.
The initial value of the level set function is set to 1.0 in the fixed design domain, which signifies that the fixed design domain consists entirely of the material domain.
The representative displacements $\bm{U}_{out}$ and $\bm{U}_{in}$ are set to the displacement of the initial structure at $\Gamma_{out}$ and $\Gamma_{in}$, respectively.
We define the evaluation functions $U_o$ and $U_i$ to evaluate the displacement at the output and input ports as follows:
\begin{eqnarray}
\begin{split}
U_o := \frac{\int _{\Gamma _{out}} \bm{e} \cdot \bm{u} \, \mathrm{d\Gamma}}{\int _{\Gamma _{out}} \mathrm{d\Gamma}} ,\\
U_i := \frac{\int _{\Gamma _{in}} \bm{t} \cdot \bm{u} \, \mathrm{d\Gamma}}{ \int _{\Gamma _{in}} \sqrt{\bm{t} \cdot \bm{t}} \, \mathrm{d\Gamma}}.
\end{split}
\end{eqnarray}

\begin{figure*}[htbp]
	\begin{center}
		\includegraphics[height=4.0cm]{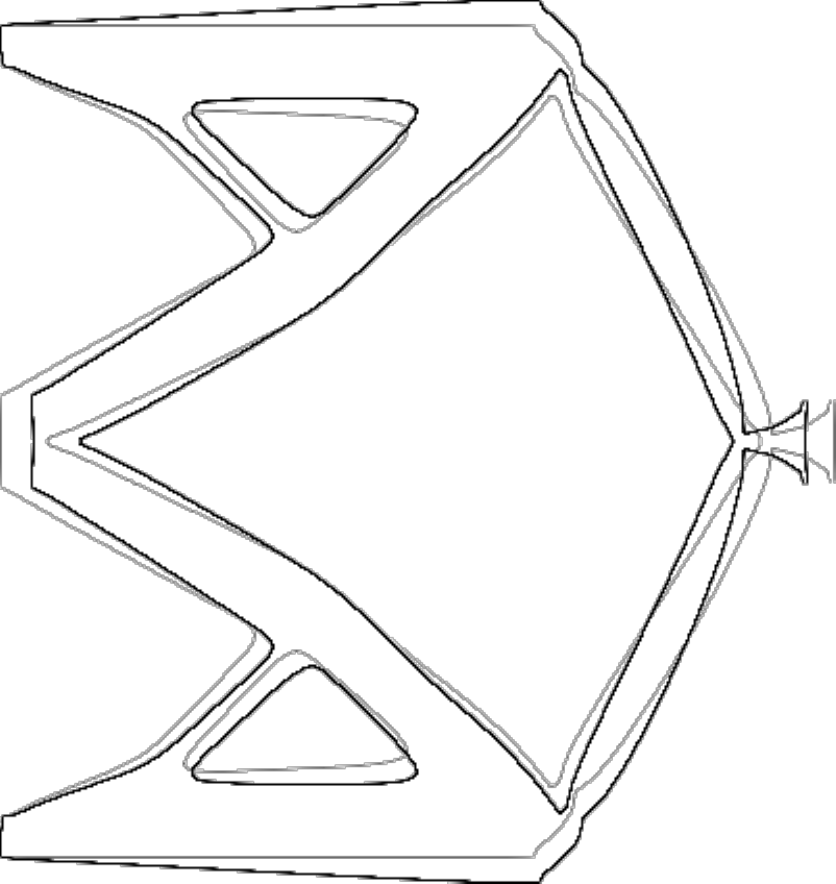}\qquad \qquad
		\includegraphics[height=4.0cm]{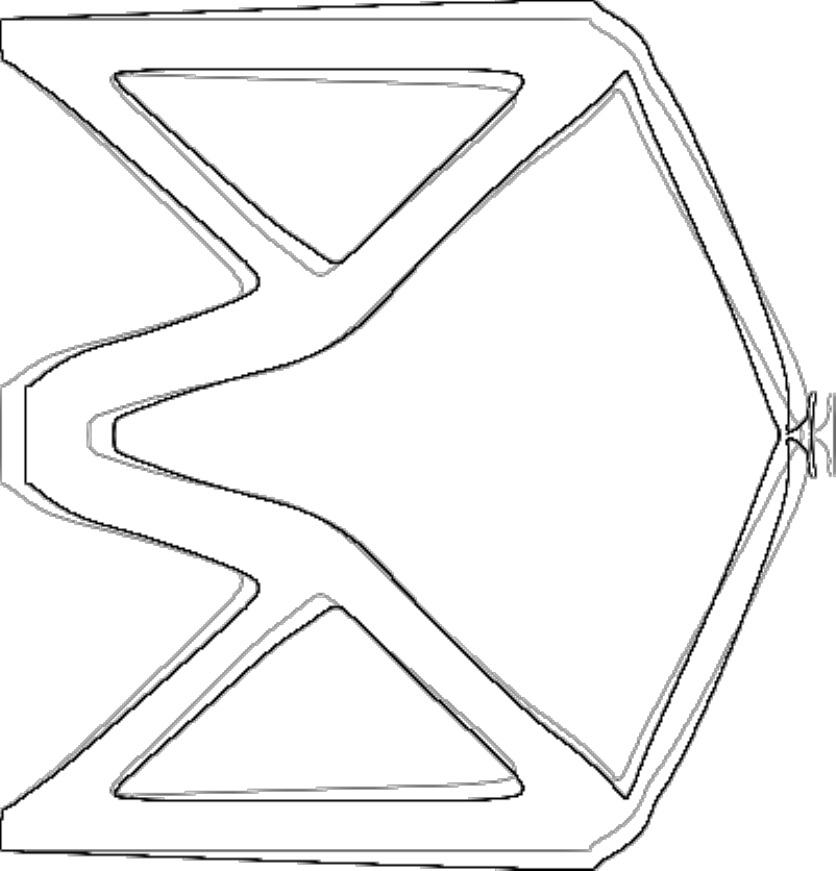}
		\caption{Deformation of displacement inverter obtained in condition (m) (left) and (o) (right) (enhanced by a factor of 500)}
		\label{fig:def_inv_mu}
	\end{center}
\end{figure*}

\begin{figure}[htbp]
\begin{center}
  \begin{tabular}{ccccc}
   && $\mu = 0\quad$ & $\mu = 0.1\quad$ & $\mu = 0.3\quad$ \\
	\rotatebox[origin=r]{90}{$\alpha = 1.0 \qquad$}
	&
	\rotatebox[origin=r]{90}{$\beta = 0 \qquad \;\;\:$}
	&
	\begin{minipage}[t]{0.2\hsize}
	\subfigure[]{\includegraphics[height=1.9cm]{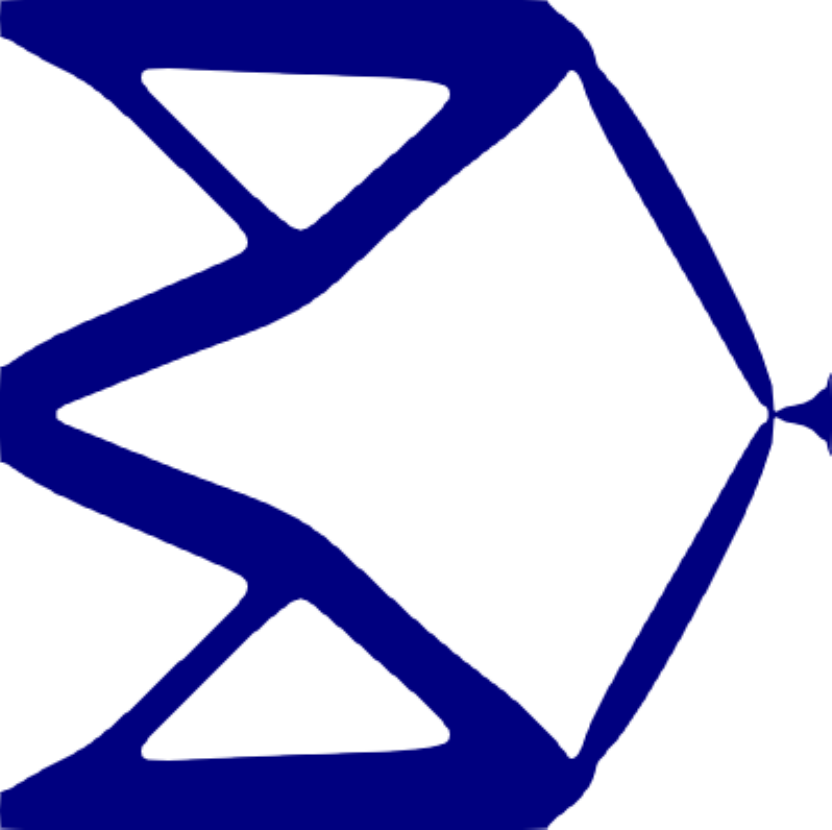} }
	\end{minipage}&
	\begin{minipage}[t]{0.2\hsize}
	\subfigure[]{\includegraphics[height=1.9cm]{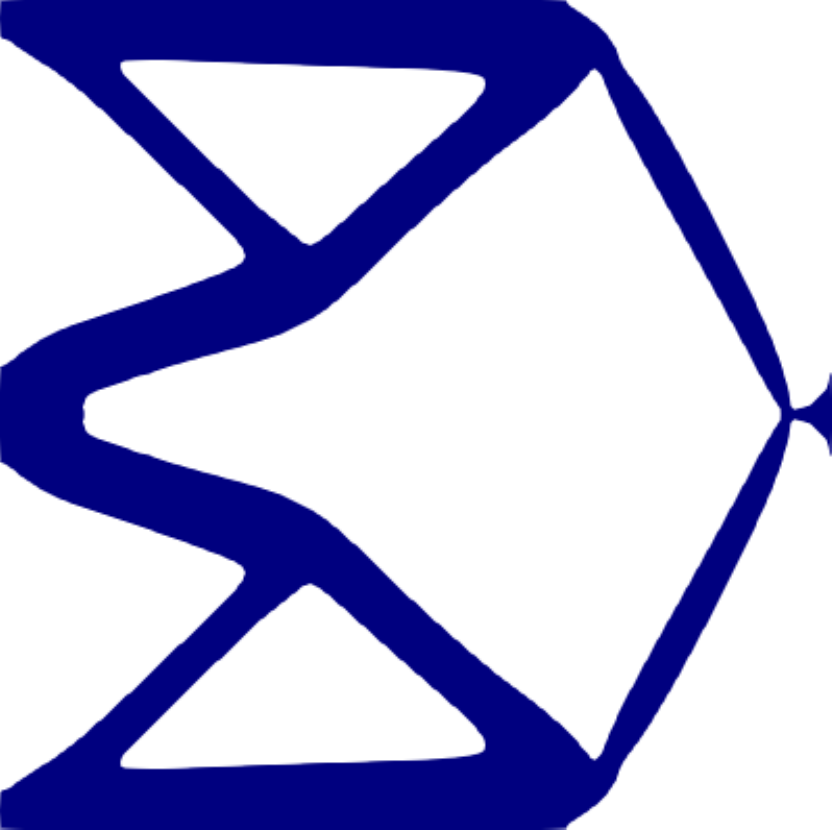} }
	\end{minipage}&
	\begin{minipage}[t]{0.2\hsize}
	\subfigure[]{\includegraphics[height=1.9cm]{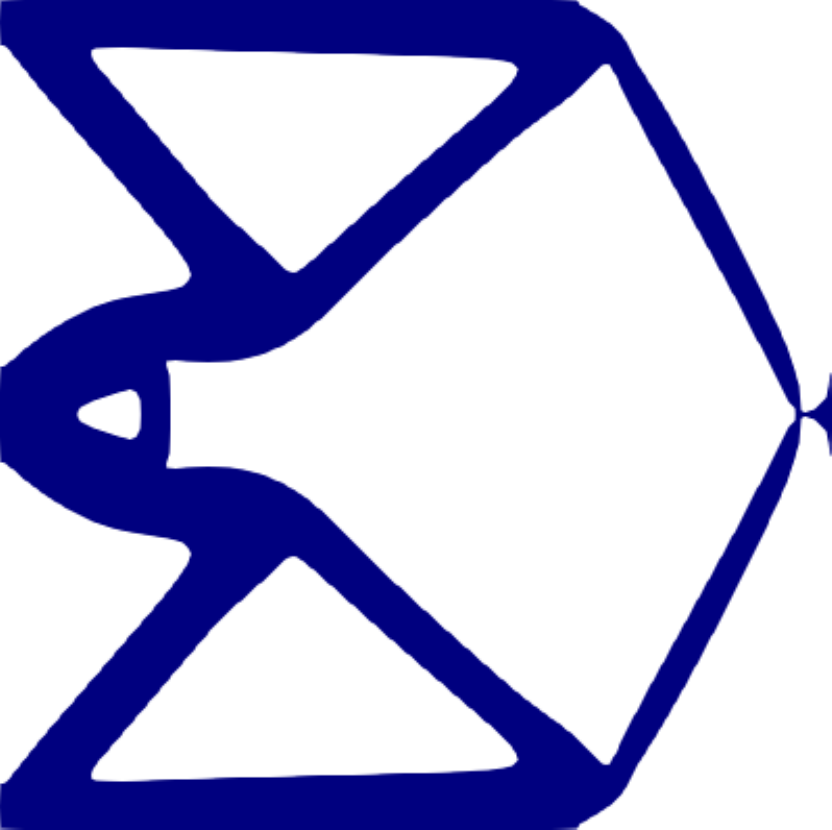} }
	\end{minipage}
	\\
	\rotatebox[origin=r]{90}{$\alpha = 1.0\qquad$}
	&
	\rotatebox[origin=r]{90}{$\beta = 0.5\qquad$}
	&
  	\begin{minipage}[t]{0.2\hsize}
	\subfigure[]{\includegraphics[height=1.9cm]{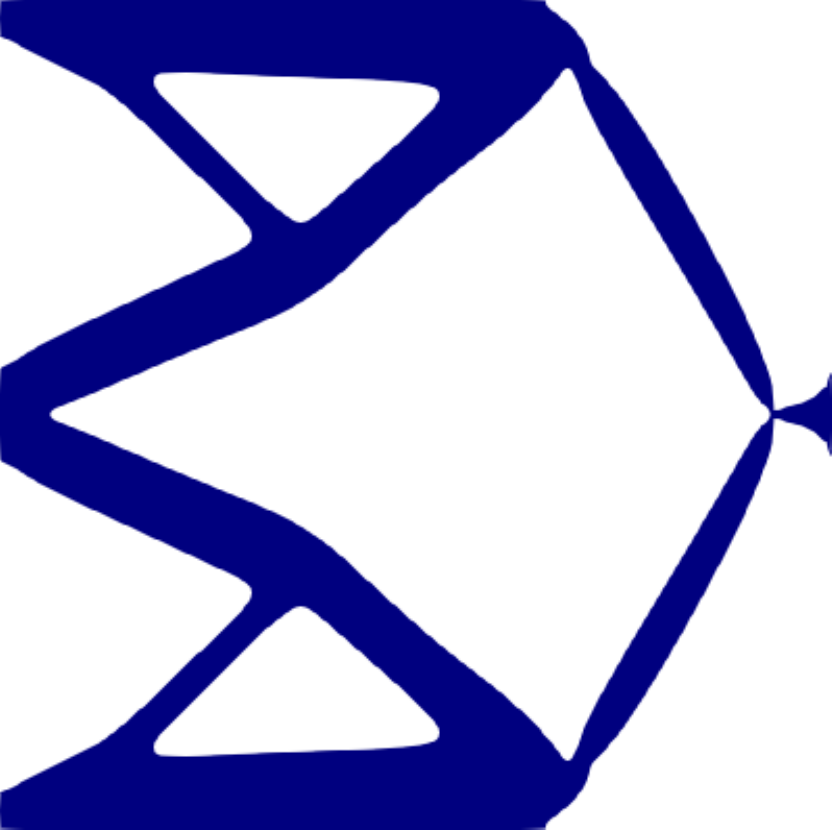} }
	\end{minipage}&
	\begin{minipage}[t]{0.2\hsize}
	\subfigure[]{\includegraphics[height=1.9cm]{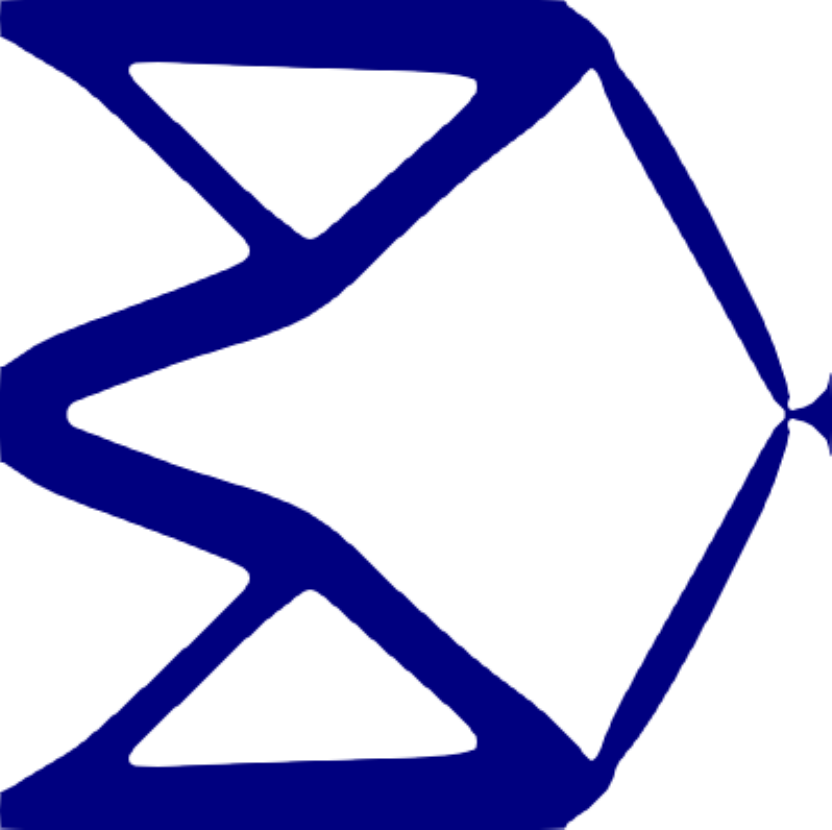} }
	\end{minipage}&
	\begin{minipage}[t]{0.2\hsize}
	\subfigure[]{\includegraphics[height=1.9cm]{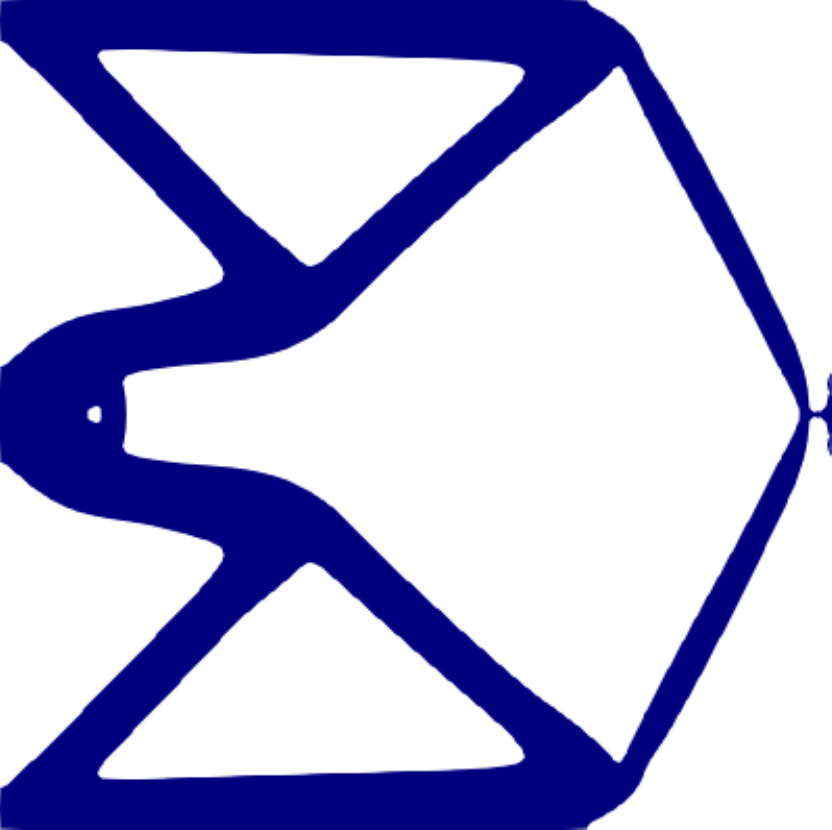} }
	\end{minipage}
	\\
	\rotatebox[origin=r]{90}{$\alpha = 1.0\qquad$}
	&
	\rotatebox[origin=r]{90}{$\beta = 1.0\qquad$}
	&
  	\begin{minipage}[t]{0.2\hsize}
	\subfigure[]{\includegraphics[height=1.9cm]{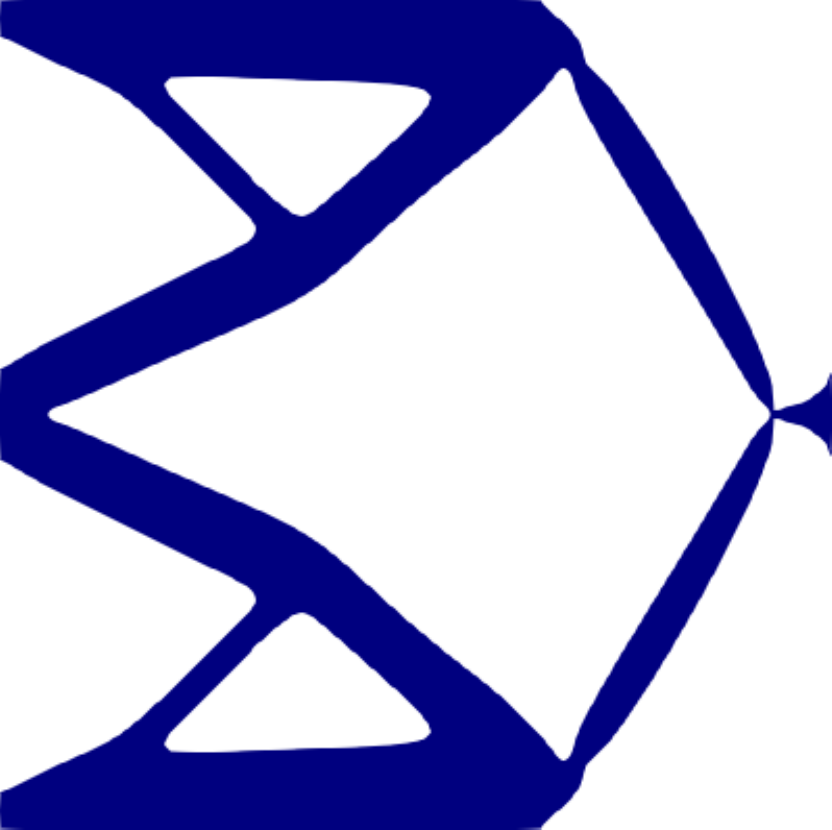} }
	\end{minipage}&
	\begin{minipage}[t]{0.2\hsize}
	\subfigure[]{\includegraphics[height=1.9cm]{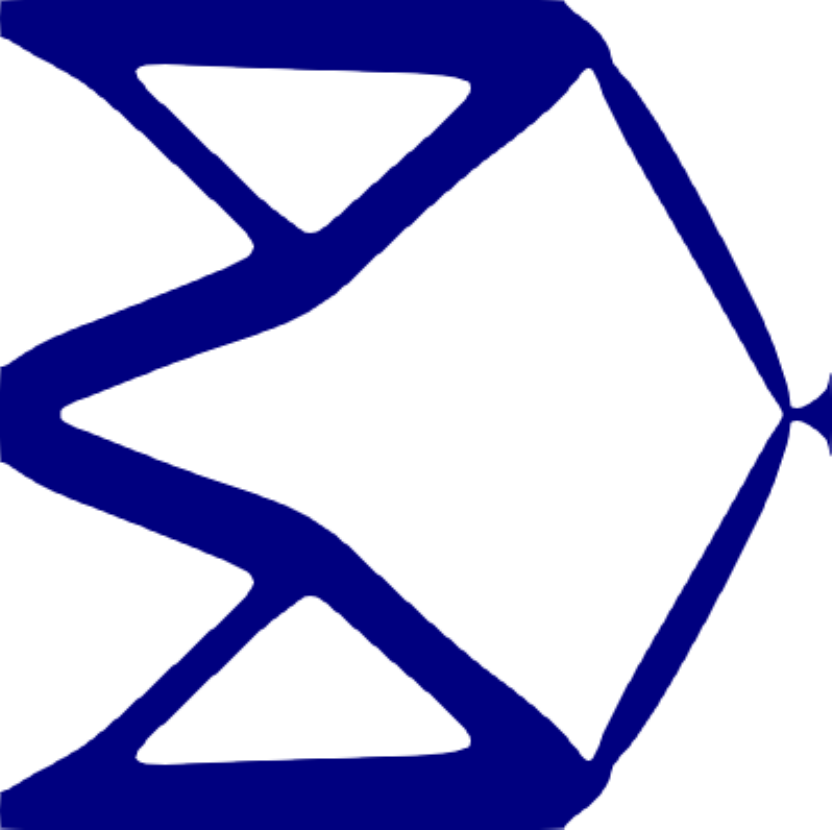} }
	\end{minipage}&
	\begin{minipage}[t]{0.2\hsize}
	\subfigure[]{\includegraphics[height=1.9cm]{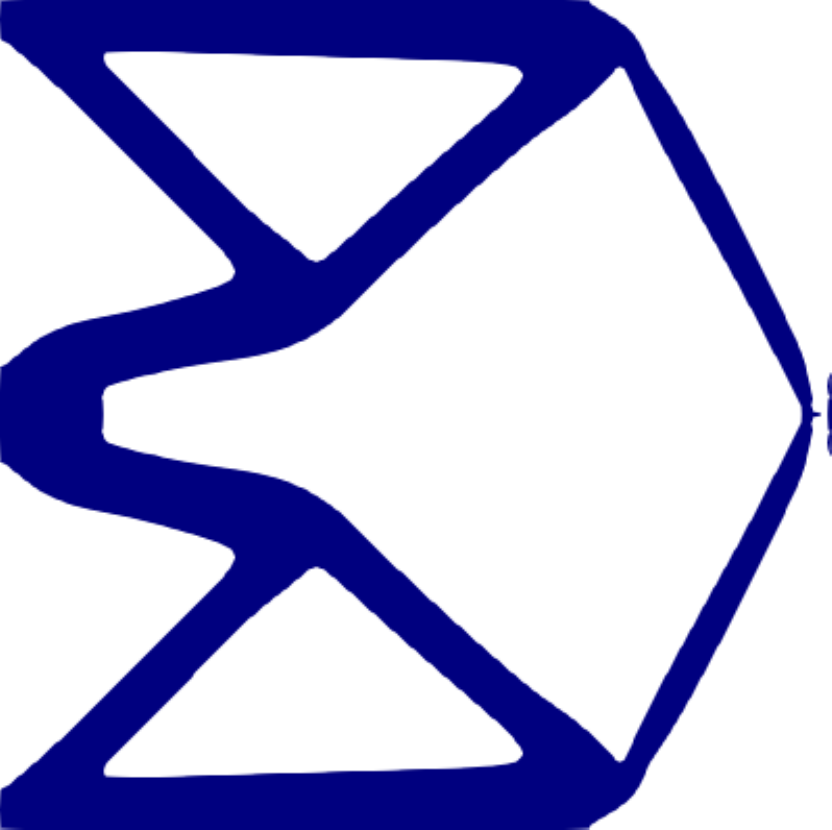} }
	\end{minipage}
	\\
	\rotatebox[origin=r]{90}{$\alpha = 0.5\qquad$}
	&
	\rotatebox[origin=r]{90}{$\beta = 1.0\qquad$}
	&
	\begin{minipage}[t]{0.2\hsize}
	\subfigure[]{\includegraphics[height=1.9cm]{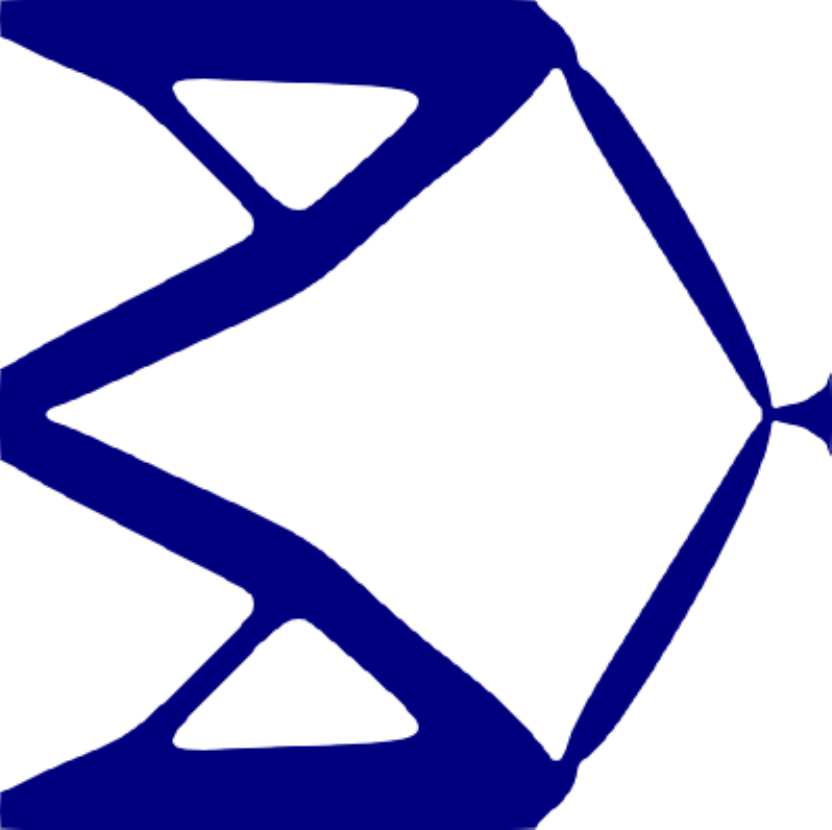} }
	\end{minipage}&
	\begin{minipage}[t]{0.2\hsize}
	\subfigure[]{\includegraphics[height=1.9cm]{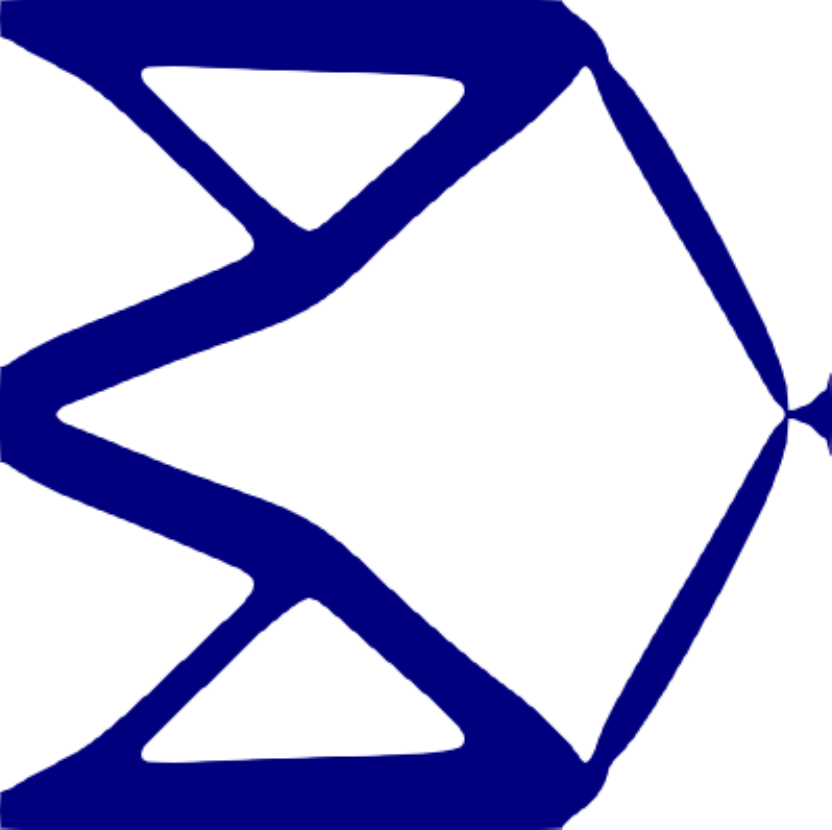} }
	\end{minipage}&
	\begin{minipage}[t]{0.2\hsize}
	\subfigure[]{\includegraphics[height=1.9cm]{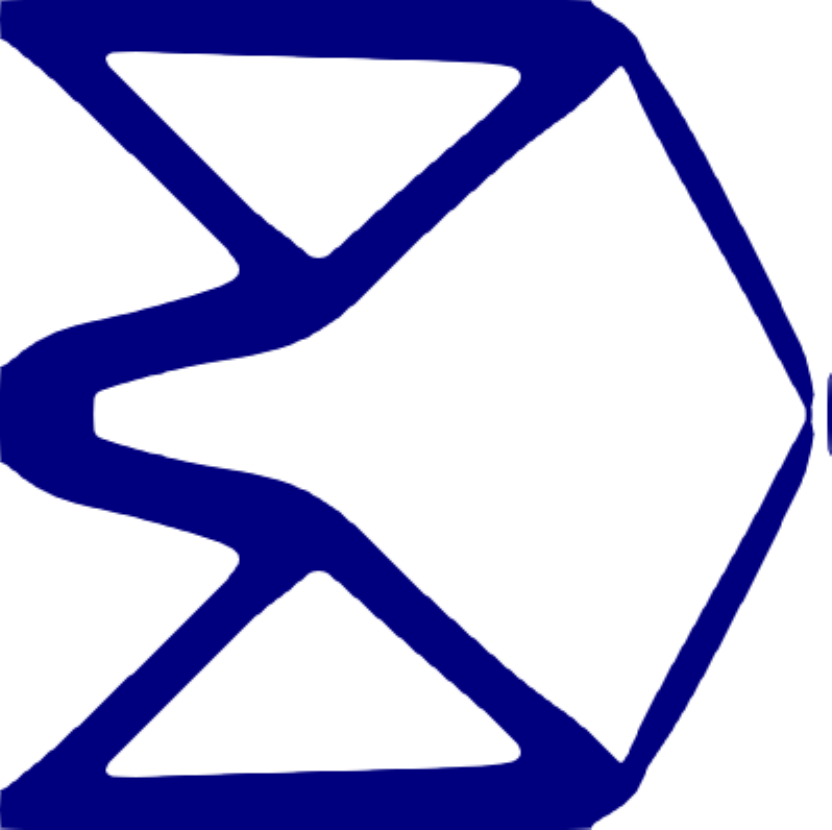} }
	\end{minipage}
	\\
	\rotatebox[origin=r]{90}{$\alpha = 0\qquad\;\;\:$}
	&
	\rotatebox[origin=r]{90}{$\beta = 1.0\qquad$}
	&
  	\begin{minipage}[t]{0.2\hsize}
	\subfigure[]{\includegraphics[height=1.9cm]{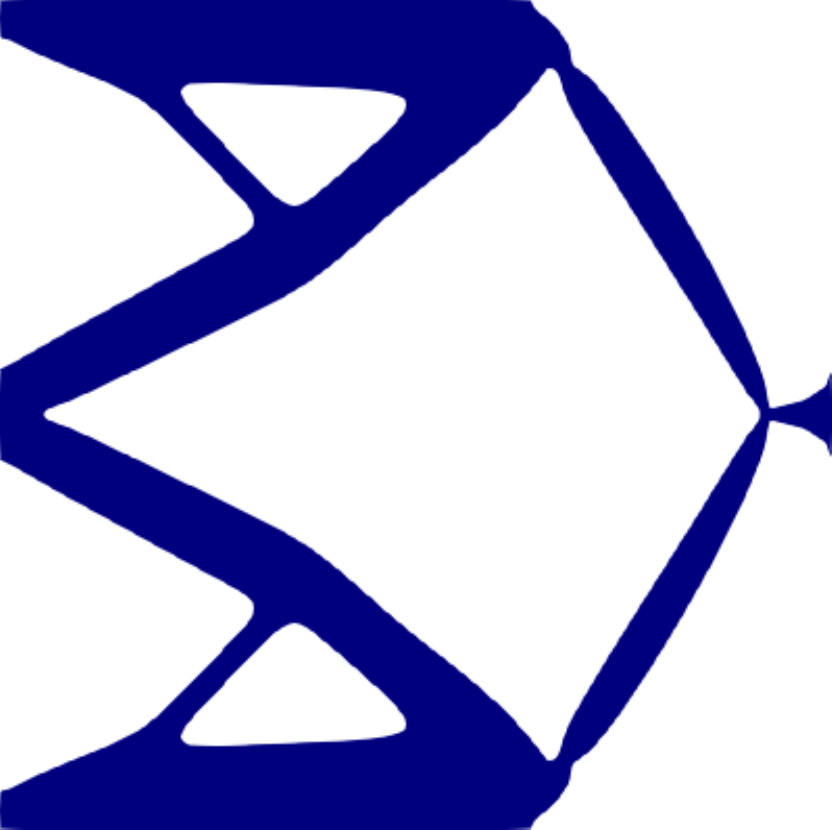} }
	\end{minipage}&
	\begin{minipage}[t]{0.2\hsize}
	\subfigure[]{\includegraphics[height=1.9cm]{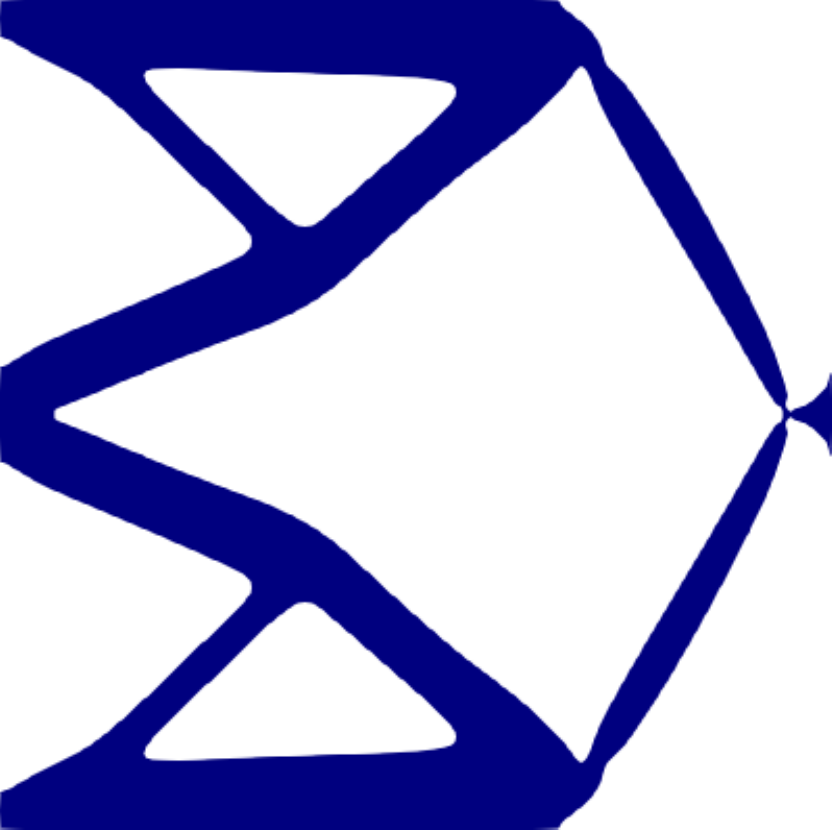} }
	\end{minipage}&
	\begin{minipage}[t]{0.2\hsize}
	\subfigure[]{\includegraphics[height=1.9cm]{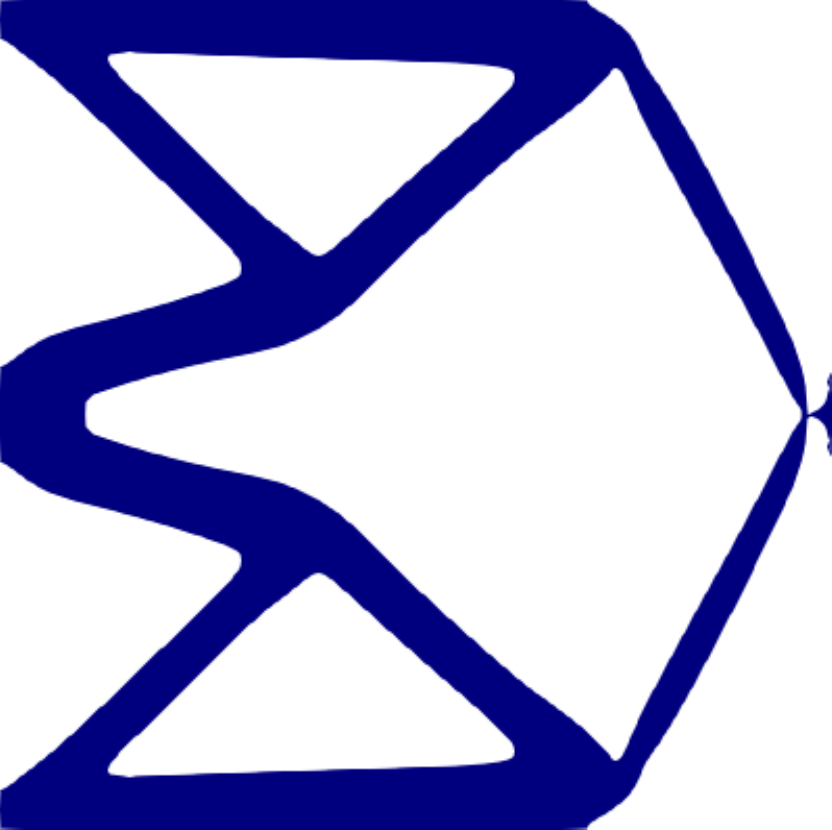} }
	\end{minipage}
  \end{tabular}
		\caption{Optimal configuration of displacement inverter for conditions (a)--(o) listed in Table \ref{tab:inv_disp}}
		\label{fig:inv_mu}
\end{center}
\end{figure}

\begin{figure}[htpb]
\begin{center}
  \begin{tabular}{cccccc}
     & & $\mu = 0\quad$ & $\mu = 0.1\quad$ & $\mu = 0.3\quad$ & \\
	\rotatebox[origin=r]{90}{$\alpha = 1.0 \qquad$}
	&
	\rotatebox[origin=r]{90}{$\beta = 0 \qquad \;\;\:$}
	&
	\begin{minipage}[t]{0.2\hsize}
	\subfigure[]{\includegraphics[height=1.9cm]{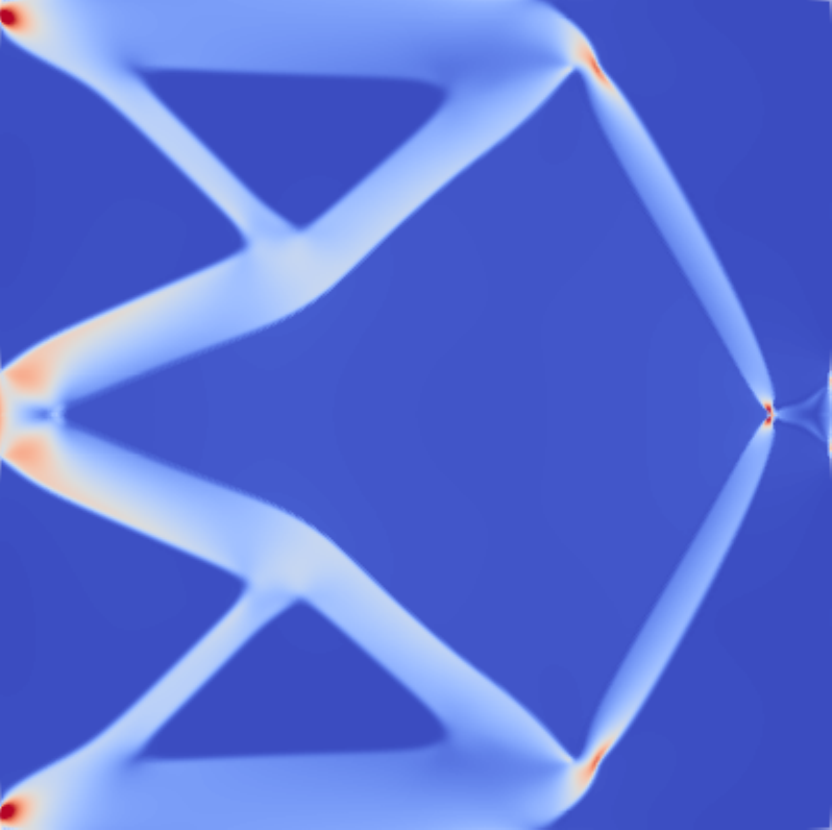} }
	\end{minipage}&
	\begin{minipage}[t]{0.2\hsize}
	\subfigure[]{\includegraphics[height=1.9cm]{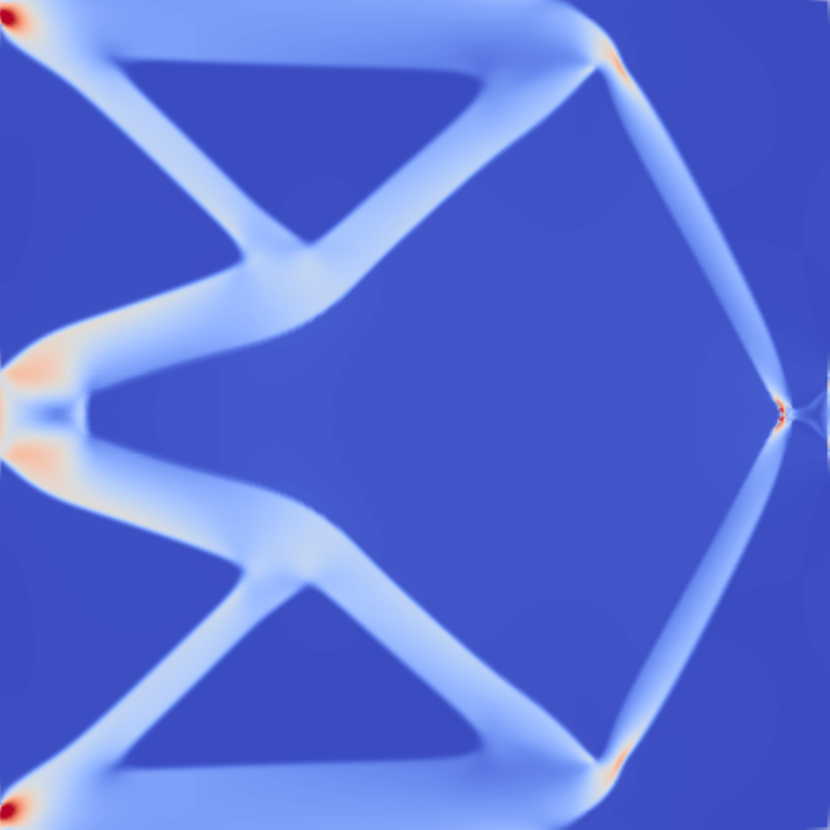} }
	\end{minipage}&
	\begin{minipage}[t]{0.2\hsize}
	\subfigure[]{\includegraphics[height=1.9cm]{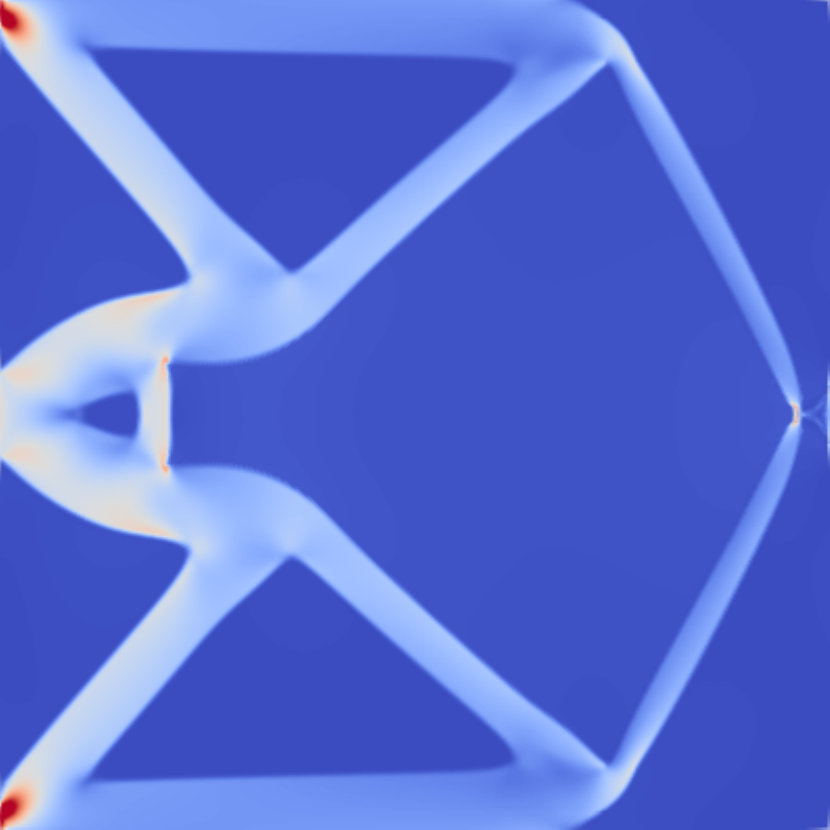} }
	\end{minipage}
	&
	\\
	\rotatebox[origin=r]{90}{$\alpha = 1.0 \qquad$}
	&
	\rotatebox[origin=r]{90}{$\beta = 0.5 \qquad$}
	&
  	\begin{minipage}[t]{0.2\hsize}
	\subfigure[]{\includegraphics[height=1.9cm]{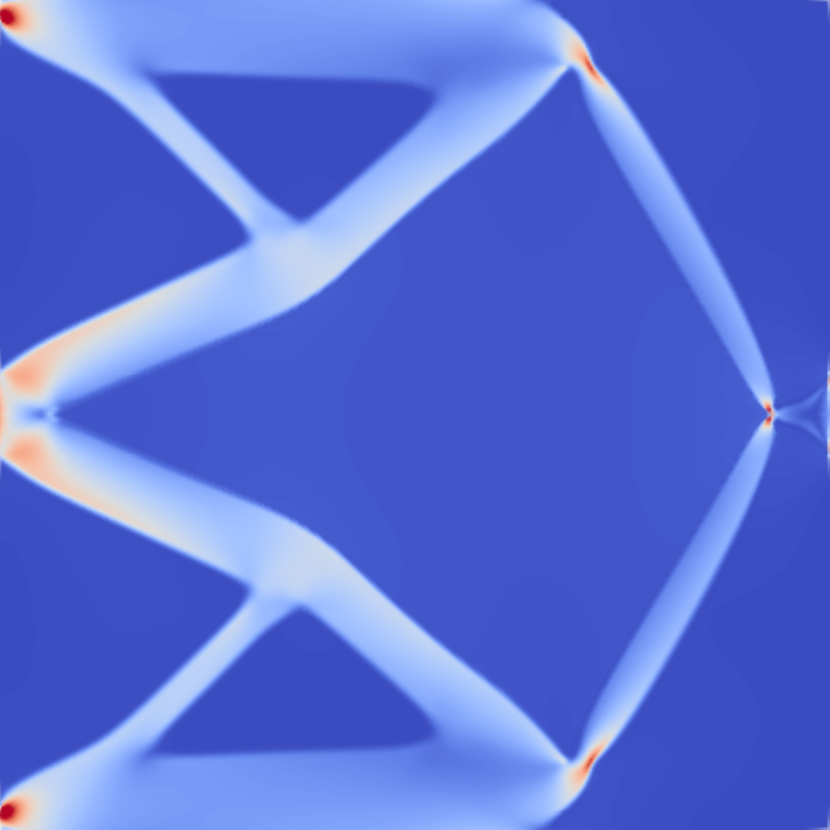} }
	\end{minipage}&
	\begin{minipage}[t]{0.2\hsize}
	\subfigure[]{\includegraphics[height=1.9cm]{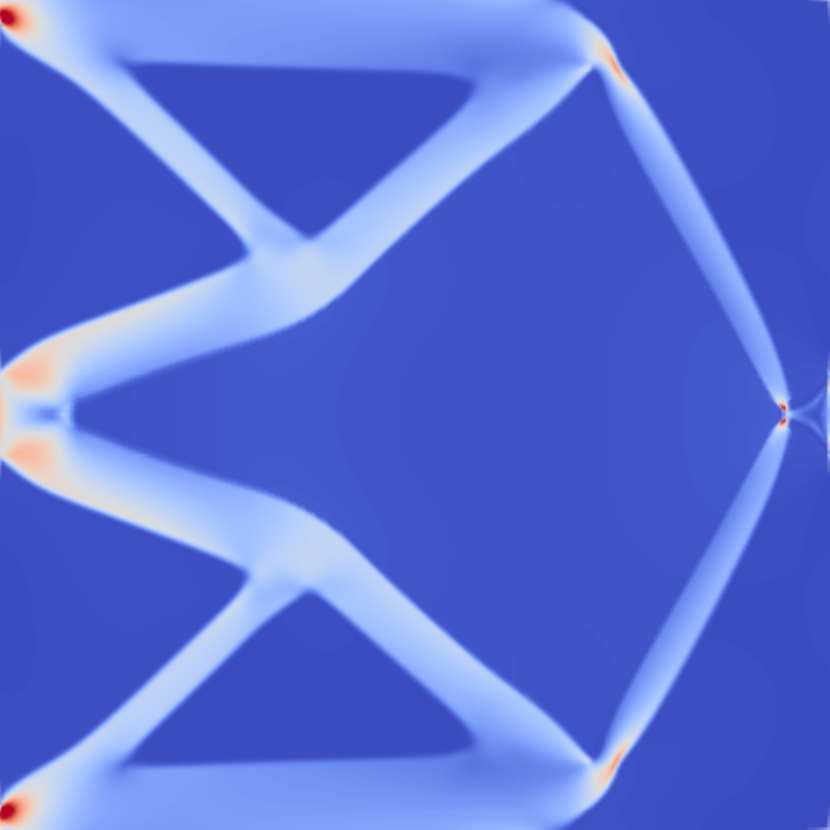} }
	\end{minipage}&
	\begin{minipage}[t]{0.2\hsize}
	\subfigure[]{\includegraphics[height=1.9cm]{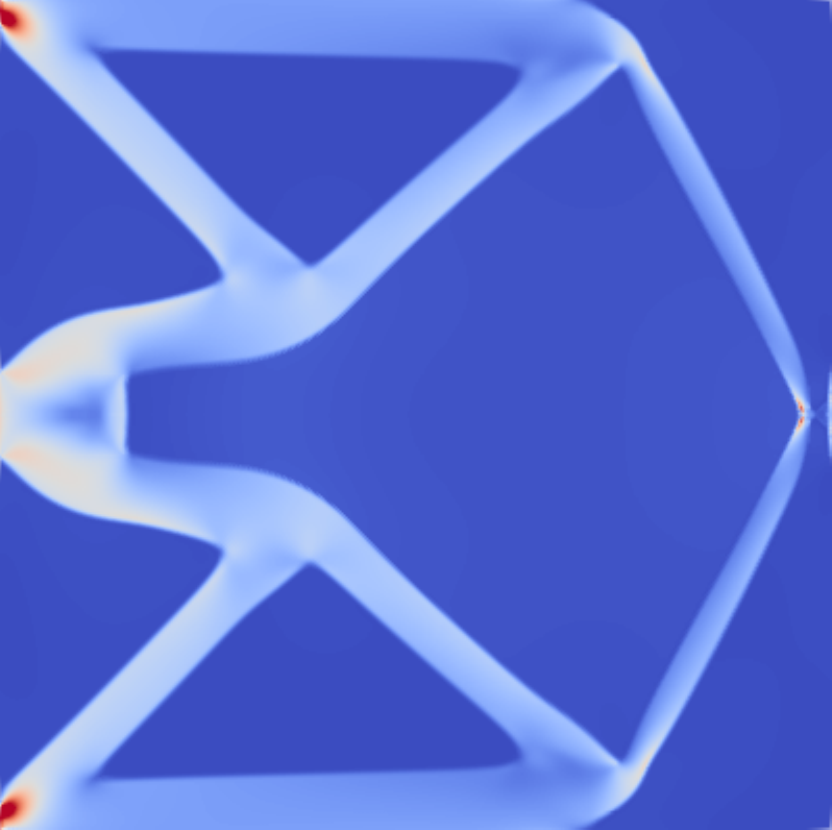} }
	\end{minipage}
	&
	\\
	\rotatebox[origin=r]{90}{$\alpha = 1.0 \qquad$}
	&
	\rotatebox[origin=r]{90}{$\beta = 1.0 \qquad$}
	&
  	\begin{minipage}[t]{0.2\hsize}
	\subfigure[]{\includegraphics[height=1.9cm]{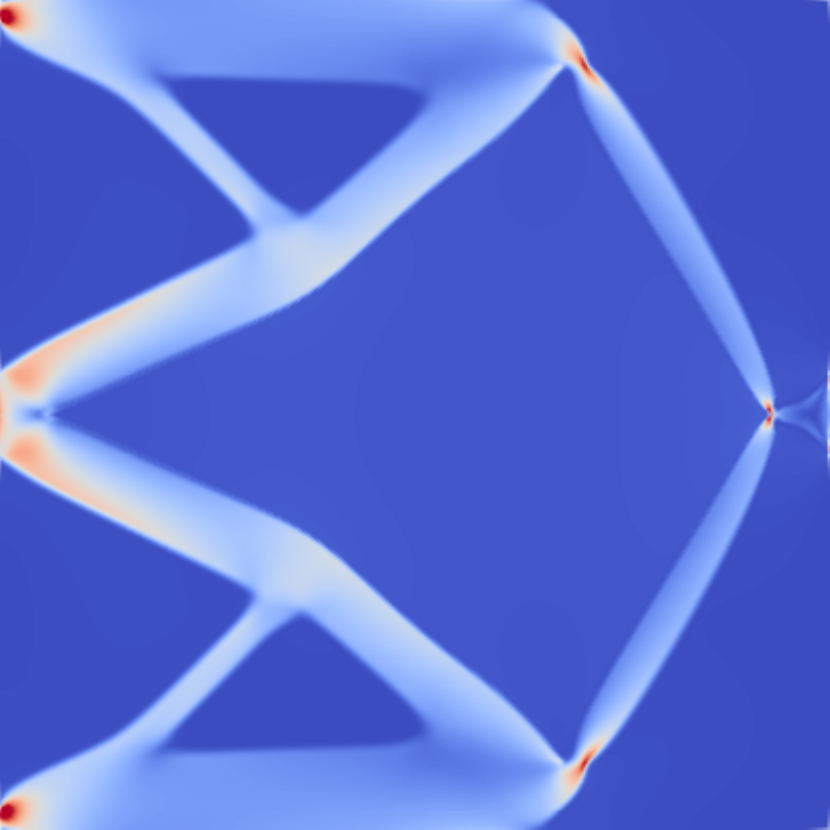} }
	\end{minipage}&
	\begin{minipage}[t]{0.2\hsize}
	\subfigure[]{\includegraphics[height=1.9cm]{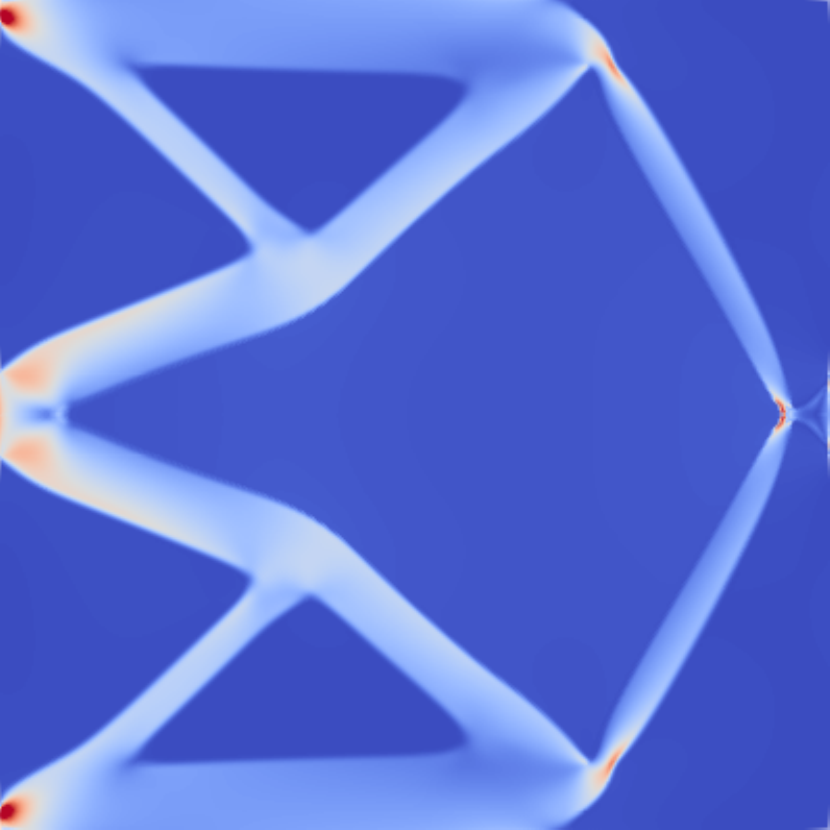} }
	\end{minipage}&
	\begin{minipage}[t]{0.2\hsize}
	\subfigure[]{\includegraphics[height=1.9cm]{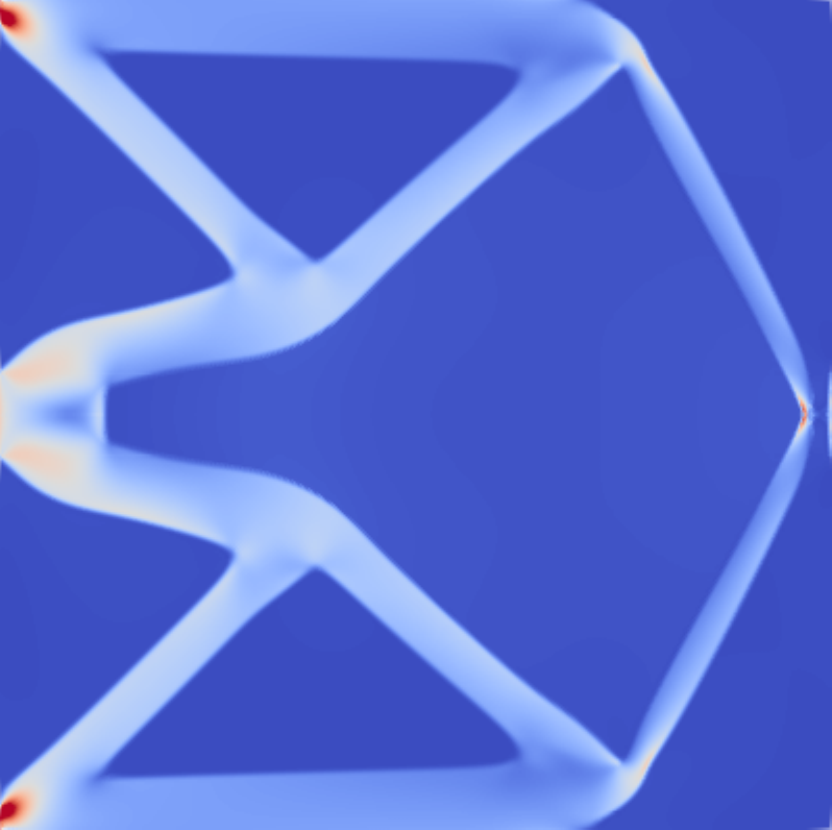} }
	\end{minipage}&
	\multirow{5}{*}{
	\begin{minipage}[t]{0.2\hsize}
  		\includegraphics[height=4.0cm]{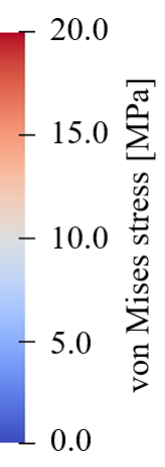}
	\end{minipage}}
	\\
	\rotatebox[origin=r]{90}{$\alpha = 0.5 \qquad$}
	&
	\rotatebox[origin=r]{90}{$\beta = 1.0 \qquad$}
	&
  	\begin{minipage}[t]{0.2\hsize}
	\subfigure[]{\includegraphics[height=1.9cm]{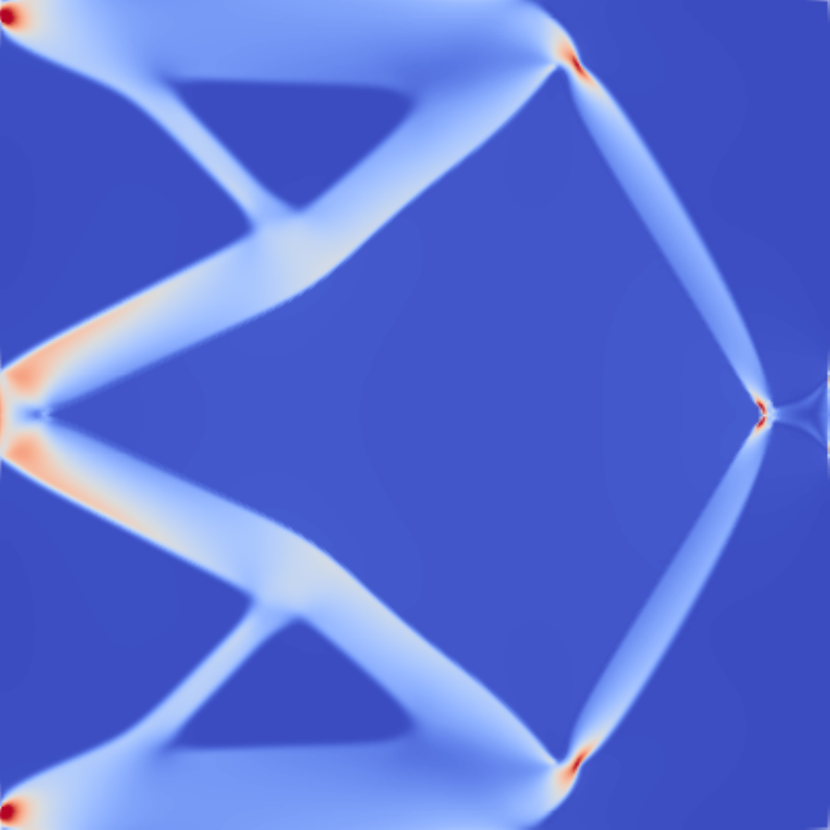} }
	\end{minipage}&
	\begin{minipage}[t]{0.2\hsize}
	\subfigure[]{\includegraphics[height=1.9cm]{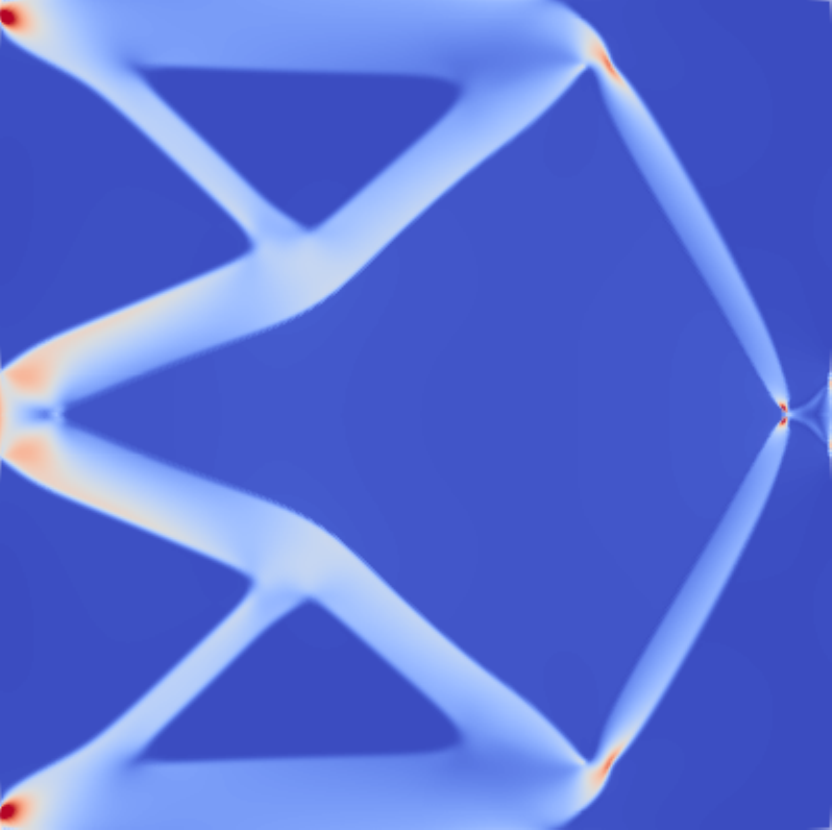} }
	\end{minipage}&
	\begin{minipage}[t]{0.2\hsize}
	\subfigure[]{\includegraphics[height=1.9cm]{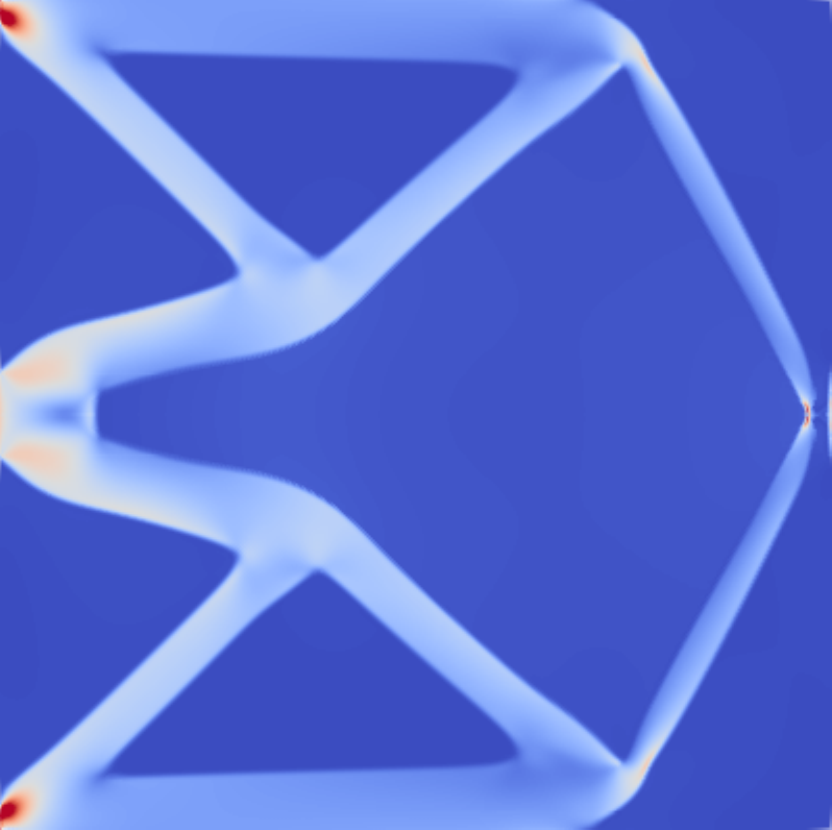} }
	\end{minipage}
	&
	\\
	\rotatebox[origin=r]{90}{$\alpha = 0 \qquad \;\;\:$}
	&
	\rotatebox[origin=r]{90}{$\beta = 1.0 \qquad$}
	&
  	\begin{minipage}[t]{0.2\hsize}
	\subfigure[]{\includegraphics[height=1.9cm]{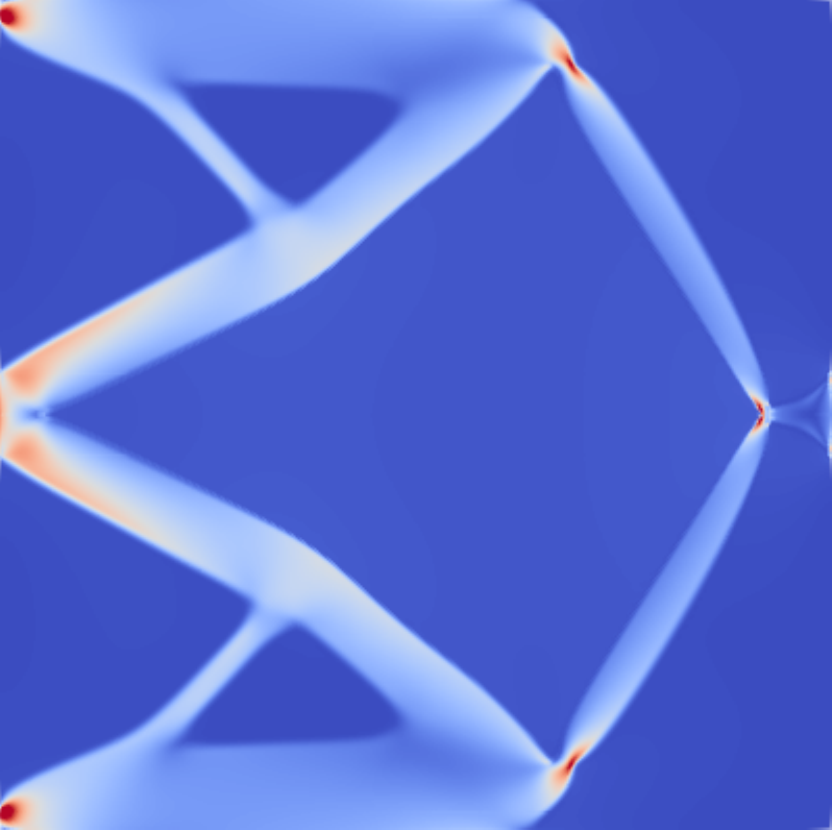} }
	\end{minipage}&
	\begin{minipage}[t]{0.2\hsize}
	\subfigure[]{\includegraphics[height=1.9cm]{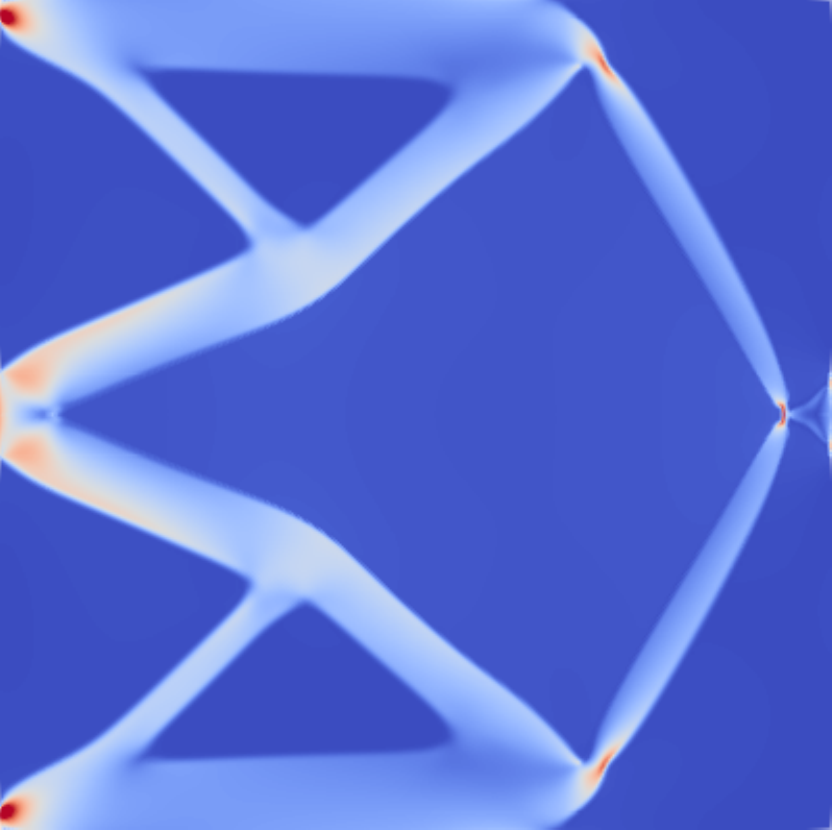} }
	\end{minipage}&
	\begin{minipage}[t]{0.2\hsize}
	\subfigure[]{\includegraphics[height=1.9cm]{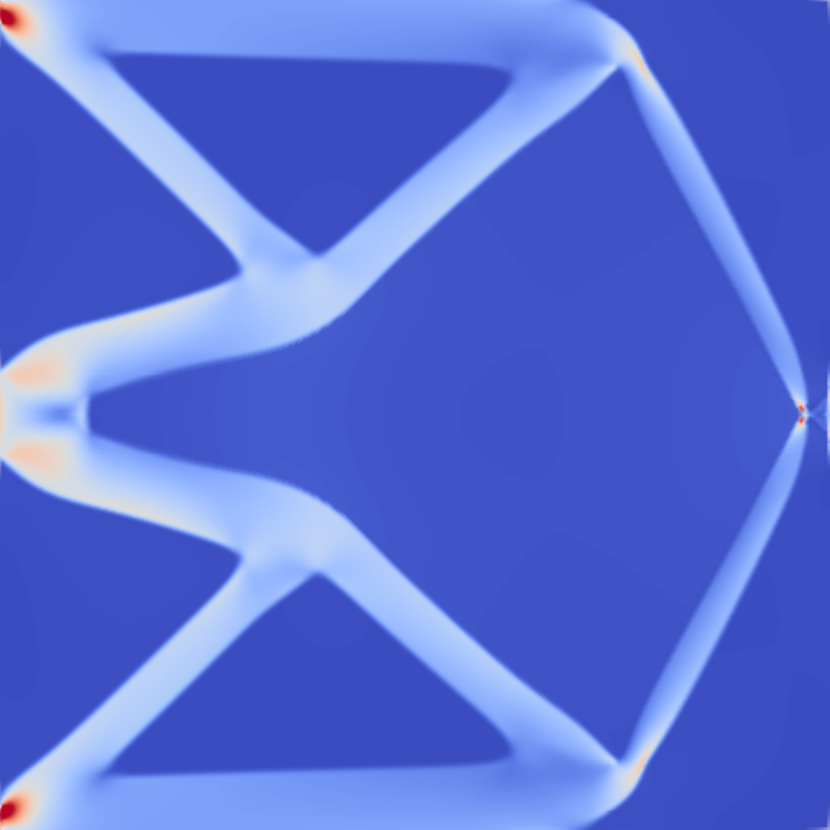} }
	\end{minipage}&
  \end{tabular}
		\caption{Von Mises stress of displacement inverter for conditions (a)--(o) listed in Table \ref{tab:inv_disp}}
		\label{fig:mises_inv_mu}
\end{center}
\end{figure}

\begin{figure}
\begin{center}
\includegraphics[height = 3.0cm]{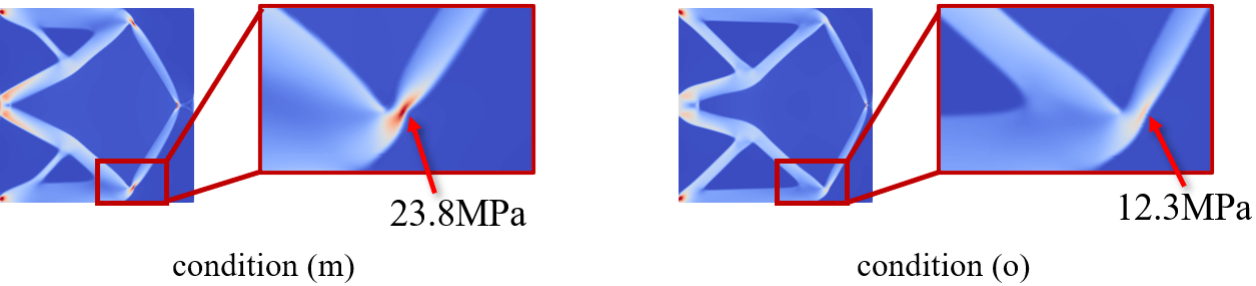}
\caption{Enlarged view of the von Mises stress concentration area of displacement inverters obtained in conditions (m)(left) and (o)(right)}
\label{fig:inv_conce}
\end{center}
\end{figure}

In this subsection, we examine the effect of the Lagrange multiplier $\mu$ on the stress constraint and the parameters $\alpha$ and $\beta$ in the objective function on the resulting optimal configurations.
We set 15 conditions as listed in Table \ref{tab:inv_disp}.
Figure \ref{fig:def_inv_mu} presents the deformation diagram of conditions (m) and (o), while Figure \ref{fig:inv_mu} illustrates the optimal configuration of the displacement inverter for each condition.
In Figure \ref{fig:inv_mu}, the dark blue domain represents the material domain.
Figure \ref{fig:mises_inv_mu} presents the distribution of the von Mises stress for each condition.
Figure \ref{fig:inv_conce} presents the enlarged view of the distribution of the von Mises stress at the same area in conditions (m) and (o).

As illustrated in Table \ref{tab:inv_disp}, $U_o$ is positive in all conditions, whereas it is negative in the initial structure.
This indicates that output port $\Gamma_{out}$ is displaced in the direction of output vector $\bm{e}$.
First, we discuss the effects of $\alpha$ and $\beta$.
For the same value of $\mu$, the values of $U_o$ and $U_i$ are smaller when $\alpha > \beta$ than when $\alpha = \beta$.
This indicates that the condition $\alpha > \beta$ increases the denominator of the objective function; that is, the optimization procedure tends to maximize the stiffness against the input force.
Furthermore, for the same value of $\mu$, $U_o$ and $U_i$ increasing when $\beta > \alpha$.
This signifies that the condition $\beta > \alpha$ tends to increase the numerator of the objective function; that is, the optimization procedure tends to maximize the output displacement.
The above results indicate that the effects of parameters $\alpha$ and $\beta$ are as assumed in Section \ref{subsec:prop}.

Next, we discuss the effect of $\mu$.
As illustrated in Figure \ref{fig:mises_inv_mu} and \ref{fig:inv_conce}, in conditions (a), (d), (g), (j), and (m), where $\mu$ is set to 0, the von Mises stress is concentrated near the upper and lower central parts of the structure because the stress is not constrained in these conditions.
However, as $\mu$ increases, the stress concentration in the upper and lower central parts is no longer observed due to the increasing influence of the stress constraints.
This indicates that proposed stress constraint is effective. 
In addition, as illustrated in Figure \ref{fig:def_inv_mu}, both values of $U_o$ and $U_i$ are smaller in condition (o) than in condition (m).
The value of $\mu$ in condition (o) is larger than in condition (m). Other parameters are the same between condition (m) and (o).
This indicates that $U_o$ and $U_i$ decrease as $\mu$ increases.
The above illustrates that imposing a stress constraint results in increasing the stiffness of the obtained structure.

For displacement inverter problem, $\beta$ should be 1.0 to obtain a larger output displacement, and $\mu$ should be greater than 0.3 to avoid hinging. 
Furthermore, if it is necessary to obtain stiffness, $\alpha$ should be increased; otherwise, $\alpha$ should be reduced.

\begin{figure}[htbp]
\begin{center}
\includegraphics[height=11.5cm]{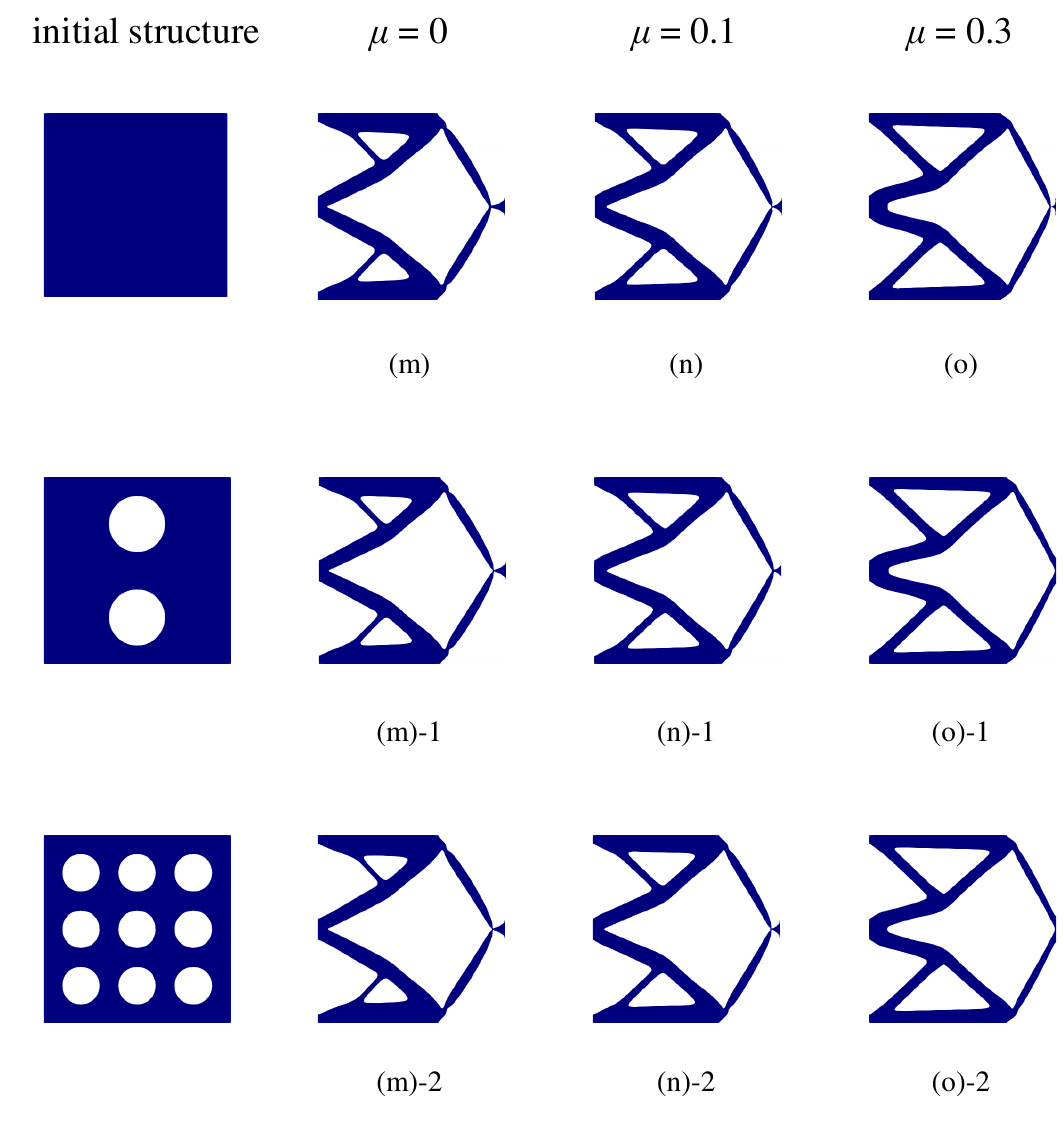}
		\caption{Initial structure dependence on the optimal configuration of displacement inverter ($\alpha = 0, \beta = 1.0$)}
		\label{fig:inv_init}
\end{center}
\end{figure}

Third, we discuss the dependence on the initial structure.
Figure \ref{fig:inv_init}  illustrates the optimal configuration of the displacement inverter obtained in conditions (m), (n), and (o) from several different initial structures.
As illustrated in Figure \ref{fig:inv_init}, the obtained structures are more dependent on the initial structure when $\mu$ becomes greater. 
However, the initial dependence on the displacement inverter problem is quite small.

\begin{table}[H]
 \caption{Displacement at output and input ports of displacement inverters}
 \label{tab:inv_disp}
 \centering
 \begin{tabular}{cccccc}
 \hline
 & & & & \multicolumn{2}{c}{displacement [$\mu$m]}\\
 condition&$\alpha$& $\beta$& $\mu$ & $\;\;\ U_{o}\;\;$ &$U_{i}$\\
 \hline
 initial structure & & & & $-$4.7 & 8.6 \\
 a &1.0&0& 0 & 30.4 & 35.6\\
 b &1.0&0& 0.1 & 26.9 & 32.7\\
 c &1.0&0& 0.3 & 19.8 & 27.2\\
 d &1.0&0.5& 0 & 31.5 & 36.9\\
 e &1.0&0.5& 0.1 & 28.7 & 34.0\\
 f &1.0&0.5& 0.3 & 22.4 & 29.5\\
 g &1.0&1.0& 0 & 32.5 & 38.0\\
 h &1.0&1.0& 0.1 & 29.1 & 34.7\\
 i &1.0&1.0& 0.3 & 23.6 & 30.5\\
 j &0.5&1.0& 0 & 32.7 & 38.7\\
 k &0.5&1.0& 0.1 & 30.2 & 35.4\\
 l &0.5&1.0& 0.3 & 24.5 & 31.1\\
 m &0&1.0& 0 & 33.4 & 39.6\\
 n &0&1.0& 0.1 & 30.8 & 36.1\\
 o &0&1.0& 0.3 & 25.2 & 31.5\\
 \hline
 \end{tabular}
\end{table}

\subsection{Displacement magnification mechanisms}

\begin{figure*}[htbp]
	\begin{center}
		\includegraphics[height=5.5cm]{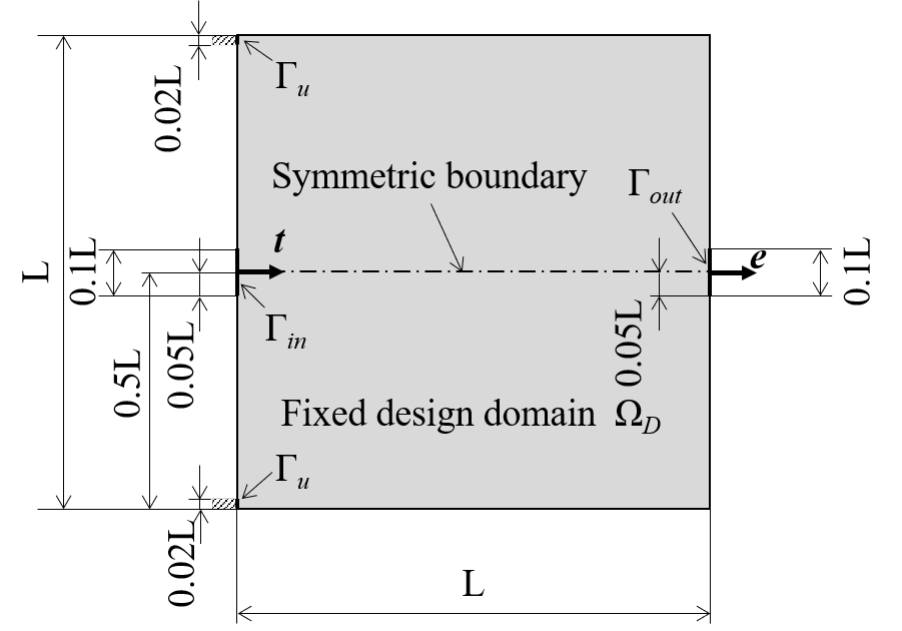}
		\caption{Design settings for displacement magnification mechanism}
		\label{fig:magni_cond}
	\end{center}
\end{figure*}

Figure \ref{fig:magni_cond} displays the fixed design domain and boundary conditions of the displacement magnification problem.
Parameters for optimization are listed in Table \ref{tab:magni_param}.
The output vector $\bm{e}$ is a right--direction vector of size $1.0$.
The fixed design domain is discretized using a structural mesh and three-node triangular plane stress elements whose length is $5.0 \times 10^{-3}$ L.
The other conditions are the same as in the inverter problem.

\begin{table}[htbp]
 \caption{Parameters for the design of displacement magnification mechanisms}
 \label{tab:magni_param}
 \centering
 \begin{tabular}{cccccccc}
 \hline
$K$ & $C$ & $\tau$ & $p$ & $d$ & $w_p$& $V_{max}$&$\sigma_{max}$[Pa]\\
\hline
1.0 &  0.8 & 5.0$\times 10^{-5}$ & 2.0 & 0.01 & 0.9 &0.3 & $2.0\times 10^7$\\ 
 \hline
 \end{tabular}
\end{table}

\begin{figure*}[hbtp]
	\begin{center}
			\includegraphics[height=4.0cm]{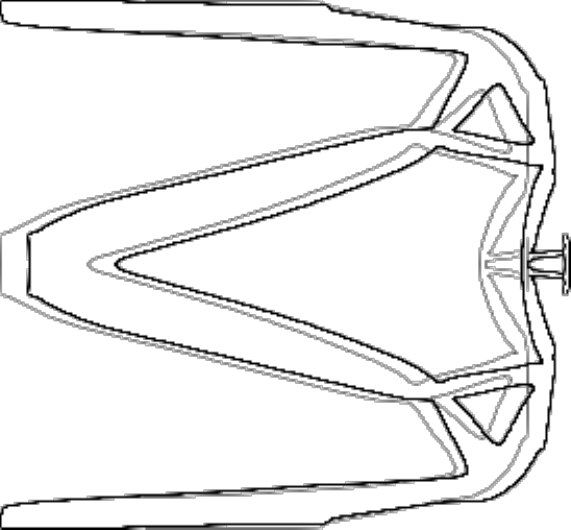}\qquad
			\includegraphics[height=4.0cm]{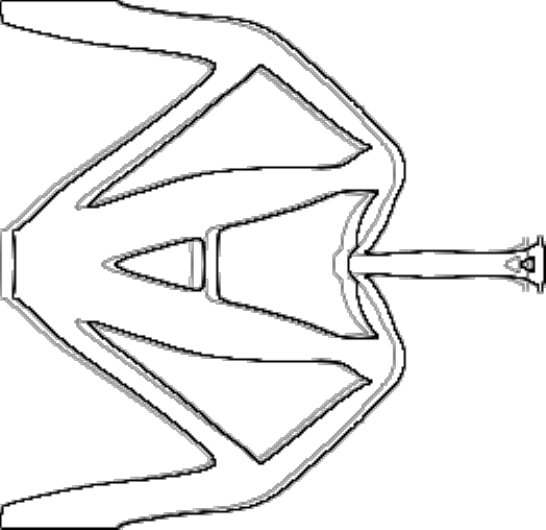}
		\caption{Deformation of displacement magnification mechanisms obtained in condition (d) (left) and (f) (right) (enhanced by a factor of 500)}
		\label{fig:def_magni_mu}
	\end{center}
\end{figure*}

\begin{figure}[hbtp]
\begin{center}
  \begin{tabular}{ccccc}
   && $\mu = 0\quad$ & $\mu = 0.1\quad$ & $\mu = 0.3\quad$ \\
	\rotatebox[origin=r]{90}{$\alpha = 1.0 \qquad$}
	&
	\rotatebox[origin=r]{90}{$\beta = 0 \qquad \;\;\:$}
	&
	\begin{minipage}[t]{0.2\hsize}
	\subfigure[]{\includegraphics[height=1.9cm]{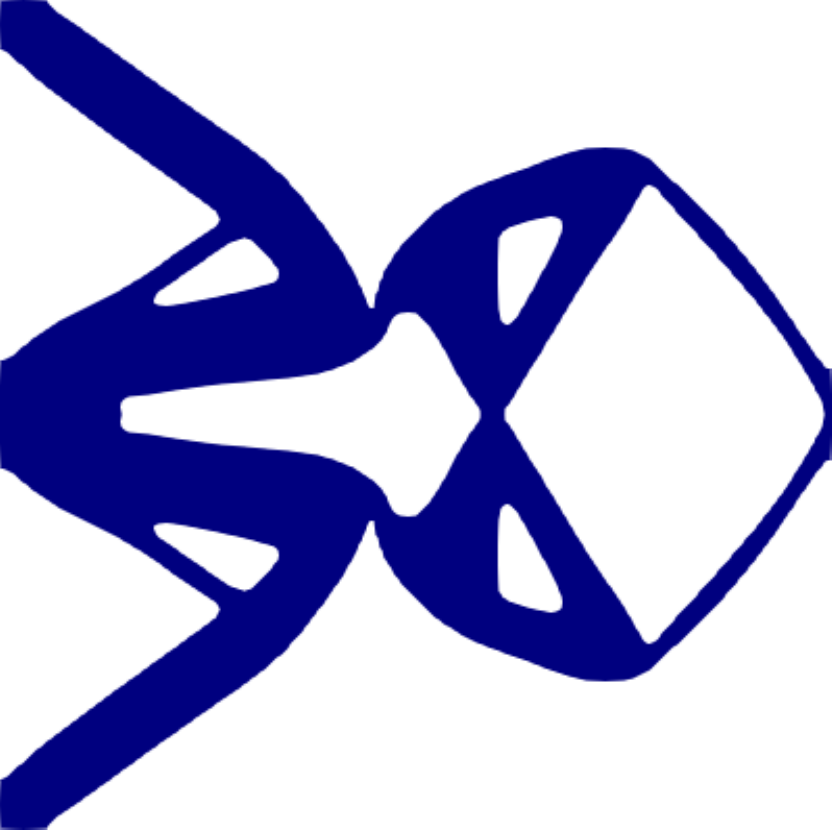} }
	\end{minipage}&
	\begin{minipage}[t]{0.2\hsize}
	\subfigure[]{\includegraphics[height=1.9cm]{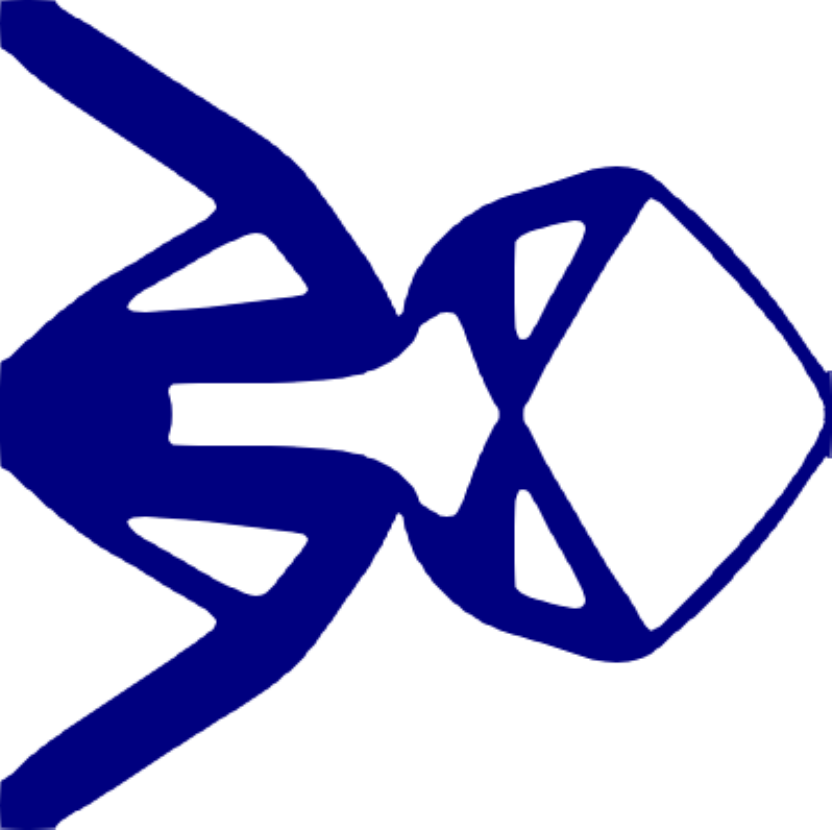} }
	\end{minipage}&
	\begin{minipage}[t]{0.2\hsize}
	\subfigure[]{\includegraphics[height=1.9cm]{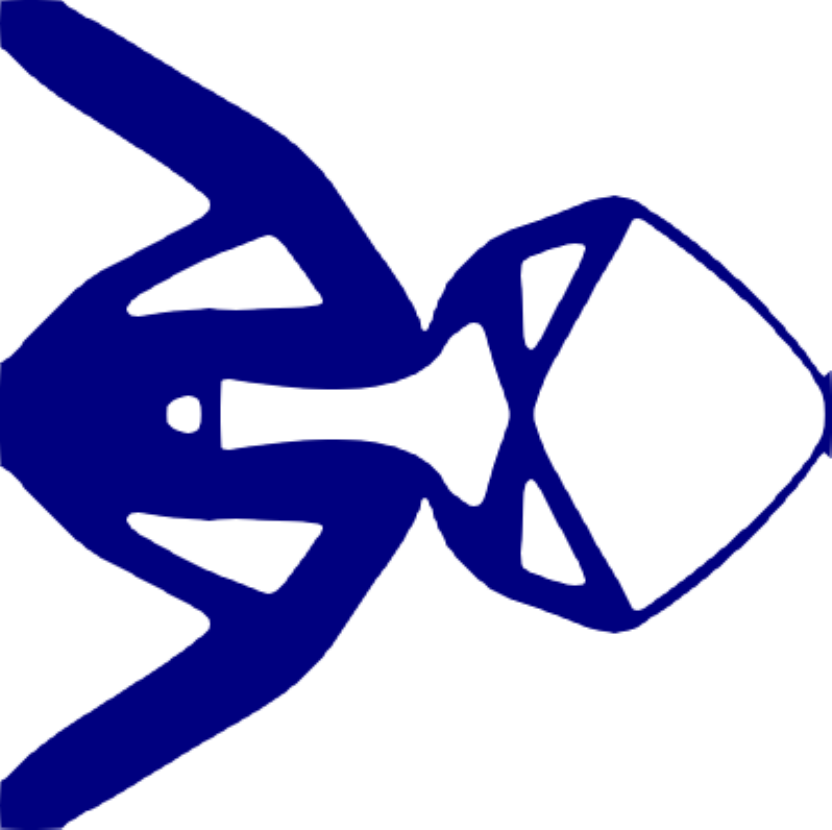} }
	\end{minipage}
	\\
	\rotatebox[origin=r]{90}{$\alpha = 1.0 \qquad$}
	&
	\rotatebox[origin=r]{90}{$\beta = 0.5 \qquad$}
	&  
  	\begin{minipage}[t]{0.2\hsize}
	\subfigure[]{
	\includegraphics[height=1.9cm]{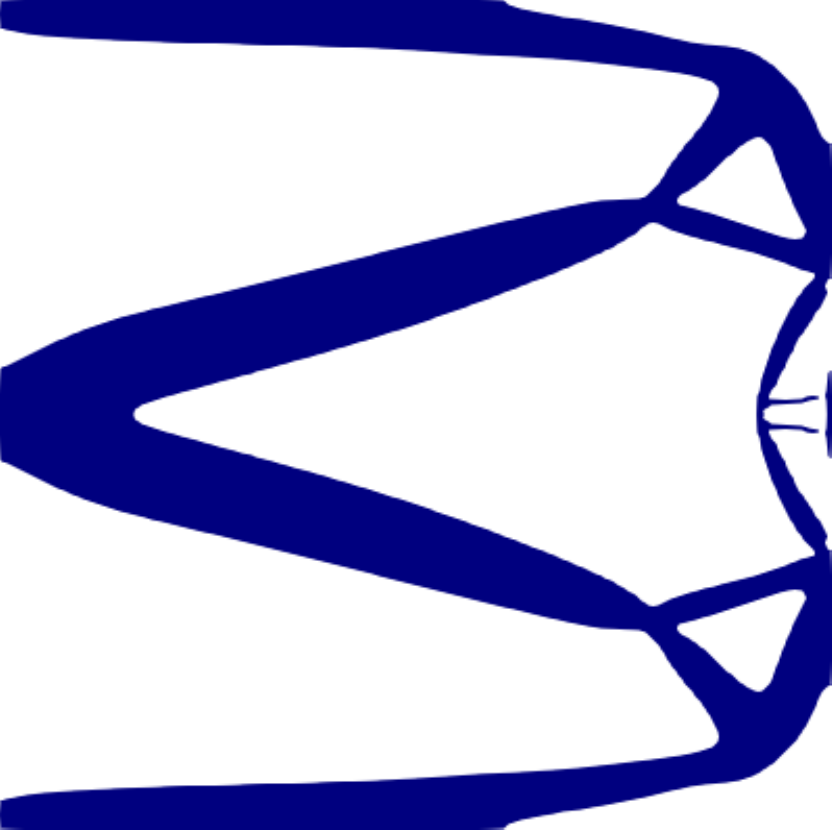} 
	\label{fig:magni_mu=0}
	}
	\end{minipage}&
	\begin{minipage}[t]{0.2\hsize}
	\subfigure[]{\includegraphics[height=1.9cm]{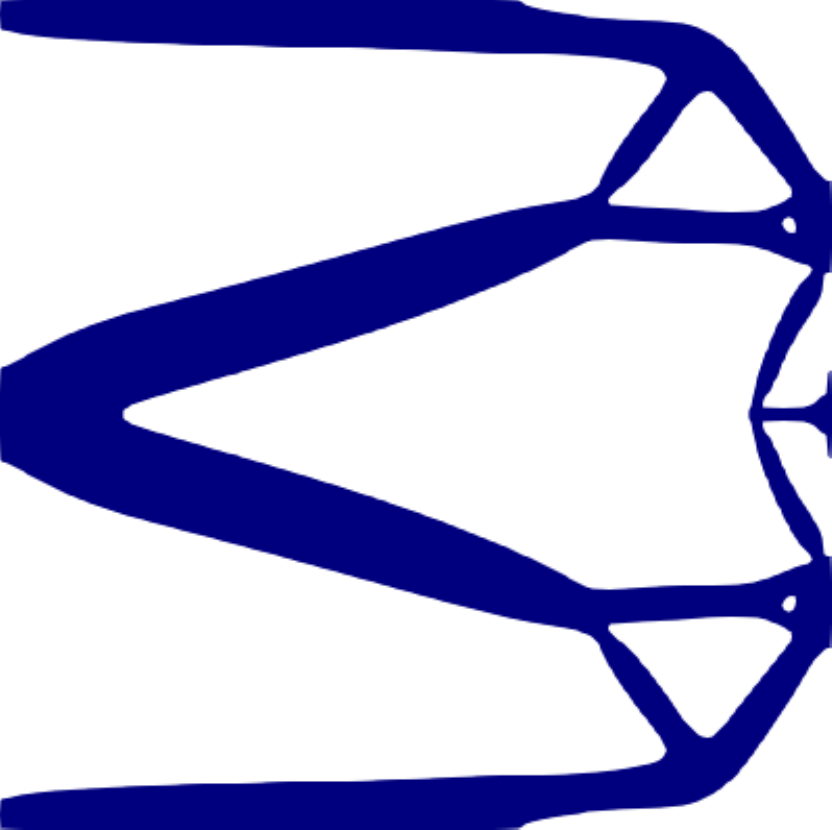} }
	\end{minipage}&
	\begin{minipage}[t]{0.2\hsize}
	\subfigure[]{
	\includegraphics[height=1.9cm]{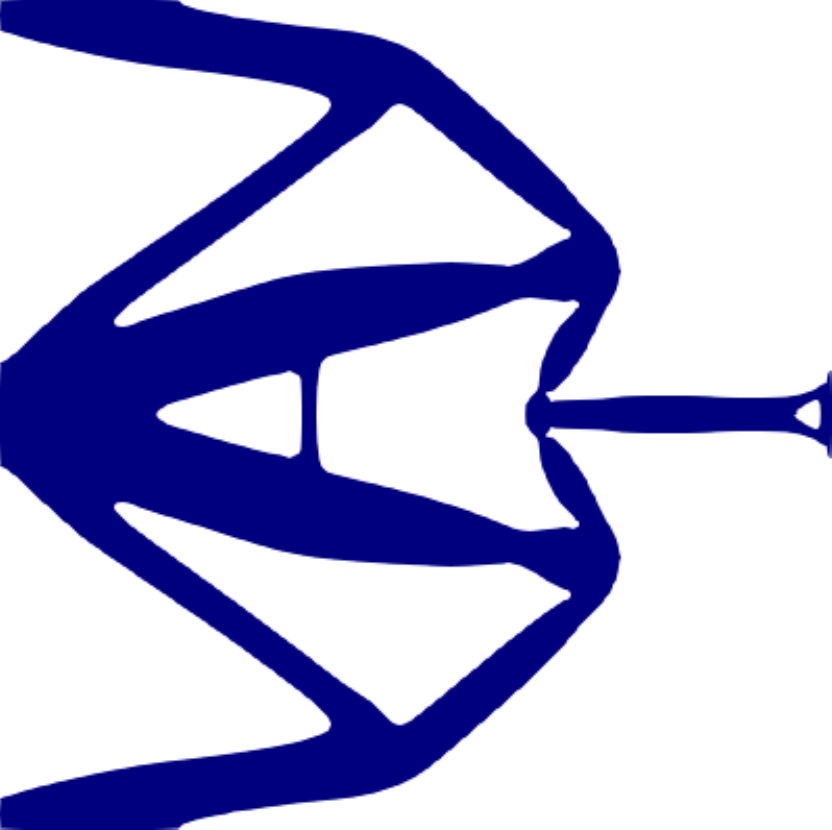} 
	\label{fig:magni_mu=0.3}
	}
	\end{minipage}
	\\
	\rotatebox[origin=r]{90}{$\alpha = 1.0 \qquad$}
	&
	\rotatebox[origin=r]{90}{$\beta = 1.0 \qquad$}
	&  
  	\begin{minipage}[t]{0.2\hsize}
	\subfigure[]{\includegraphics[height=1.9cm]{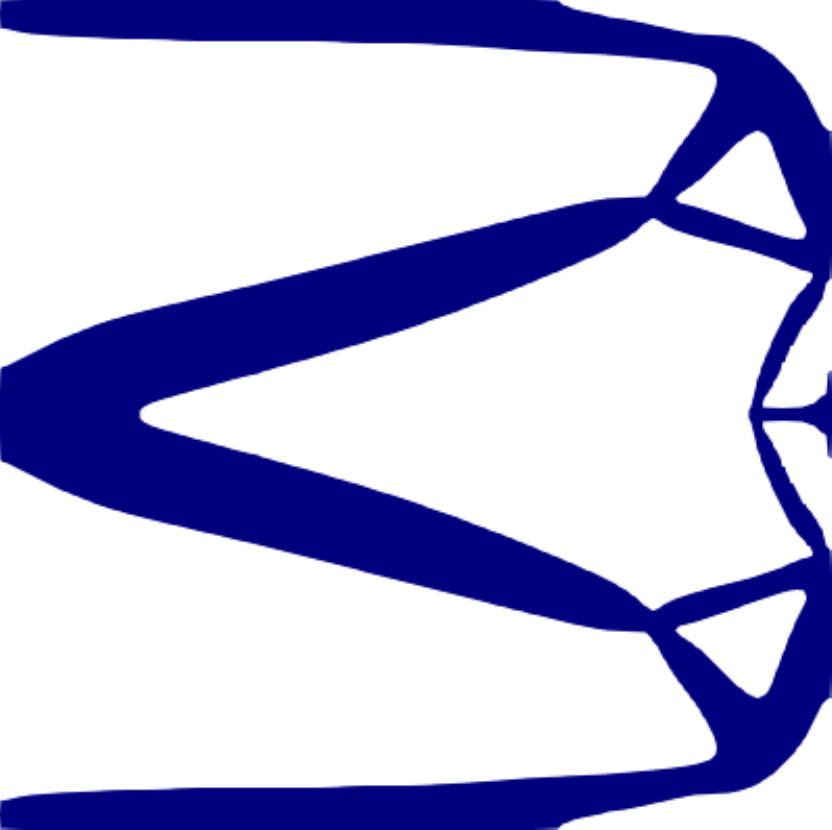} }
	\end{minipage}&
	\begin{minipage}[t]{0.2\hsize}
	\subfigure[]{\includegraphics[height=1.9cm]{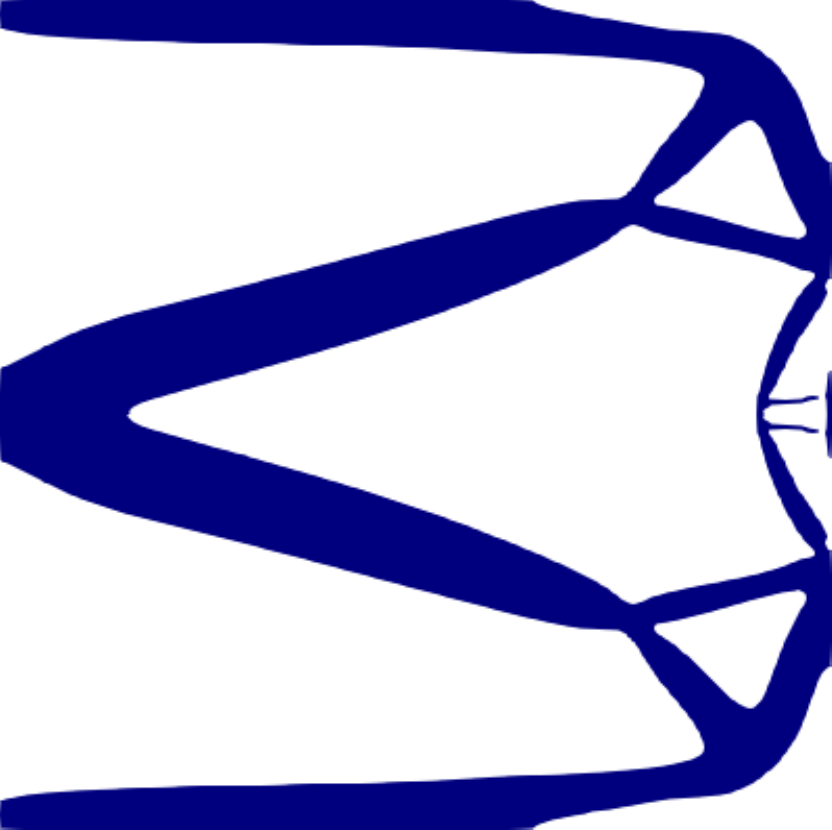} }
	\end{minipage}&
	\begin{minipage}[t]{0.2\hsize}
	\subfigure[]{\includegraphics[height=1.9cm]{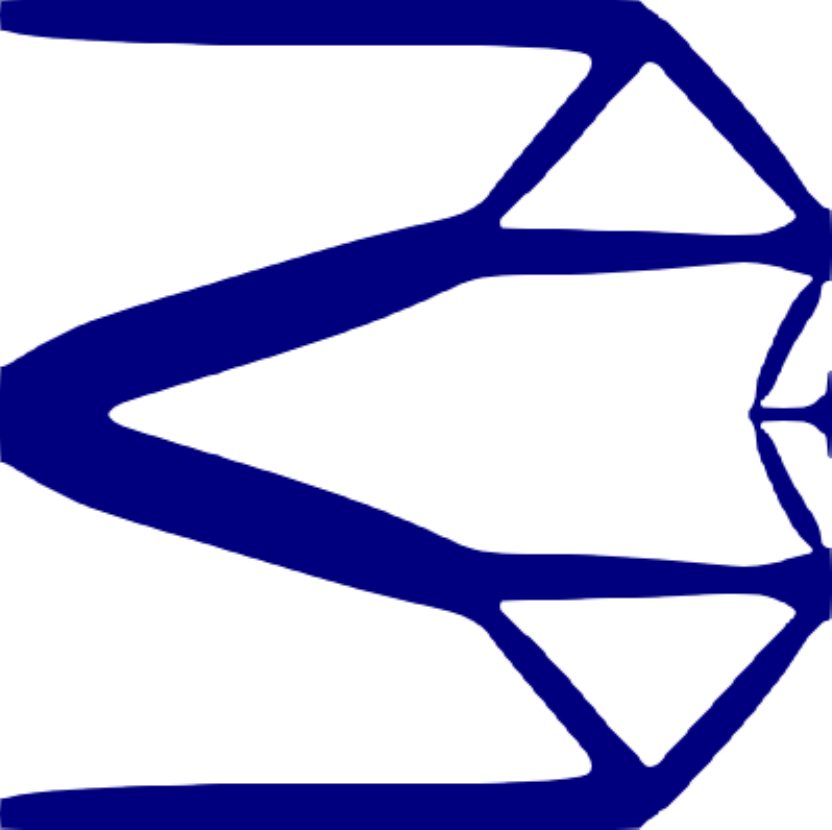} }
	\end{minipage}
	\\
	\rotatebox[origin=r]{90}{$\alpha = 0.5 \qquad$}
	&
	\rotatebox[origin=r]{90}{$\beta = 1.0 \qquad$}
	&  
  	\begin{minipage}[t]{0.2\hsize}
	\subfigure[]{\includegraphics[height=1.9cm]{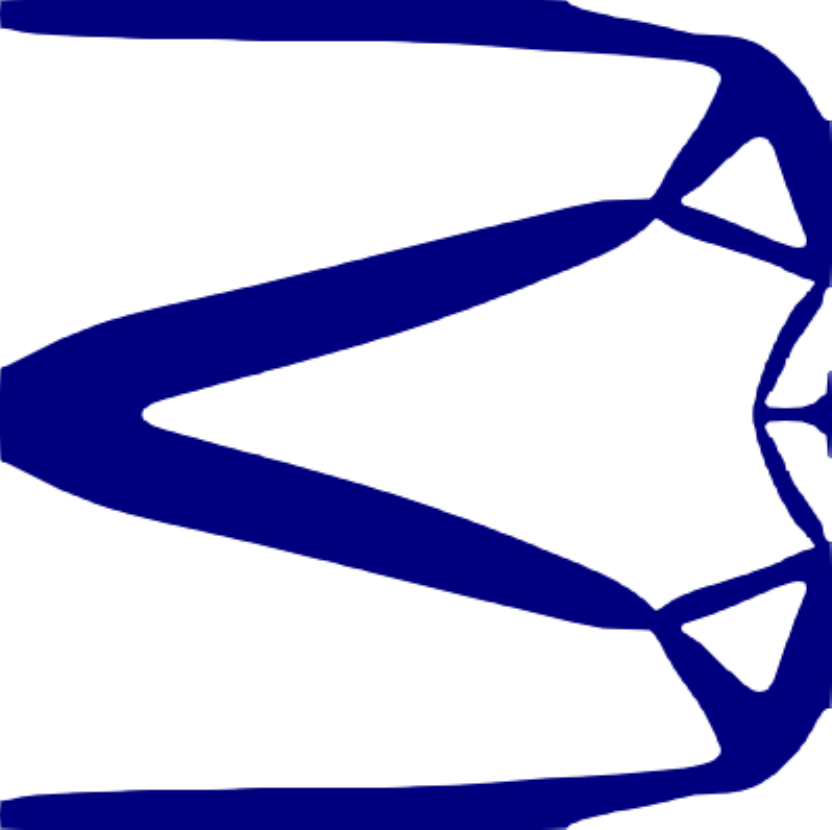} }
	\end{minipage}&
	\begin{minipage}[t]{0.2\hsize}
	\subfigure[]{\includegraphics[height=1.9cm]{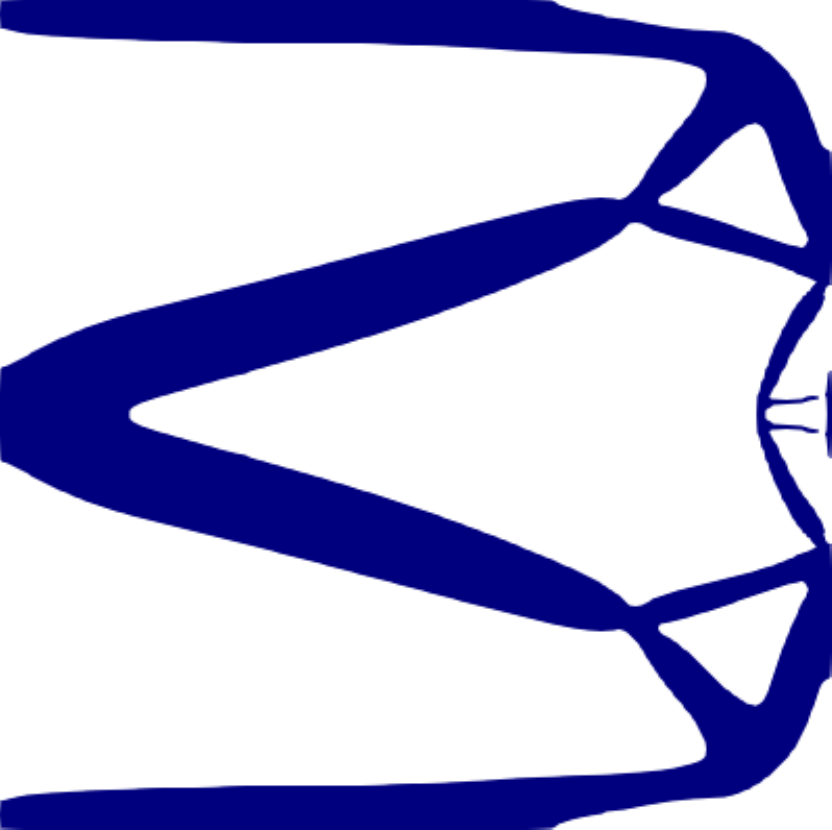} }
	\end{minipage}&
	\begin{minipage}[t]{0.2\hsize}
	\subfigure[]{\includegraphics[height=1.9cm]{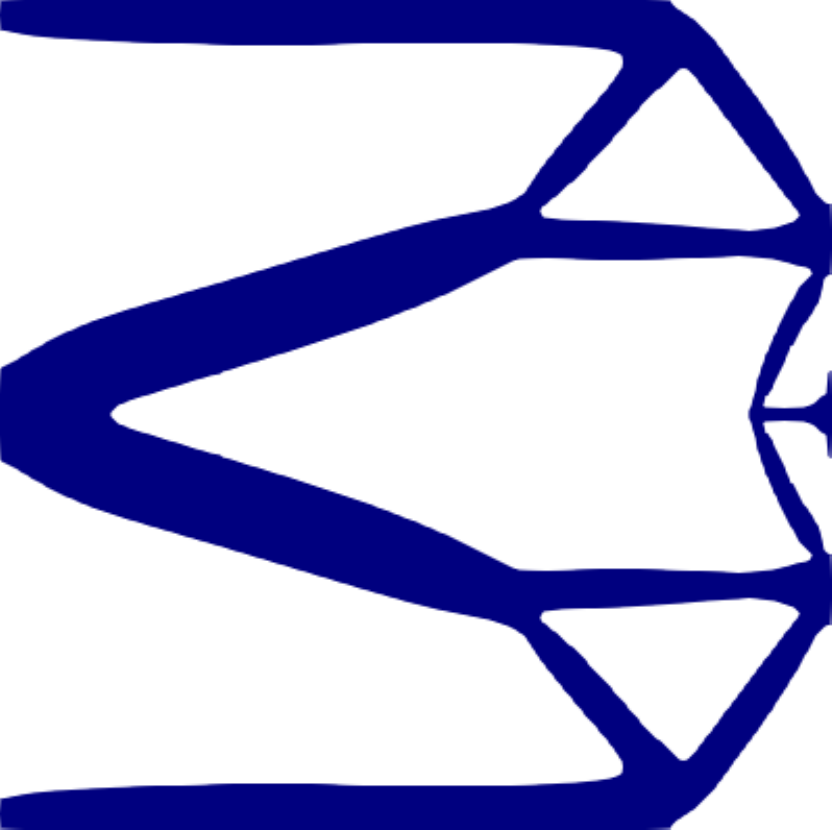} }
	\end{minipage}
	\\
	\rotatebox[origin=r]{90}{$\alpha = 0 \qquad \;\;\:$}
	&
	\rotatebox[origin=r]{90}{$\beta = 1.0 \qquad$}
	&  
  	\begin{minipage}[t]{0.2\hsize}
	\subfigure[]{\includegraphics[height=1.9cm]{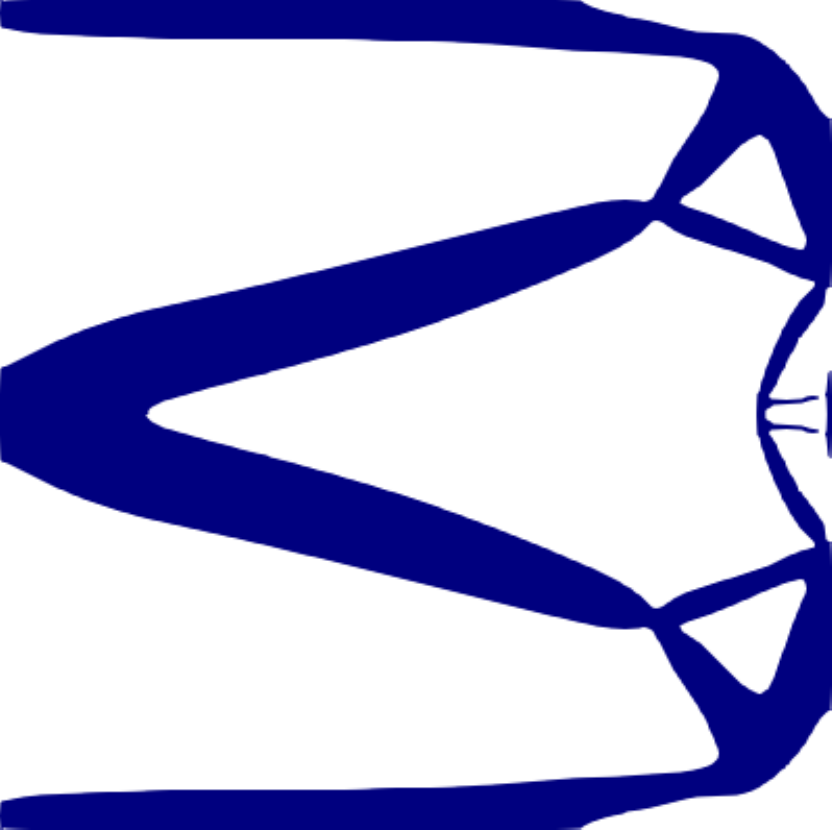} }
	\end{minipage}&
	\begin{minipage}[t]{0.2\hsize}
	\subfigure[]{\includegraphics[height=1.9cm]{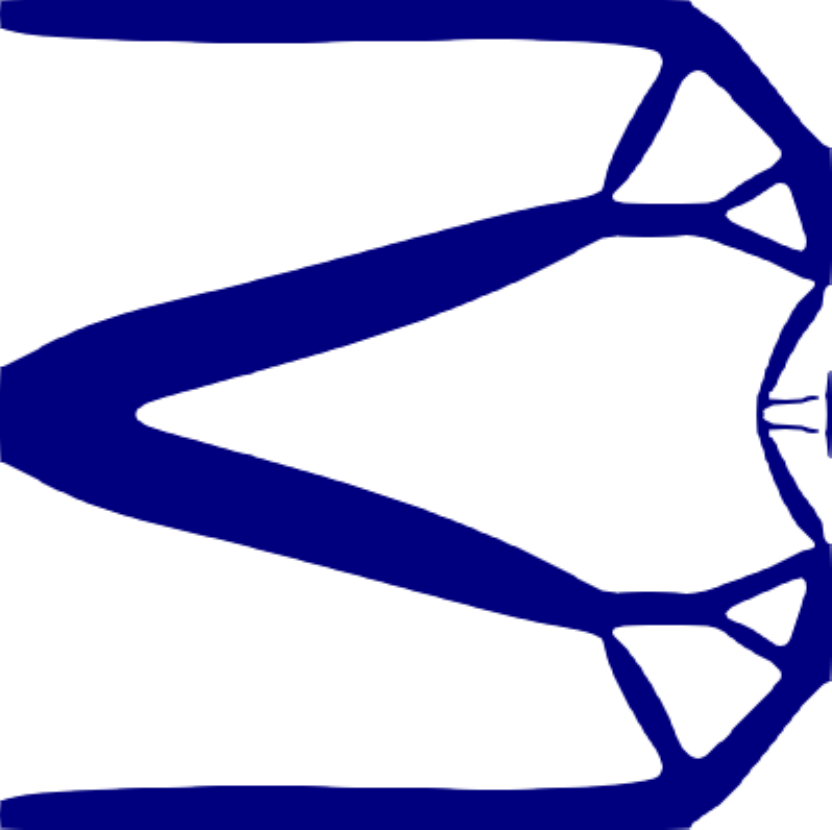} }
	\end{minipage}&
	\begin{minipage}[t]{0.2\hsize}
	\subfigure[]{\includegraphics[height=1.9cm]{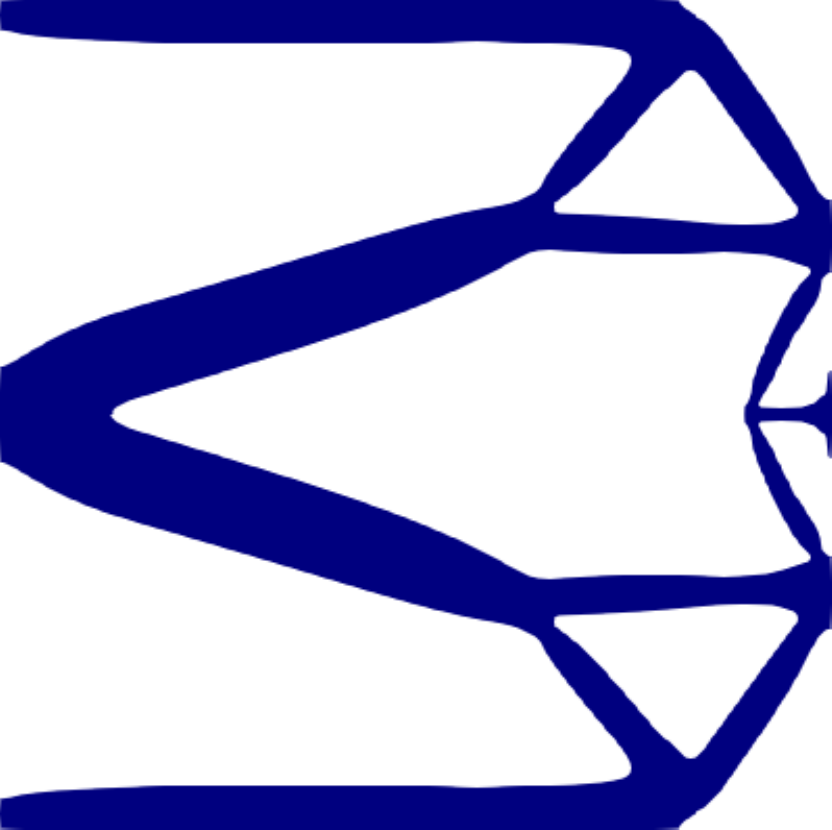} }
	\end{minipage}
  \end{tabular}
		\caption{Optimal configuration of displacement magnification mechanisms for conditions (a)--(o) listed in Table \ref{tab:magni_disp}}
		\label{fig:shape_magni_mu}
\end{center}
\end{figure}

\begin{figure}[hbtp]
\begin{center}
  \begin{tabular}{cccccc}
   && $\mu = 0\quad$ & $\mu = 0.1\quad$ & $\mu = 0.3\quad$ & \\
	\rotatebox[origin=r]{90}{$\alpha = 1.0 \qquad$}
	&
	\rotatebox[origin=r]{90}{$\beta = 0 \qquad \;\;\:$}
	&  
	\begin{minipage}[t]{0.2\hsize}
	\subfigure[]{\includegraphics[height=1.9cm]{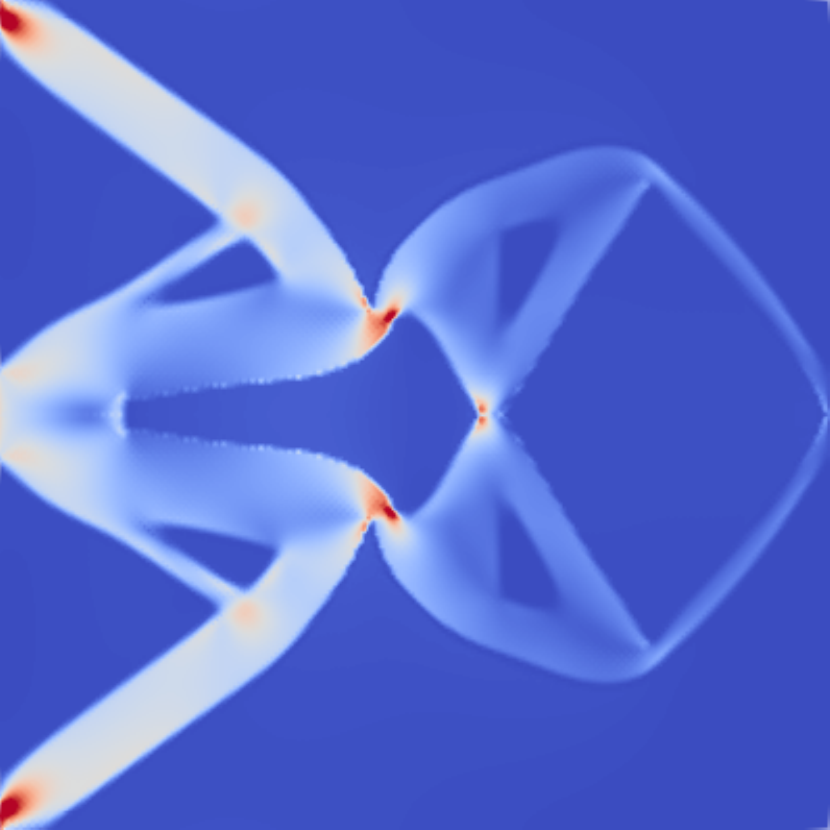} }
	\end{minipage}&
	\begin{minipage}[t]{0.2\hsize}
	\subfigure[]{\includegraphics[height=1.9cm]{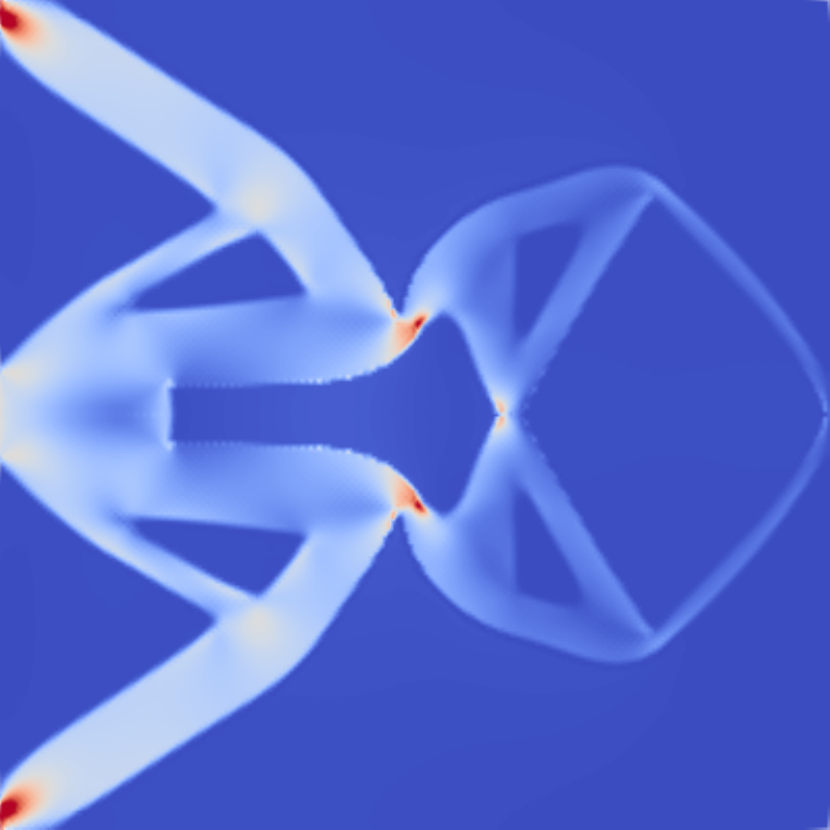} }
	\end{minipage}&
	\begin{minipage}[t]{0.2\hsize}
	\subfigure[]{\includegraphics[height=1.9cm]{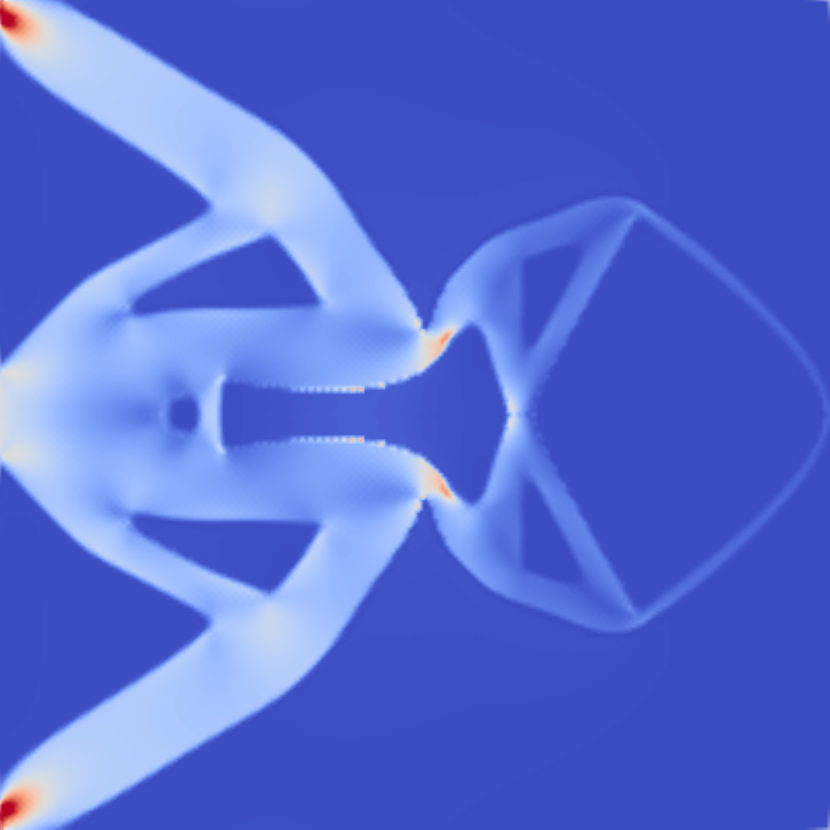} }
	\end{minipage}
	&
	\\
	\rotatebox[origin=r]{90}{$\alpha = 1.0 \qquad$}
	&
	\rotatebox[origin=r]{90}{$\beta = 0.5 \qquad$}
	&
  	\begin{minipage}[t]{0.2\hsize}
	\subfigure[]{\includegraphics[height=1.9cm]{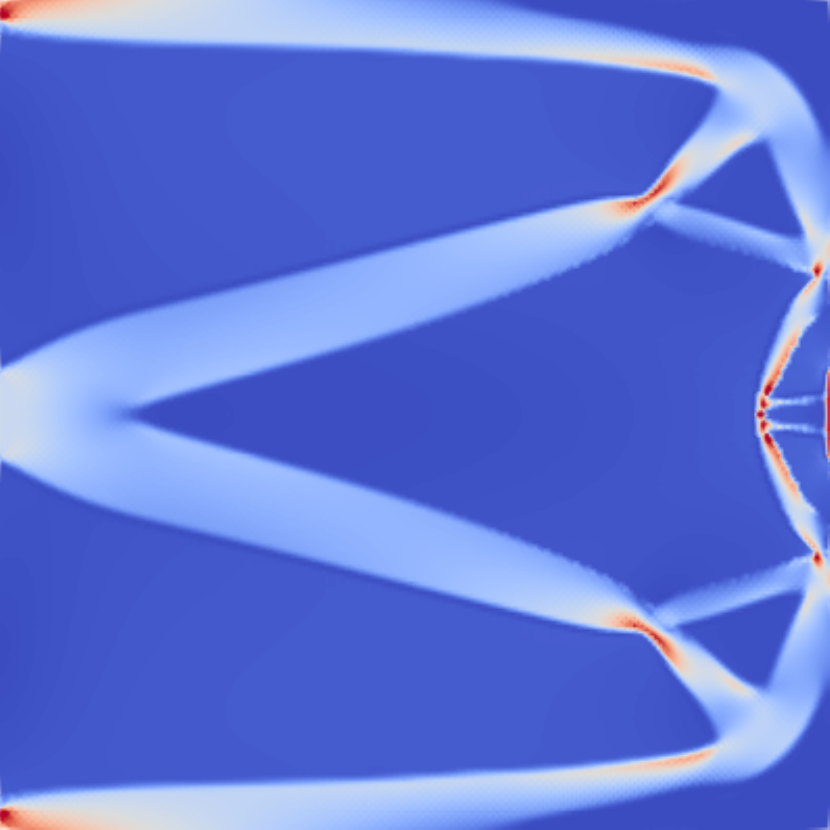} }
	\end{minipage}&
	\begin{minipage}[t]{0.2\hsize}
	\subfigure[]{\includegraphics[height=1.9cm]{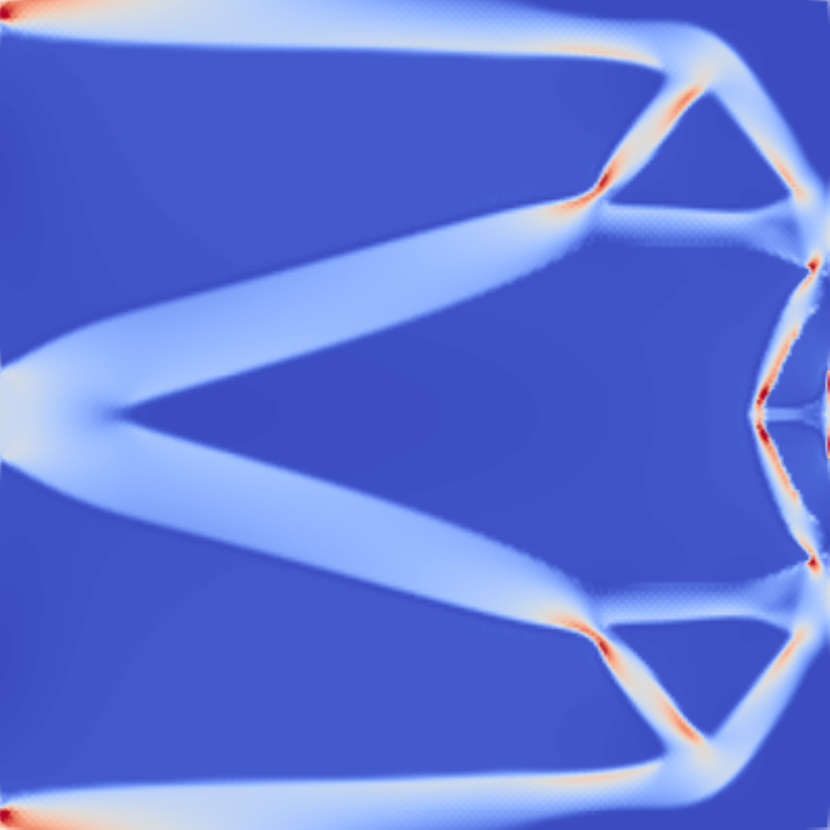} }
	\end{minipage}&
	\begin{minipage}[t]{0.2\hsize}
	\subfigure[]{\includegraphics[height=1.9cm]{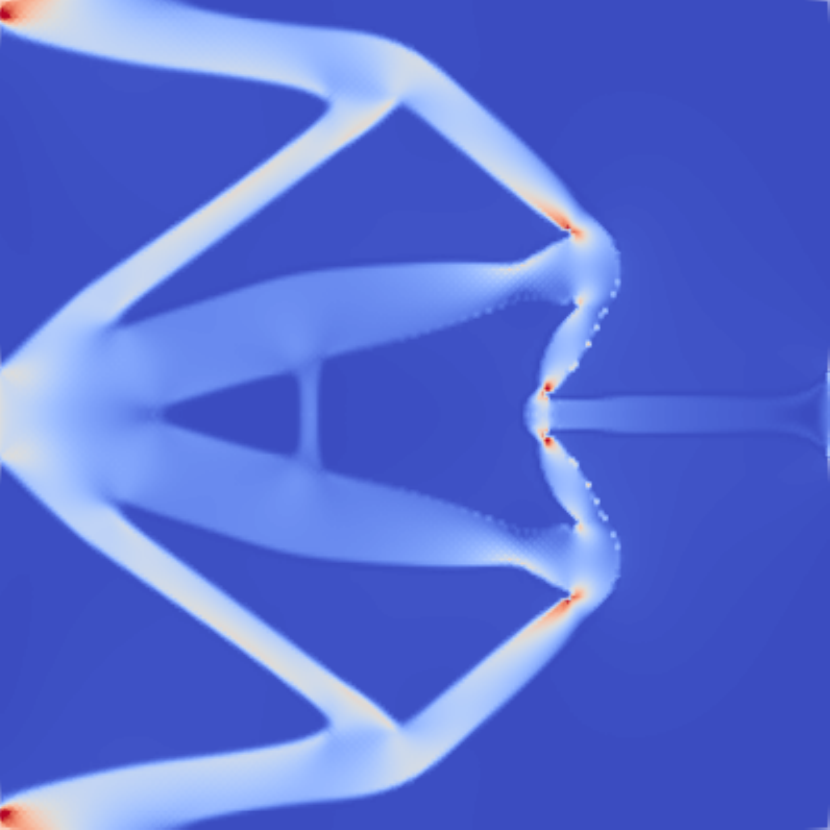} }
	\end{minipage}
	&
	\\
	\rotatebox[origin=r]{90}{$\alpha = 1.0 \qquad$}
	&
	\rotatebox[origin=r]{90}{$\beta = 1.0 \qquad$}
	&
  	\begin{minipage}[t]{0.2\hsize}
	\subfigure[]{\includegraphics[height=1.9cm]{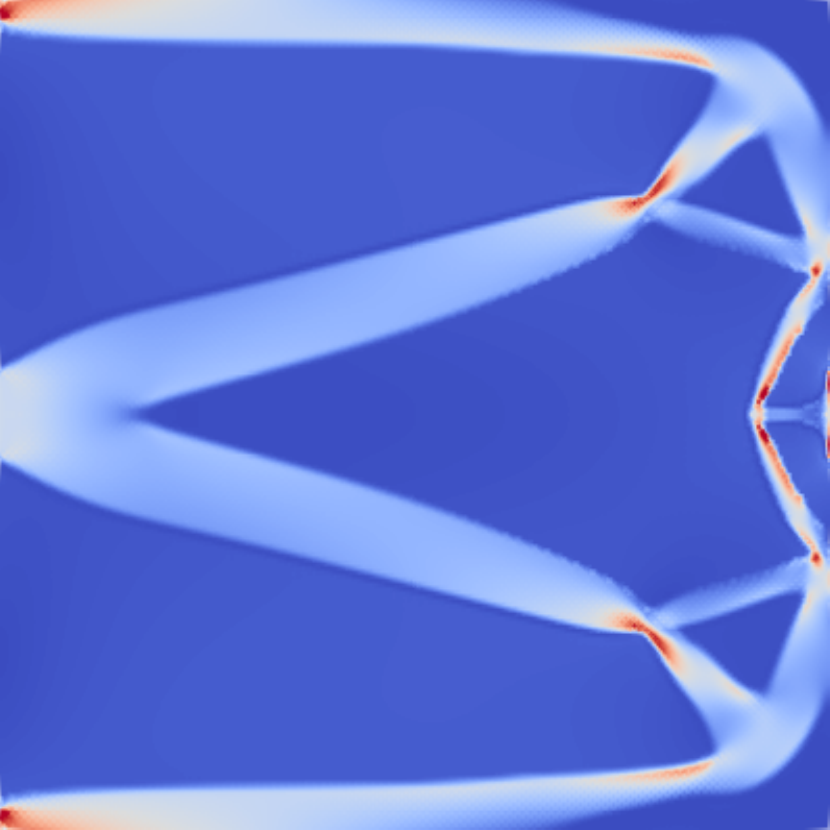} }
	\end{minipage}&
	\begin{minipage}[t]{0.2\hsize}
	\subfigure[]{\includegraphics[height=1.9cm]{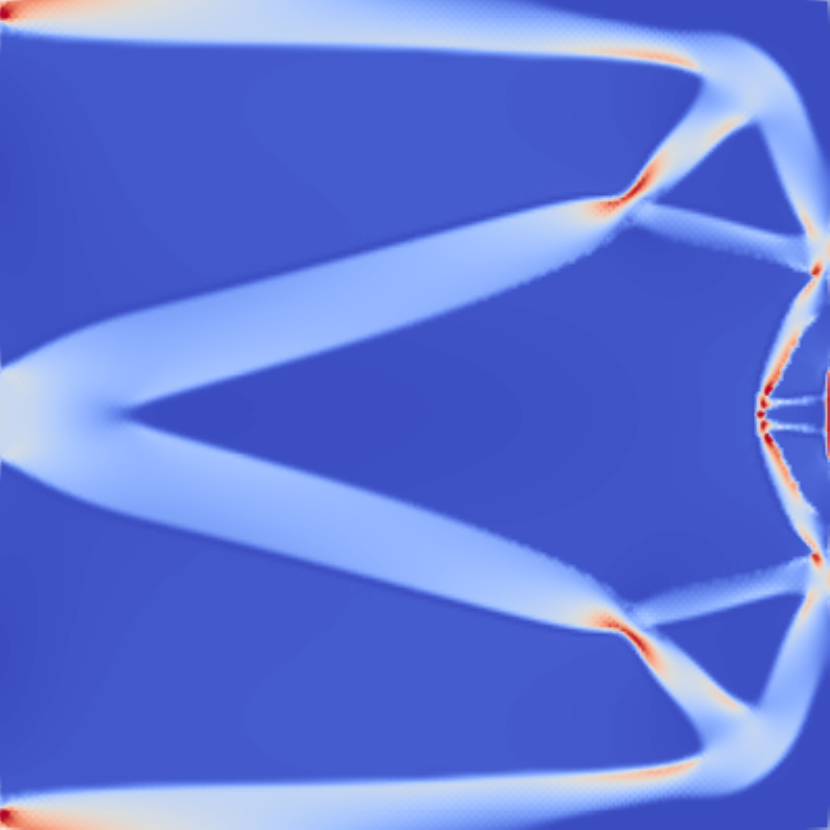} }
	\end{minipage}&
	\begin{minipage}[t]{0.2\hsize}
	\subfigure[]{\includegraphics[height=1.9cm]{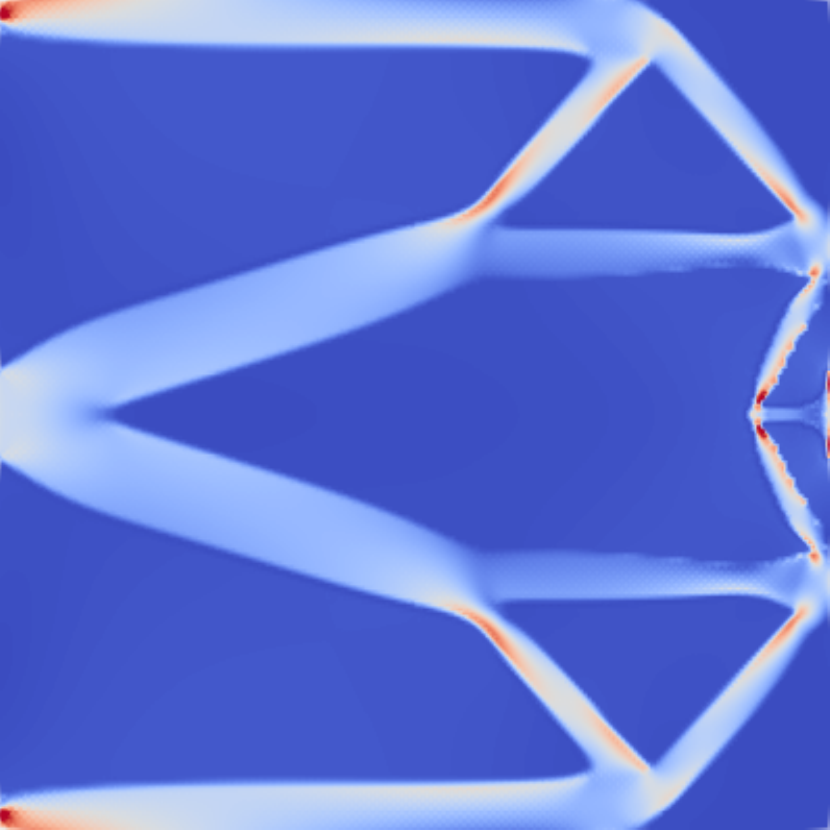} }
	\end{minipage}&
	\multirow{5}{*}{
	\begin{minipage}[t]{0.2\hsize}
  		\includegraphics[height=4.0cm]{Fig/colorbar_MPa.pdf}
	\end{minipage}}
	\\
	\rotatebox[origin=r]{90}{$\alpha = 0.5 \qquad$}
	&
	\rotatebox[origin=r]{90}{$\beta = 1.0 \qquad$}
	&
  	\begin{minipage}[t]{0.2\hsize}
	\subfigure[]{\includegraphics[height=1.9cm]{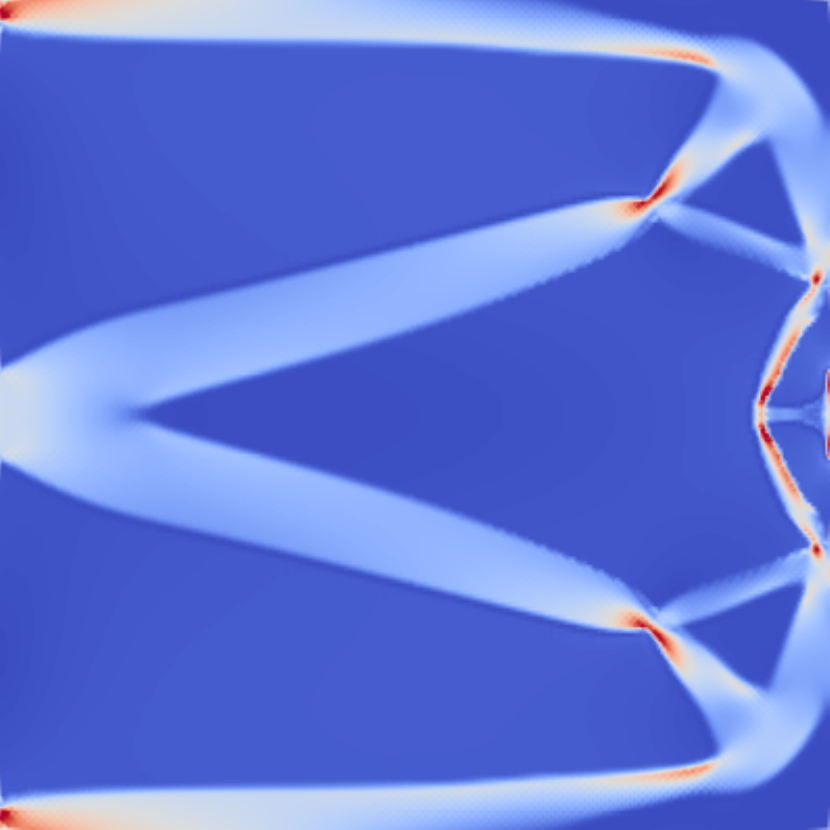} }
	\end{minipage}&
	\begin{minipage}[t]{0.2\hsize}
	\subfigure[]{\includegraphics[height=1.9cm]{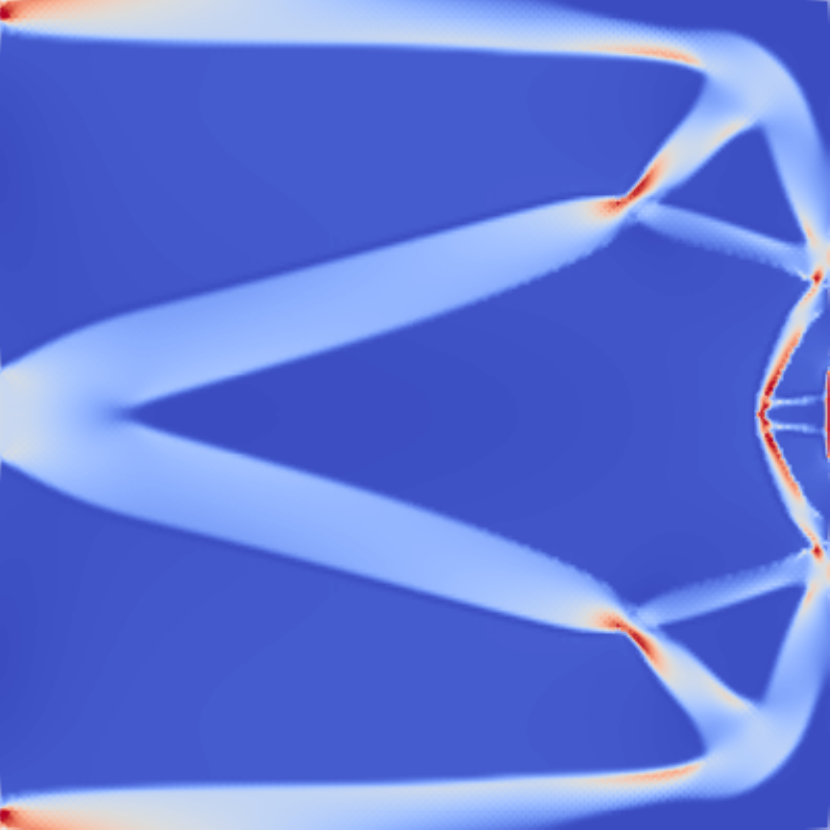} }
	\end{minipage}&
	\begin{minipage}[t]{0.2\hsize}
	\subfigure[]{\includegraphics[height=1.9cm]{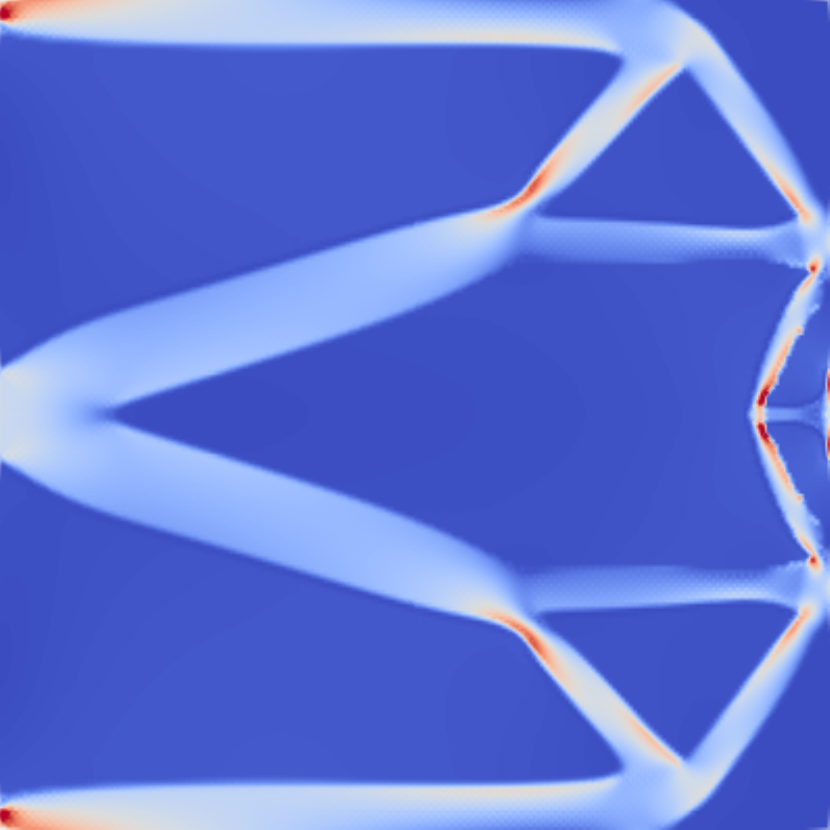} }
	\end{minipage}
	&
	\\
	\rotatebox[origin=r]{90}{$\alpha = 0 \qquad \;\;\:$}
	&
	\rotatebox[origin=r]{90}{$\beta = 1.0 \qquad$}
	&
  	\begin{minipage}[t]{0.2\hsize}
	\subfigure[]{\includegraphics[height=1.9cm]{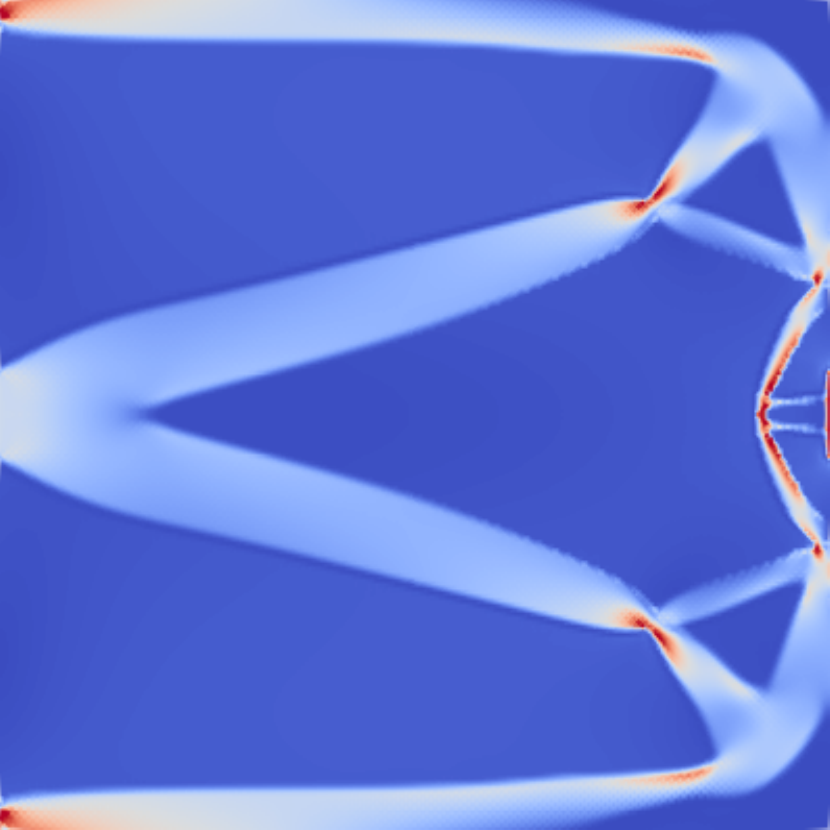} }
	\end{minipage}&
	\begin{minipage}[t]{0.2\hsize}
	\subfigure[]{\includegraphics[height=1.9cm]{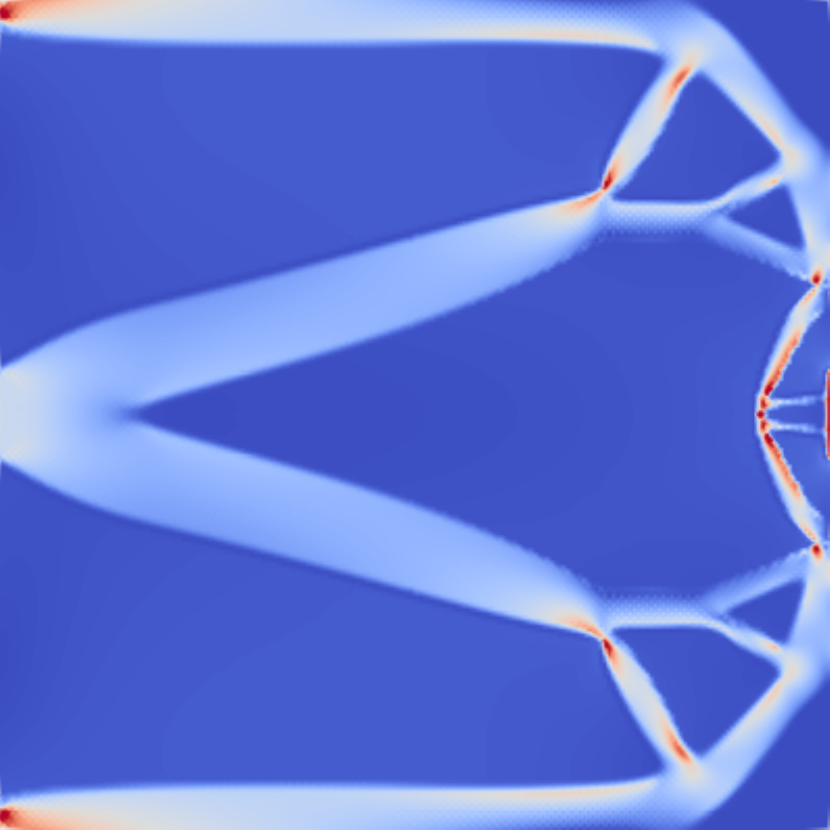} }
	\end{minipage}&
	\begin{minipage}[t]{0.2\hsize}
	\subfigure[]{\includegraphics[height=1.9cm]{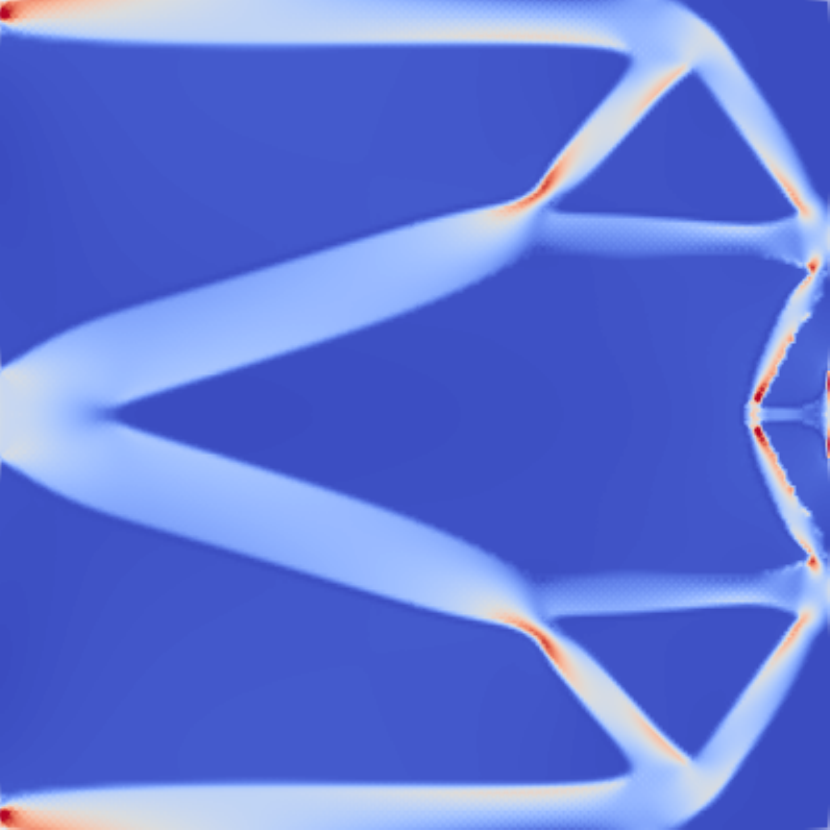} }
	\end{minipage}&
  \end{tabular}
		\caption{Von Mises stress of displacement magnification mechanisms for conditions (a)--(o) listed in Table \ref{tab:magni_disp}}
		\label{fig:mises_magni_mu}
\end{center}
\end{figure}

\begin{figure}
\begin{center}
\includegraphics[height = 3.5cm]{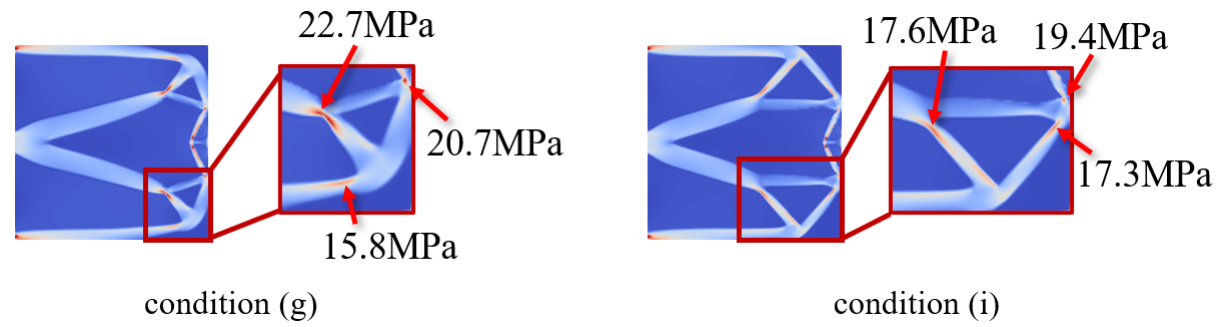}
\caption{Enlarged view of the von Mises stress concentration area of displacement magnification mechanisms obtained in conditions (g)(left) and (i)(right)}
\label{fig:magni_conce}
\end{center}
\end{figure}

As in Section \ref{subsec:inv}, we examine the effect of the Lagrange multiplier $\mu$ on the stress constraint and the parameters $\alpha$ and $\beta$ in the objective function on the resulting optimal configurations.
We set 15 conditions as listed in Table \ref{tab:magni_disp}.
In all conditions, $\alpha$, $\beta$ and $\mu$ are set to the same values as mentioned Section \ref{subsec:inv}. 
Table \ref{tab:magni_disp} presents the values of the evaluation functions for each condition, while Figure \ref{fig:def_magni_mu} presents the deformation diagram of conditions (d) and (f). 
Figure \ref{fig:shape_magni_mu} displays the optimal configuration of the displacement magnification mechanism for each condition, while Figure \ref{fig:mises_magni_mu} presents the distribution of the von Mises stress for each condition.
Figure \ref{fig:magni_conce} presents the enlarged view of the von Mises stress concentration area in conditions (g) and (i).

As illustrated in Figures \ref{fig:shape_magni_mu} (a), (b), and (c), the result is similar to the structure obtained by maximizing the stiffness against the input force as a result of applying strong stress constraints and increasing the stiffness against the input force.
To avoid this result, it is necessary to adjust the proposed parameters $\alpha$ and $\beta$.
As illustrated in Table \ref{tab:magni_disp},  the value of $U_o/U_i$ is larger than 1.0 in all conditions.
This signifies that the input displacement is magnified as the output displacement.

\begin{table}[H]
 \caption{Displacement at the output and input ports of the displacement magnification mechanisms}
 \label{tab:magni_disp}
 \centering
 \begin{tabular}{ccccccc}
 \hline
 & & & & \multicolumn{2}{c}{displacement [$\mu$m]}&\\
 condition&$\alpha$& $\beta$& $\mu$ & $\;\;\ U_{o}\;\;$ &$U_{i}$&  $U_o / U_i $\\
 \hline
 initial structure & & & & 4.6 & 8.5 & 0.54\\
 a &1.0&0& 0 & 31.0 & 26.3 & 1.18\\
 b &1.0&0& 0.1 & 26.3 & 23.0 & 1.14\\
 c &1.0&0& 0.3 & 20.9 & 19.7 & 1.06\\
 d &1.0&0.5& 0 & 77.4 & 52.7 & 1.47\\
 e &1.0&0.5& 0.1 & 70.9 & 48.5 & 1.46\\
 f &1.0&0.5& 0.3 & 42.0 & 33.5 & 1.25\\
 g &1.0&1.0& 0 & 82.8 & 56.3 & 1.47\\
 h &1.0&1.0& 0.1 & 76.1 & 51.8 & 1.47\\
 i &1.0&1.0& 0.3 & 60.6 & 43.4 & 1.40\\
 j &0.5&1.0& 0 & 84.3 & 57.6 & 1.46\\
 k &0.5&1.0& 0.1 & 78.7 & 53.5 & 1.47\\
 l &0.5&1.0& 0.3 & 63.3 & 44.7 & 1.42\\
 m &0&1.0& 0 & 86.1 & 58.6 & 1.47\\
 n &0&1.0& 0.1 & 79.5 & 53.9 & 1.48\\
 o &0&1.0& 0.3 & 65.8 & 46.0 & 1.43\\
 \hline
 \end{tabular}
\end{table}

As in the displacement inverter problem, for the same value of $\mu$, the values of $U_o$ and $U_i$ are smaller when $\alpha > \beta$ than when $\alpha = \beta$, and increase when $\beta > \alpha$.
The above results indicate that the effects of parameters $\alpha$ and $\beta$ are as assumed in the formulation for the displacement magnification mechanism problem.

As illustrated in Figure \ref{fig:mises_magni_mu} and \ref{fig:magni_conce}, in conditions (d), (g), (j), and (m), the von Mises stress is concentrated in some parts of the structure.
For the same values of $\alpha$ and $\beta$, the hinge-like structure becomes thicker as $\mu$ increases.
In particular, when $\mu = 0.5$, the number of hinge-like structures is reduced, and the stiffness against the input load is increased by concentrating the material close to the input port.
%Thus, as illustrated in Table \ref{tab:magni_disp} and Figure \ref{fig:def_magni_mu}, as $\mu$ increases, the displacement at the input and output ports decreases.
In addition, as illustrated in Table \ref{tab:magni_disp} and Figure \ref{fig:def_magni_mu}, both values of $U_o$ and $U_i$ are smaller in condition (f) than in condition (d).
The value of $\mu$ in condition (f) is larger than in condition (d). Other parameters are the same between condition (f) and (d).
This indicates that the displacement at the input and output ports decreases as $\mu$ increases.
However, the magnification ratio increases without the stress constraint.

For the displacement magnification problem, $\beta$ should be set to 1.0, and $\alpha$ should be set to a smaller value to obtain a high magnification ratio. 
However, $\mu$ should be set to greater than 0.1 to avoid hinging.
If it is necessary to increase the stiffness, $\alpha$ should be set to 1.0.

\begin{figure}[htbp]
\begin{center}
\includegraphics[height=11.5cm]{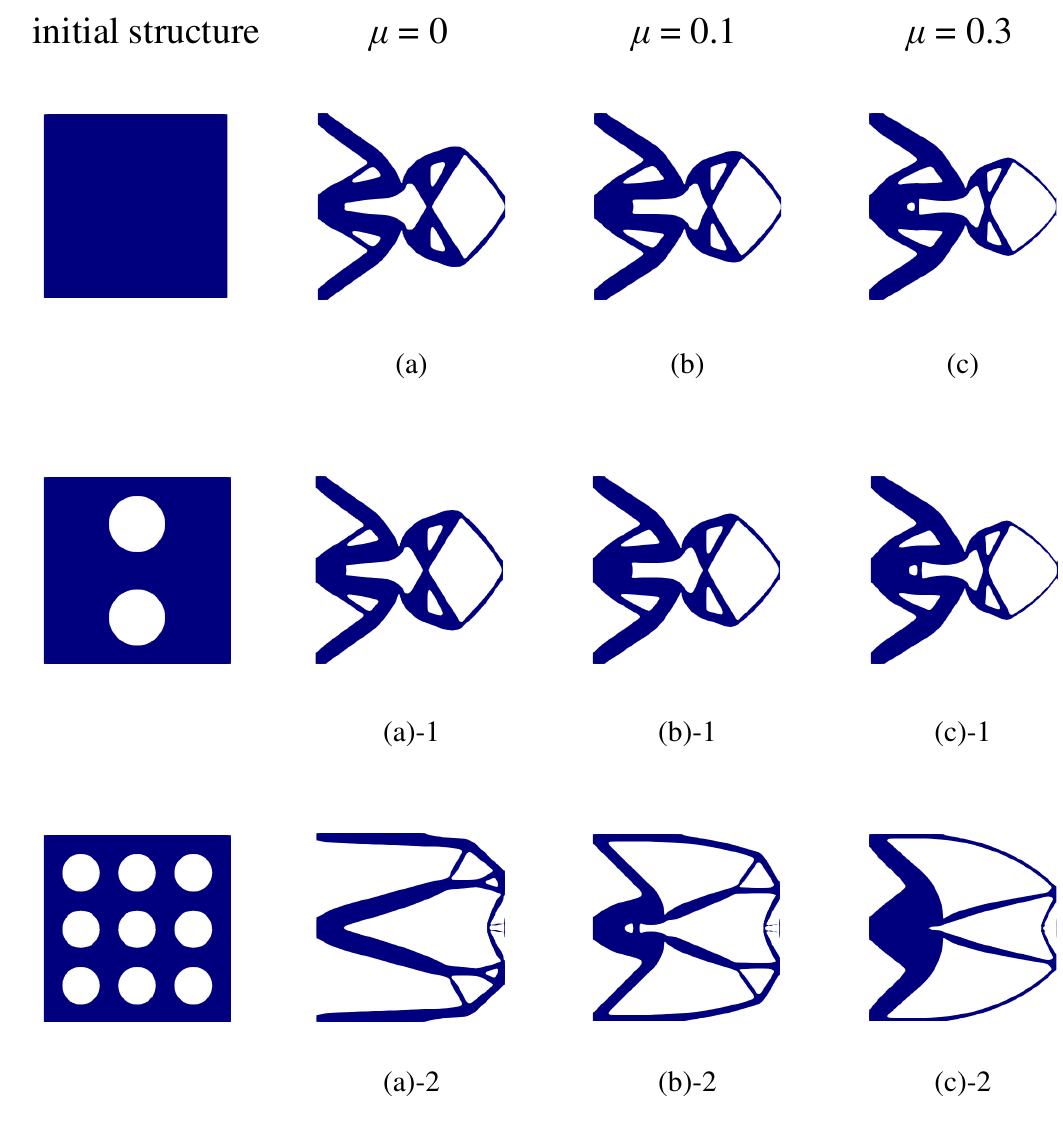}
		\caption{Initial structure dependence on the optimal configuration of displacement magnification mechanisms ($\alpha = 1.0, \beta = 0$)}
		\label{fig:magni_init_a1b0}
\end{center}
\end{figure}

\begin{figure}[htbp]
\begin{center}
\includegraphics[height=11.5cm]{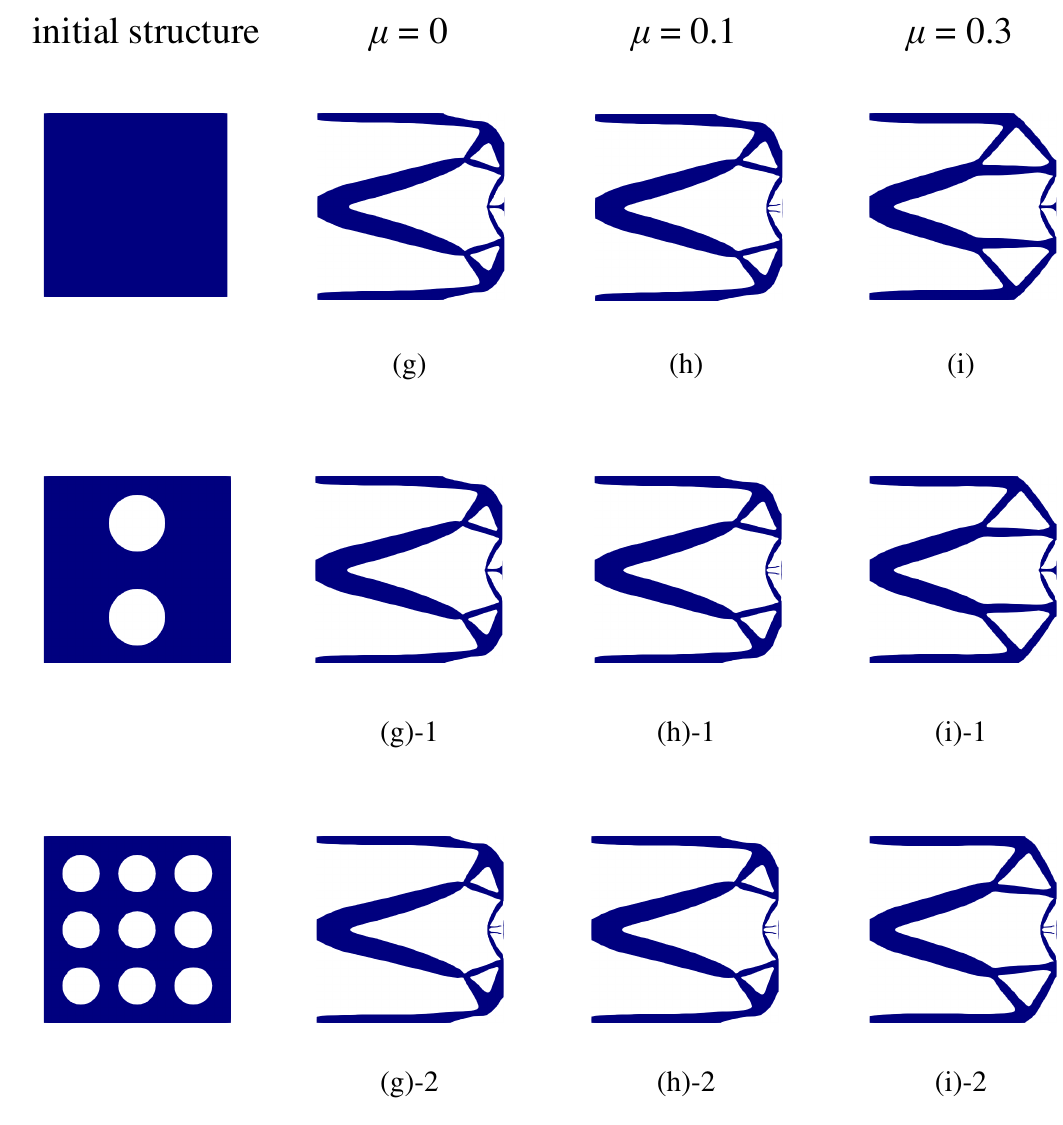}
		\caption{Initial structure dependence on the optimal configuration of displacement magnification mechanisms ($\alpha = 1.0, \beta = 1.0$)}
		\label{fig:magni_init_a1b1}
\end{center}
\end{figure}

Next, we discuss on initial structure dependence on optimal configurations.
Figure \ref{fig:magni_init_a1b0} and \ref{fig:magni_init_a1b1} present optimal configurations obtained from several different initial structures.
Figure \ref{fig:magni_init_a1b0} presents configurations obtained in condition (a), (b), and (c), whereas Figure \ref{fig:magni_init_a1b1} presents configurations in condition (g), (h), and (i).
In the displacement magnification problem, the initial structure dependence of optimal configuration is greater than in the displacement inverter problem.
The tendency about $\alpha$, $\beta$, and $\mu$ is the same regardless of the initial structure.

\begin{figure*}[htbp]
	\begin{center}
		\includegraphics[height=11.5cm]{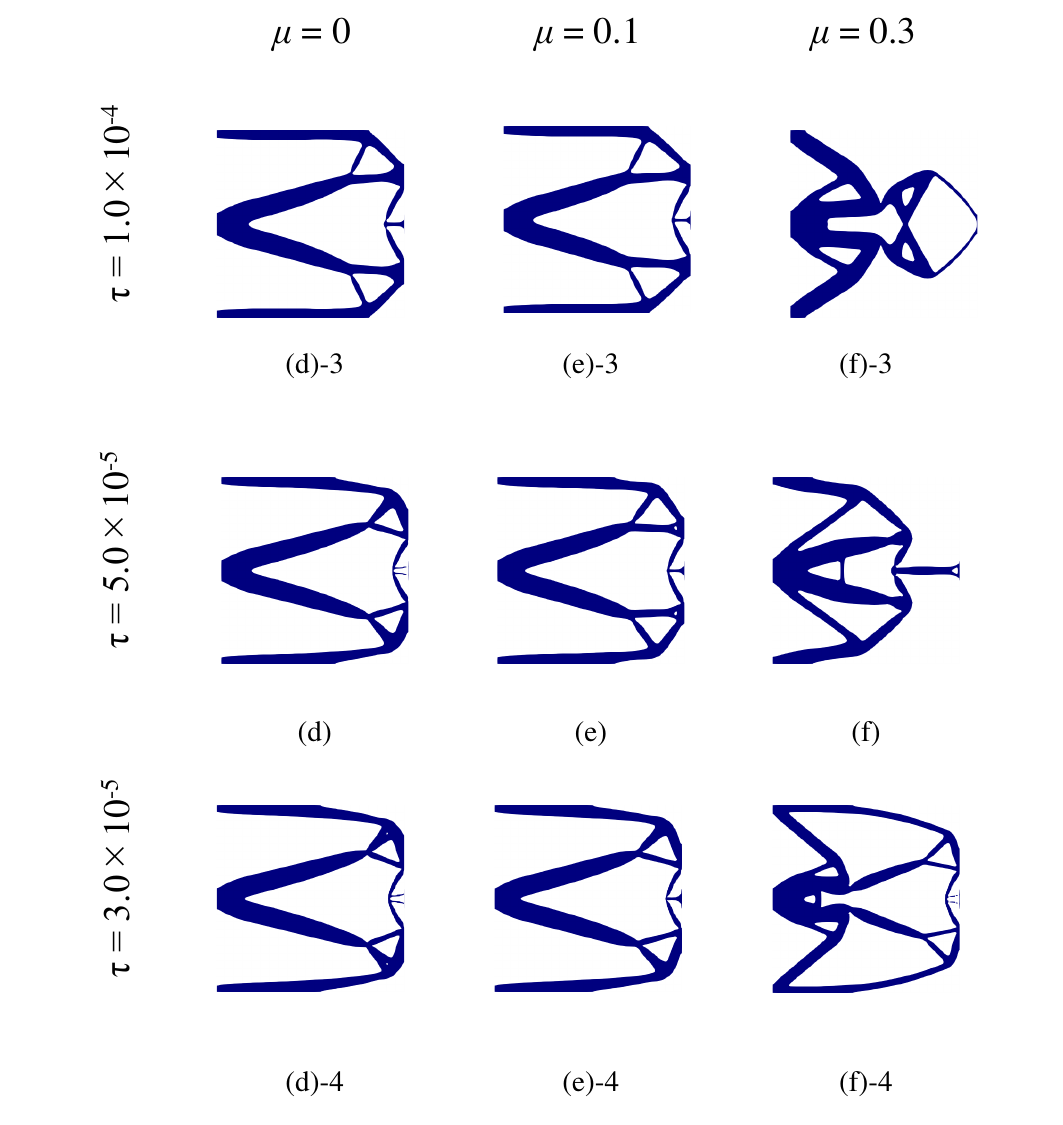}
		\caption{Optimal configuration of displacement magnification mechanisms in various setting of parameter $\tau$ ($\alpha = 1.0, \beta = 0.5$)}
		\label{fig:magni_tau}
	\end{center}
\end{figure*}

Lastly, we discuss effect of parameter $\tau$.
Figure \ref{fig:magni_tau} illustrates optimal configuration of displacement magnification mechanisms in various setting of parameter $\tau$.
We set $\tau$ to $1.0\times 10^{-4}$, $5.0\times10^{-5}$(same as in Figure \ref{fig:shape_magni_mu}), and $3.0\times10^{-5}$.
As illustrated in Figure \ref{fig:magni_tau}, the optimal configurations become complex structure as the value of $\tau$ decreases when $\mu$ is 0.3.
Whereas, the dependence of the optimal configuration on the value of $\tau$ is quite small when $\mu$ is 0.1 or less.

In conclusion, the proposed method can be used to design displacement magnification mechanisms without hinges.
In \ref{proto}, we present a prototype of the displacement magnification mechanisms illustrated in Figure \ref{fig:magni_tau} (d)-3 and (f)-3 created using a 3D printer.

\section{Conclusion}
\label{sec:conclusion}

In this study, we propose a method for designing displacement magnification mechanisms based on the concept of effective energy. 
The proposed method accommodates cases in which the reaction force is not applied to the output port. Furthermore, this method can obtain structures without one-node-connected hinges. The results obtained in this study are summarized as follows.

\begin{itemize}
	\item[1.]A stress-constrained topology optimization method based on the concept of effective energy is formulated. In the numerical implementation, an optimization algorithm is constructed.
	\item[2.]Several numerical examples are provided to demonstrate the utility and validity of the proposed method. 
	The proposed method can be used to design displacement inverter mechanisms and displacement magnification mechanisms. 
	In addition, the stress concentration can be reduced by adjusting the Lagrange multiplier $\mu$.
\end{itemize}

\section*{Acknowledgments}
This work was partly supported by JSPS KAKENHI Grant Number 19H02049 and Katsu start-up fund of the University of Tokyo.

\appendix
\section{Numerical example of minimizing the p-norm of the von Mises stress and stress-constrained mean compliance minimization}
\label{Lbeam}

\setcounter{figure}{0}
\setcounter{table}{0}

\begin{figure*}[htb]
	\begin{center}
		\includegraphics[height=5.5cm]{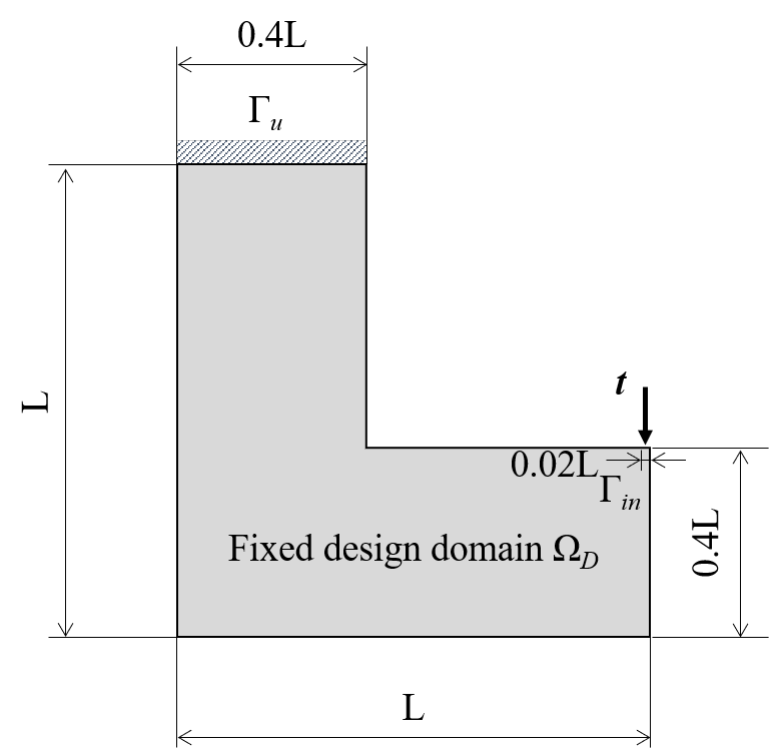}
		\caption{Design setting for minimization of p-norm}
		\label{fig:cond_Lbeam}
	\end{center}
\end{figure*}

%Several numerical examples are presented in which the p-norm of the von Mises stress defined in (\ref{eq:pn}) is minimized as the objective function to verify the validity of the stress constraint defined in (\ref{eq:dif_Gs}).
%The difference between the p-norm minimization and stress-constrained design problems is important from an engineering and design standpoint,  although from a mechanics standpoint they involve designing structures with identical characteristics.
%In this section, minimization of p-norm of the von Mises stress is described in detail.
%Numerical examples of the problem of stress-constrained mean compliance minimization are presented later in this section.
%  
%As illustrated in Figure \ref{fig:cond_Lbeam}, we consider the L-beam problem, which has been treated as a benchmark in previous studies \cite{duysinx1998topology, le2010stress, ogawa2022topology}. 
%
%We formulate the optimization problem as follows:
In this section, we first tackle the minimization of p-norm of the von Mises stress.
Next, we tackle the problem of the stress-constrained mean compliance minimization.
As illustrated in Figure \ref{fig:cond_Lbeam}, we consider the L-beam problem, which has been treated as a benchmark in previous studies \cite{duysinx1998topology, le2010stress, ogawa2022topology}. 

First, several numerical examples are presented in which the p-norm of the von Mises stress defined in (\ref{eq:pn}) is minimized as the objective function to verify the validity of the stress constraint defined in (\ref{eq:dif_Gs}).
We formulate the minimization of p-norm of the von Mises stress as follows:
\begin{eqnarray}
\begin{split}
  \underset{\phi}{\text{minimize}} \qquad \quad\tilde{\sigma}_{pn}&(\bm{u},\phi)\\
		\text{subject to} \qquad G_\phi =& 
		0
		\qquad \mathrm{in} \hspace{2mm} \Omega_D\\
 		\bm{u} =& 0 
 		\qquad\mathrm{on} \hspace{2mm} \Gamma_u\\
 		\bm{\sigma_{n}} =& \bm{t} 
 		\qquad\mathrm{on} \hspace{2mm} \Gamma_{in}\\
 		G_{V\phi} 
 		\leq& 0 .\\
\end{split}
\end{eqnarray}
The topological derivative used to update the level set function using (\ref{eq:RDE}) is expressed as follows:
\begin{eqnarray}
d_t L &= (\bm{\nabla}\bm{v})^T h(\phi) \bm{A\nabla}\bm{u} +\lambda 
	+ \frac{1}{p}	\left(\int_{\Omega} (\frac{\tilde{\sigma}_{vm}(\bm{u}) }{\Phi \sigma_{max}})^{p}  \mathrm{d\Omega}\right)^{\frac{1}{p} - 1}  
	(\frac{\tilde{\sigma}_{vm}(\bm{u} )}{\Phi \sigma_{max}})^{p} ,
\end{eqnarray}
where $\lambda$ is the Lagrange multiplier, $\bm{A}$ is defined in (\ref{eq:A}), and $\bm{v}$ is the adjoint variable, which is defined as the solution of the following equation: 
\begin{eqnarray}
\begin{split}
& \int_{\Omega \backslash \Omega_\epsilon} (\bm{\nabla} \bm{v})^T \tilde{\bm{D}}\bm{\nabla} \bm{\psi} (\bm{x}) \mathrm{d\Omega}=\\
&-	\left(\int_{\Omega} (\frac{\tilde{\sigma}_{vm}(\bm{u}) }{\Phi \sigma_{max}})^{p}  \mathrm{d\Omega}\right)^{\frac{1}{p} - 1}
	\int_{\Omega\backslash\Omega_\epsilon} (\frac{\tilde{\sigma}_{vm}(\bm{u} ) }{\Phi \sigma_{max}})^{p-1}  (\frac{h(\phi) (\bm{D}\nabla \bm{u} )^T  \bm{B} (\bm{D} \bm{\nabla} \bm{\psi} (\bm{x}) ) }	{ \sigma_{max}\sqrt{(\bm{D}\nabla \bm{u} )^T  \bm{B} (\bm{D} \bm{\nabla \bm{u}} )} })	\mathrm{d\Omega} ,
\end{split}
\end{eqnarray} 
where $\bm{\psi}(\bm{x})$ is a test function.

Figure \ref{fig:cond_Lbeam} presents the fixed design domain and boundary conditions.
The fixed design domain is an L $\times$ L domain with a non-design void of size 0.6L $\times$ 0.6L from the upper right edge. 
We set the upper boundary as the fixed boundary $\Gamma_{u}$, and the center-right boundary of width 0.02L as the input port $\Gamma_{in}$.
The input load vector $\bm{t}$ is a downward vector of size $1.0 \times 10^7$ Pa.
In these examples, the isotropic linear elastic material has a Young's modulus set to $210$ GPa and a Poisson ratio set to 0.3.
The upper limit of the volume constraint $V_{max}$ is set to 0.6, and the upper limit of the von Mises stress $\sigma_{max}$ is set to $2.0\times 10^7$ Pa.
In addition, the regularization parameter $\tau$ is set to $5.0\times 10^{-5}$, and L is set to 1 m.
The fixed design domain is discretized using a structural mesh and three-node triangular plane stress elements whose length is $5.0 \times 10^{-3}$ L.

\begin{figure*}[htbp]
	\begin{center}
		\subfigure[]{
			\includegraphics[height=2.5cm]{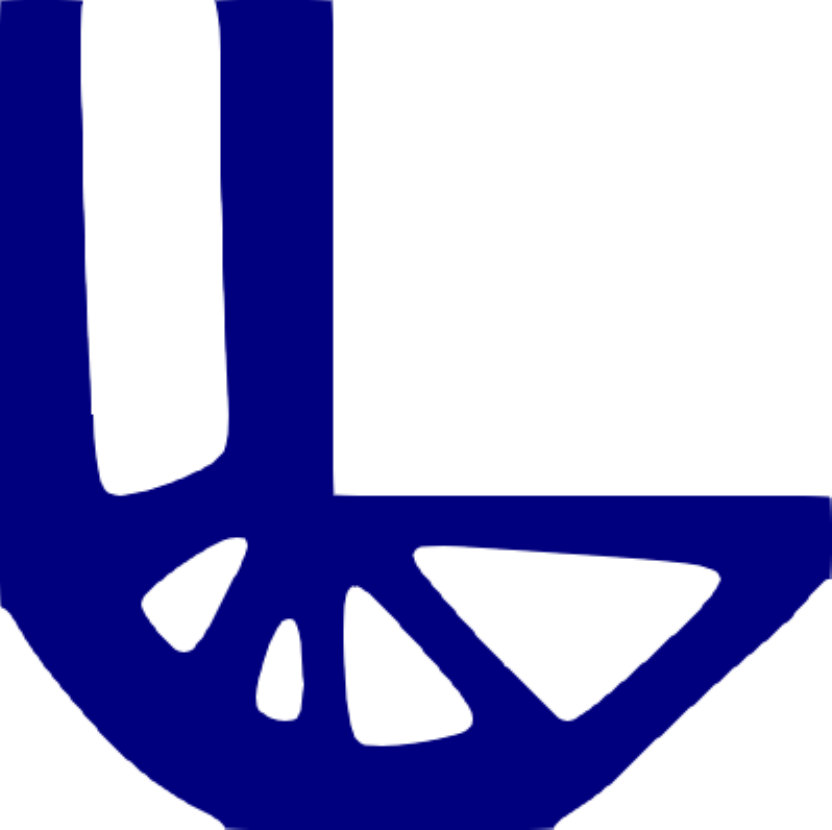}}\qquad
		\subfigure[]{
			\includegraphics[height=2.5cm]{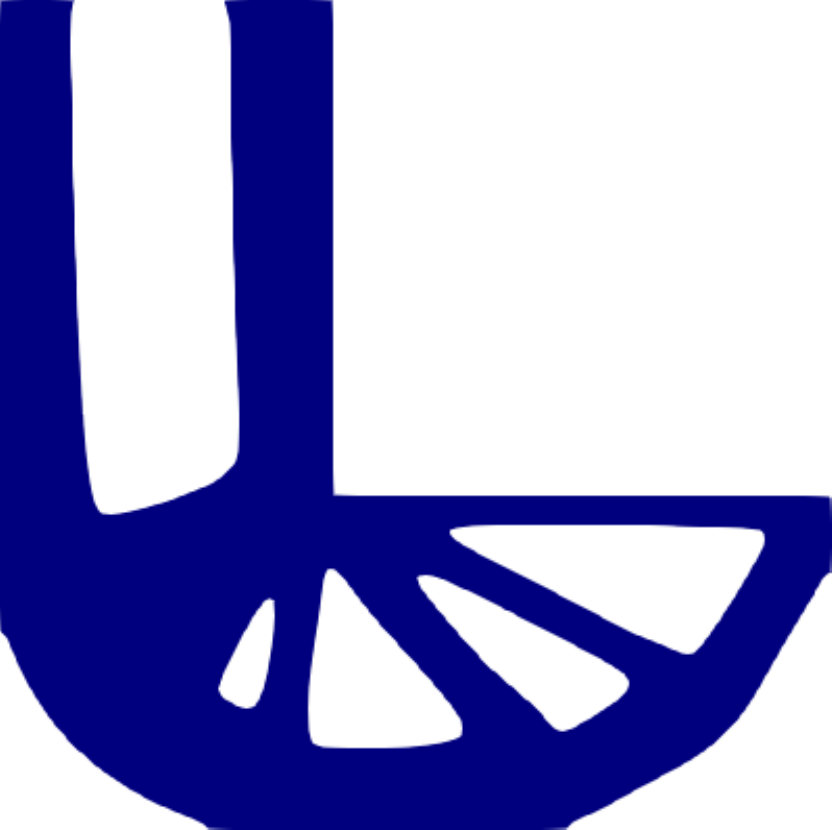}}\qquad
		\subfigure[]{
			\includegraphics[height=2.5cm]{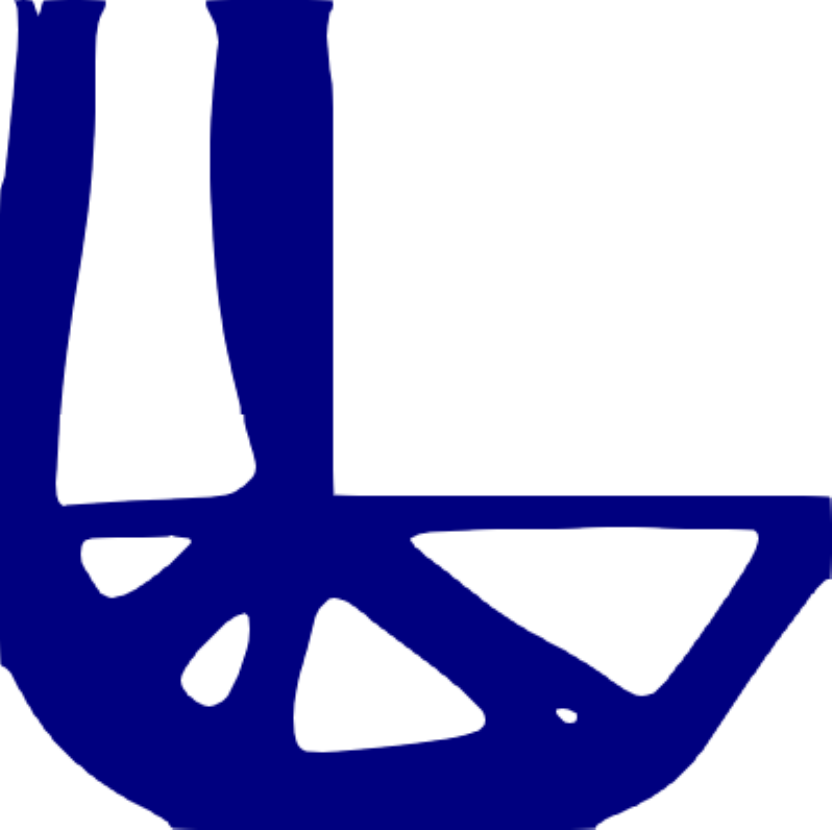}}\qquad
		\subfigure[]{
			\includegraphics[height=2.5cm]{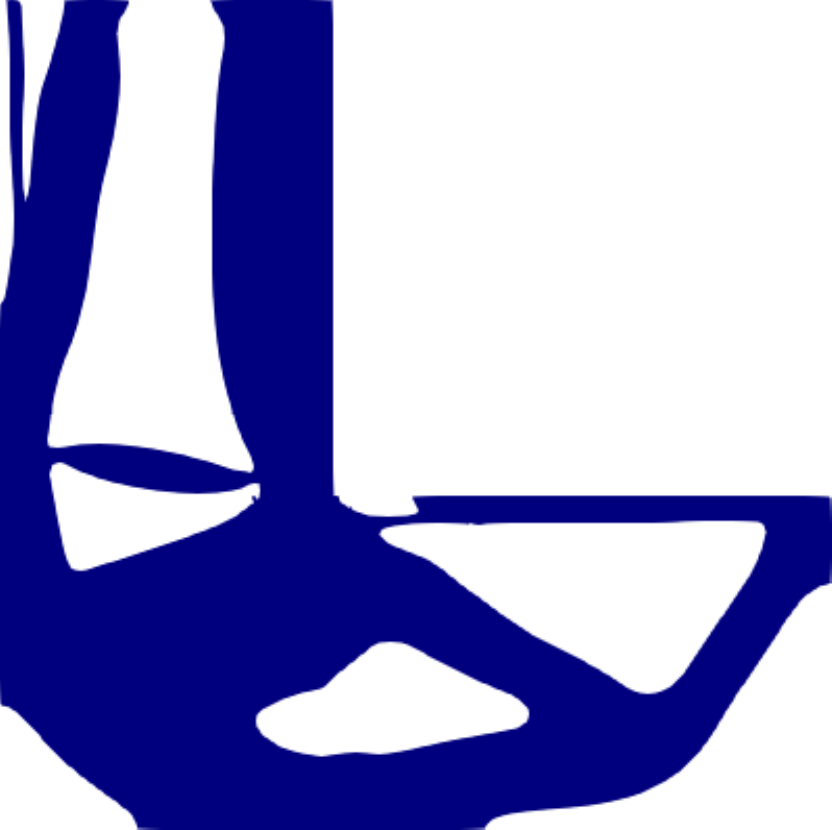}}\qquad
		\subfigure[]{
			\includegraphics[height=2.5cm]{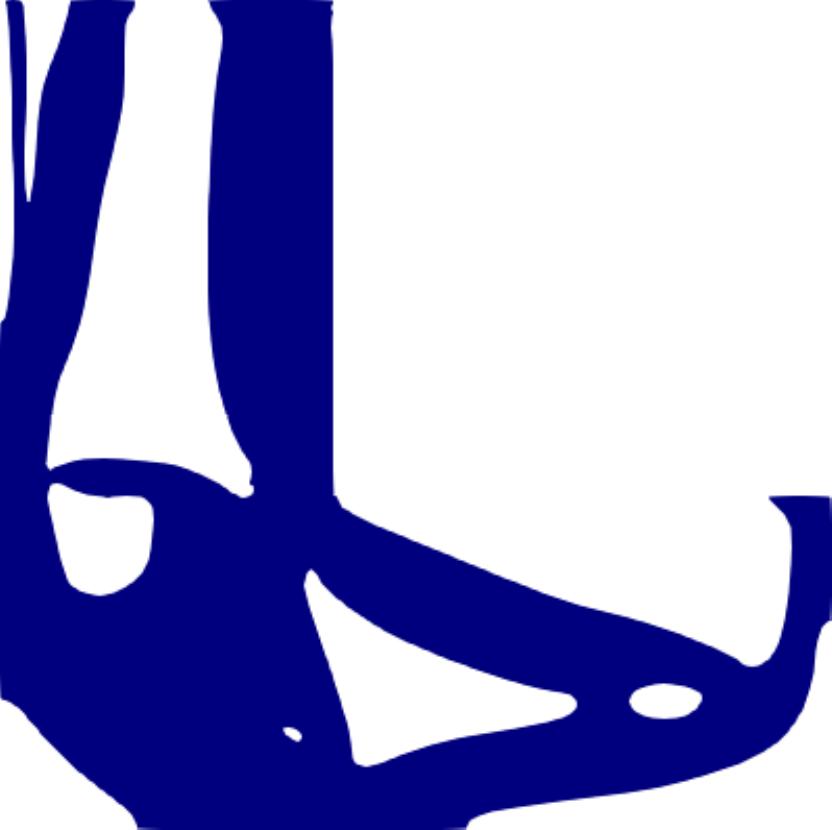}}
		\caption{Optimal configuration of L-beam for various settings of parameter $p$}
		\label{fig:shape_Lbeam}
	\end{center}
\end{figure*}

\begin{figure*}[htbp]
	\begin{center}
%	\begin{tabular}{cc}
%		\subfigure[]{
%			\includegraphics[height=2.5cm]{Fig/Lbeam_p=0_mises.pdf}}\qquad
%		\subfigure[]{
%			\includegraphics[height=2.5cm]{Fig/Lbeam_p=2_mises.pdf}}\qquad
%		\subfigure[]{
%			\includegraphics[height=2.5cm]{Fig/Lbeam_p=4_mises.pdf}}&
%		\multirow{2}{*}[6.0ex]{
%		\subfigure{
%			\includegraphics[height=4.0cm]{Fig/colorbar_MPa.pdf}}}\\
%		\subfigure[]{
%			\includegraphics[height=2.5cm]{Fig/Lbeam_p=6_mises.pdf}}\qquad
%		\subfigure[]{
%			\includegraphics[height=2.5cm]{Fig/Lbeam_p=8_mises.pdf}}&
%	\end{tabular}
		\includegraphics[height=7.144cm]{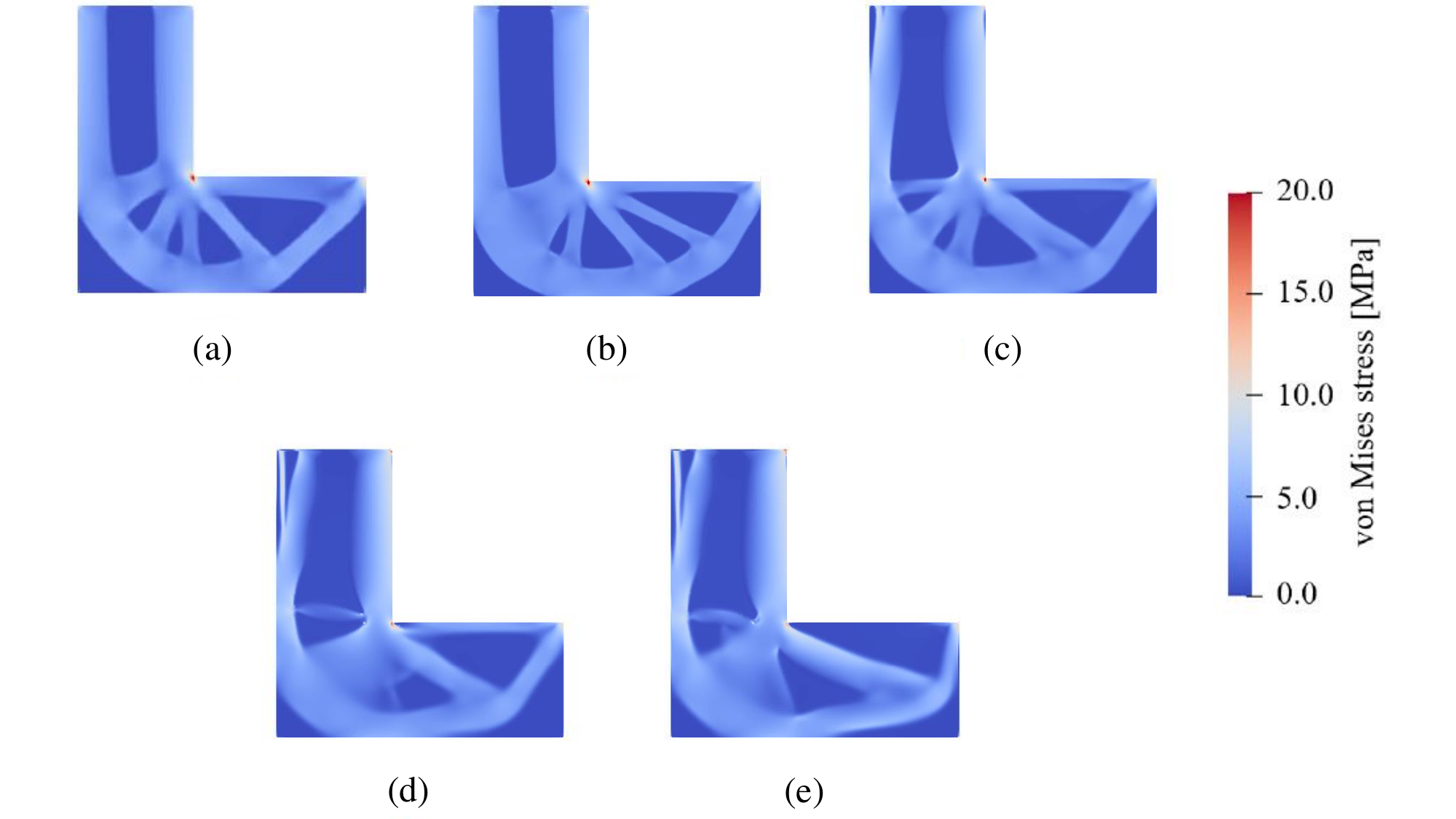}
		\caption{Von Mises stress of L-beam for various settings of parameter $p$}
		\label{fig:mises_Lbeam}
	\end{center}
\end{figure*}

We examine the effect of the multiplier $p$ of the p-norm.
We set four conditions of $p$. In condition (a), we minimize the mean compliance. In conditions (b), (c), (d), and (e), we set $p$ to 2.0, 4.0, 6.0, and 8.0, respectively.
Table \ref{tab:Lbeam} displays the maximum value of the von Mises stress for each condition.
Figure \ref{fig:shape_Lbeam} displays the optimal configuration for each condition, while Figure \ref{fig:mises_Lbeam} presents the distribution of the von Mises stress for each condition.

\begin{table}[htbp]
 \caption{Maximum value of von Mises stress for various settings of parameter $p$}
 \label{tab:Lbeam}
 \centering
 \begin{tabular}{ccc}
 \hline
  condition & $p$ & max. von Mises stress [MPa] \\
 \hline
  a &minimize mean compliance& 35.2\\
 b &2.0 & 34.1\\
 c &4.0 & 28.1\\
 d &6.0 & 18.6\\
 e &8.0 & 15.0\\
 \hline
 \end{tabular}
\end{table}

As illustrated in Figure \ref{fig:shape_Lbeam}, the structure obtained by p-norm minimization (conditions (b)--(e)) has more of a material domain clustered in the lower left corner of the figure and more of the void domain on the right side of the figure than the structure obtained by mean compliance minimization (condition (a)).
As illustrated in Figure \ref{fig:mises_Lbeam}, in conditions (a)--(c), the stress is concentrated at the inner corners and the horizontal beam is hardly stressed, whereas in conditions (d) and (e), the stress is not concentrated at the inner corners and distributed in the beam.
The maximum von Mises stress is smaller in conditions (b)--(e) than in condition (a).
These results indicate that the maximum value of the von Mises stress can be reduced by minimizing the p-norm using the proposed method.
In addition, the extent of reduction is greater as the value of $p$ increases.

Next, numerical examples of stress-constrained minimization of mean compliance are presented.
We formulate the optimization problem as follows:
\begin{eqnarray}
\begin{split}
  \underset{\phi}{\text{minimize}} \qquad \int_{\Gamma _{in}}\bm{u}\cdot&\bm{t} \dG\\
		\text{subject to} \qquad G_\phi =& 
		0
		\qquad \mathrm{in} \hspace{2mm} \Omega_D\\
 		\bm{u} =& 0 
 		\qquad\mathrm{on} \hspace{2mm} \Gamma_u\\
 		\bm{\sigma_{n}} =& \bm{t} 
 		\qquad\mathrm{on} \hspace{2mm} \Gamma_{in}\\
 		G_{V\phi} 
 		\leq& 0 .\\
 		\tilde{G}_{\sigma} \leq&0\\
\end{split}
\end{eqnarray}

\begin{figure*}[htbp]
	\begin{center}
		\subfigure[]{
			\includegraphics[height=2.5cm]{Fig/Lbeam_p=0_shape.pdf}}\qquad
		\subfigure[]{
			\includegraphics[height=2.5cm]{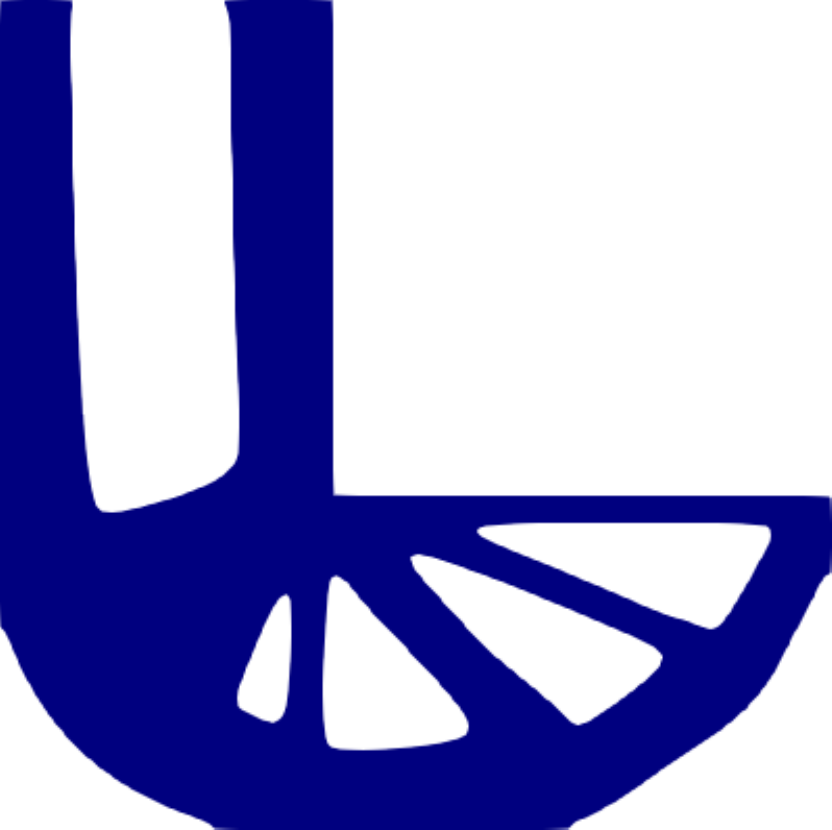}}\qquad
		\subfigure[]{
			\includegraphics[height=2.5cm]{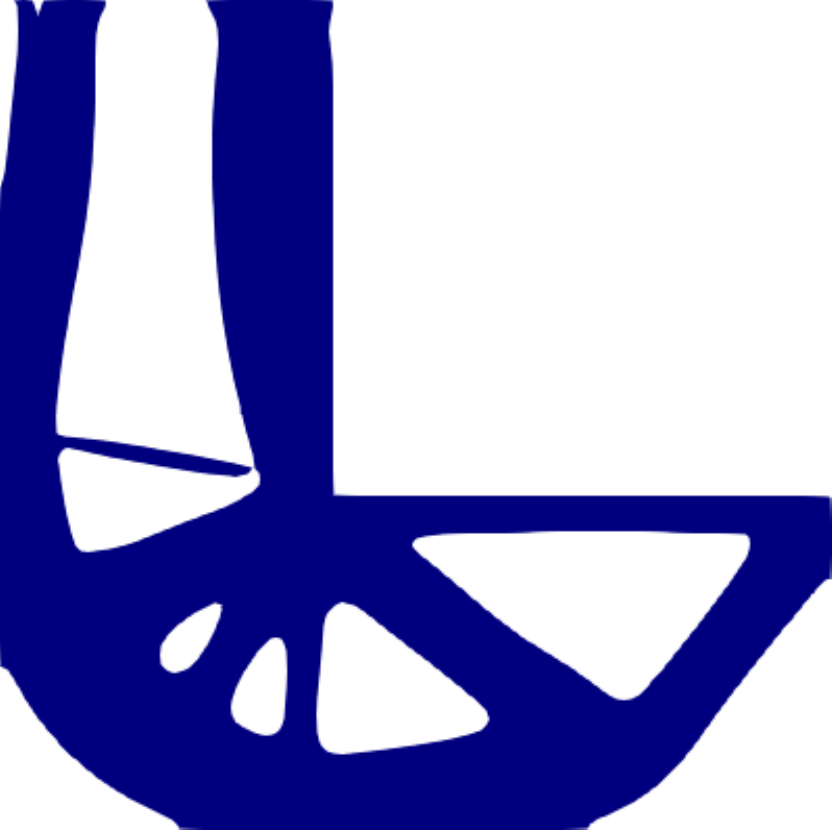}}\qquad
		\subfigure[]{
			\includegraphics[height=2.5cm]{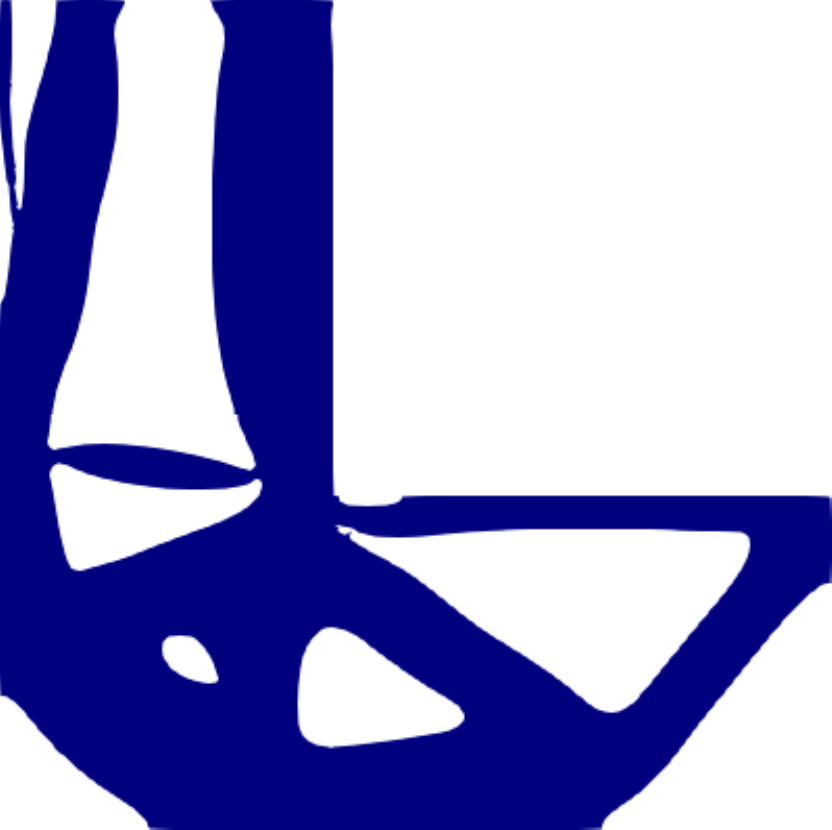}}\qquad
		\subfigure[]{
			\includegraphics[height=2.5cm]{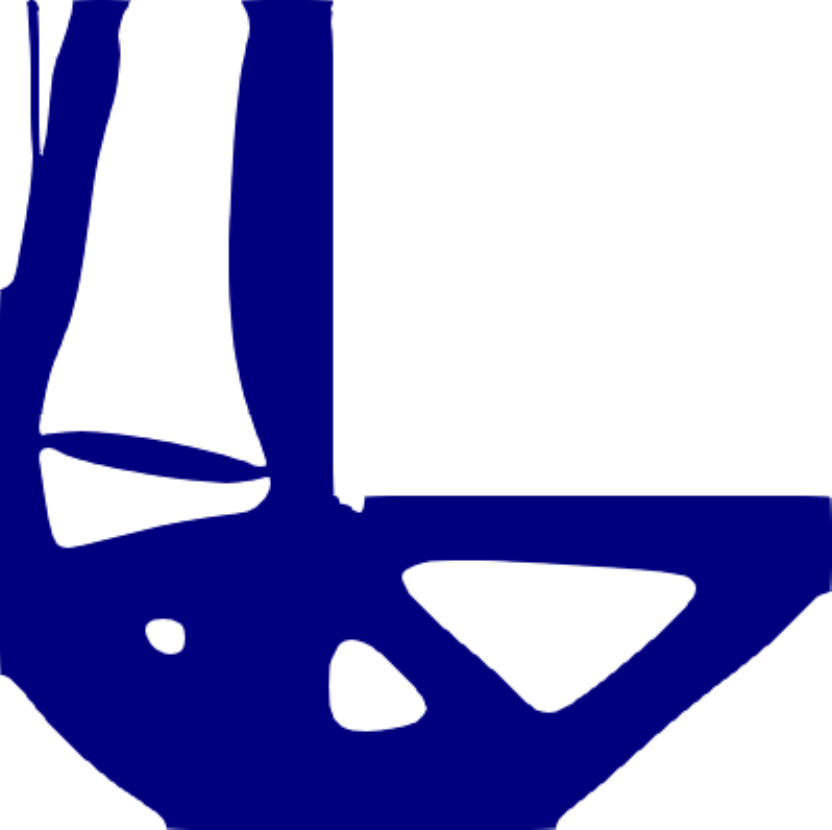}}
		\caption{Optimal configuration of stress-constrained minimization of mean compliance for various settings of parameter $p$}
		\label{fig:shape_Lbeam_const}
	\end{center}
\end{figure*}

\begin{figure*}[htbp]
	\begin{center}
%	\begin{tabular}{cc}
%		\subfigure[]{
%			\includegraphics[height=2.5cm]{Fig/Lbeam_p=0_mises.pdf}}\qquad
%		\subfigure[]{
%			\includegraphics[height=2.5cm]{Fig/L_mises_p=2_mu1e5.pdf}}\qquad
%		\subfigure[]{
%			\includegraphics[height=2.5cm]{Fig/L_mises_p=4_mu1e5.pdf}}&
%		\multirow{2}{*}[6.0ex]{
%		\subfigure{
%			\includegraphics[height=4.0cm]{Fig/colorbar_MPa.pdf}}}\\
%		\subfigure[]{
%			\includegraphics[height=2.5cm]{Fig/L_mises_p=6_mu1e5.pdf}}\qquad
%		\subfigure[]{
%			\includegraphics[height=2.5cm]{Fig/L_mises_p=8_mu1e5.pdf}}&
%	\end{tabular}
		\includegraphics[height=7.144cm]{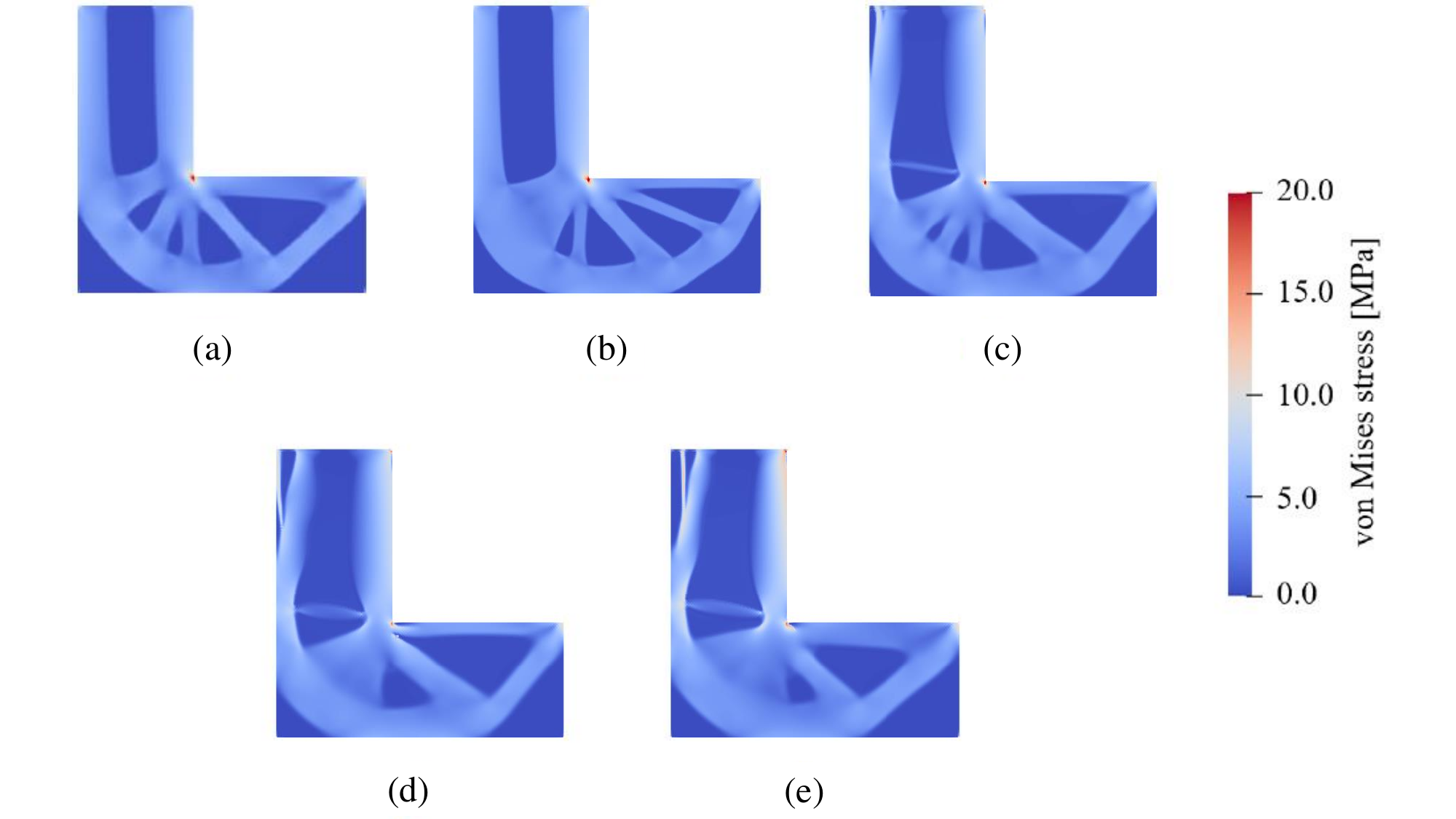}
		\caption{Von Mises stress of stress-constrained minimization of mean compliance for various settings of parameter $p$}
		\label{fig:mises_Lbeam_const}
	\end{center}
\end{figure*}

Table \ref{tab:Lbeam_const} presents the maximum value of the von Mises stress for each condition.
Figure \ref{fig:shape_Lbeam_const} displays the optimal configuration for each condition, while Figure \ref{fig:mises_Lbeam_const} presents the distribution of the von Mises stress for each condition.
Similar to the case of p-norm minimization, the maximum value of the von Mises stress can be reduced by stress-constrained minimization of mean compliance using the proposed method.

\begin{table}[htbp]
 \caption{Maximum value of the von Mises stress for various settings of parameter $p$}
 \label{tab:Lbeam_const}
 \centering
 \begin{tabular}{ccc}
 \hline
  condition & $p$ & max. von Mises stress [MPa] \\
 \hline
  a &no stress constraint& 35.2\\
 b &2.0 & 34.1\\
 c &4.0 & 27.7\\
 d &6.0 & 18.1\\
 e &8.0 & 17.6\\
 \hline
 \end{tabular}
\end{table}

\section{Comparison of displacement between proposed method and conventional topology optimization results}
\label{comp_inv}
\setcounter{figure}{0}
\setcounter{table}{0}

We compare our designs with the conventional topology optimization results in terms of delivered displacements.
The output displacement of our design is smaller than that of the conventional topology optimization results because of differences in some parameters.
Table \ref{tab:comp_param} shows some parameters and output displacement of the proposed method in condition (q) and two conventional methods.
Conventional method A \cite{sigmund1997design} is not stress-constrained.
Conventional method B \cite{emmendoerfer2020stress} is a stress-constrained method.
In both of the conventional methods, a few percent of displacement of the fixed design domain is obtained. \\
\begin{table}[htbp]
 \caption{Parameters and output displacement in the proposed method and conventional methods. $L$: Length of fixed design domain, $F_{in}$: input force, $U_{out}$: output displacement, $\sigma_{max}$: upper limit of von Mises stress}
 \label{tab:comp_param}
 \centering
 \begin{tabular}{cccccc}
 \hline
Method & Young's modulus &  $L$ &$F_{in}$ [Pa] & $U_{out}$& $\sigma_{max}$\\
\hline
Proposed method &210 GPa  & 1 m &1.0$\times10^7$ & 39.6 $\mu$m & 20 MPa\\
Conventional method A \cite{sigmund1997design} &  180 GPa & 300 $\mu$m & 4.76$\times10^7$ &7.8 $\mu$m& N/A\\
Conventional method B \cite{emmendoerfer2020stress} & 3000 MPa & 100 mm & 8.0$\times10^6$ & 1.71 mm& 40 MPa\\
 \hline
 \end{tabular}
\end{table}

For comparison, we design displacement inverters using the same parameters as the conventional method B that shown in Table \ref{tab:inv_param_convB}.
Table \ref{tab:inv_disp_convB} shows displacement at the output and input ports of the displacement inverter.
A few percent of displacement of fixed design domain is obtained similar to the conventional methods.
\begin{table}[htbp]
 \caption{Parameters for the design of the displacement inverter to compare with conventional method B}
 \label{tab:inv_param_convB}
 \centering
 \begin{tabular}{ccccccc}
 \hline
 $\tau$ & $p$ & $d$ & Young's modulus [MPa]& $L$ [mm]& $F_{in}$ [Pa] & $\sigma_{max}$[MPa]\\
\hline
 5.0$\times 10^{-5}$ & 2.0 & 0.001 & 3000 &100 & 8.0$\times10^6$ &  40 \\ 
 \hline
 \end{tabular}
\end{table}
\begin{table}[htbp]
 \caption{Displacement at the output and input ports of the displacement inverter}
 \label{tab:inv_disp_convB}
 \centering
 \begin{tabular}{ccccc}
 \hline
  & & & \multicolumn{2}{c}{displacement [mm]}\\
$\alpha$& $\beta$& $\mu$ & $\;\;\ U_o\;\;$ &$U_i$\\
 \hline
0&1.0& 0 & 1.57 & 3.28\\
0&1.0& 0.1 & 1.33 & 2.80\\
0&1.0& 0.3 & 0.99 & 2.39\\
0&1.0& 0.5 & 0.53 & 1.16\\
 \hline
 \end{tabular}
\end{table}

\section{Prototyping of compliant displacement magnification mechanism}
\label{proto}

\setcounter{figure}{0}
\setcounter{table}{0}	

\begin{figure*}[htb]
	\begin{center}
		\subfigure{\includegraphics[height=3.0cm]{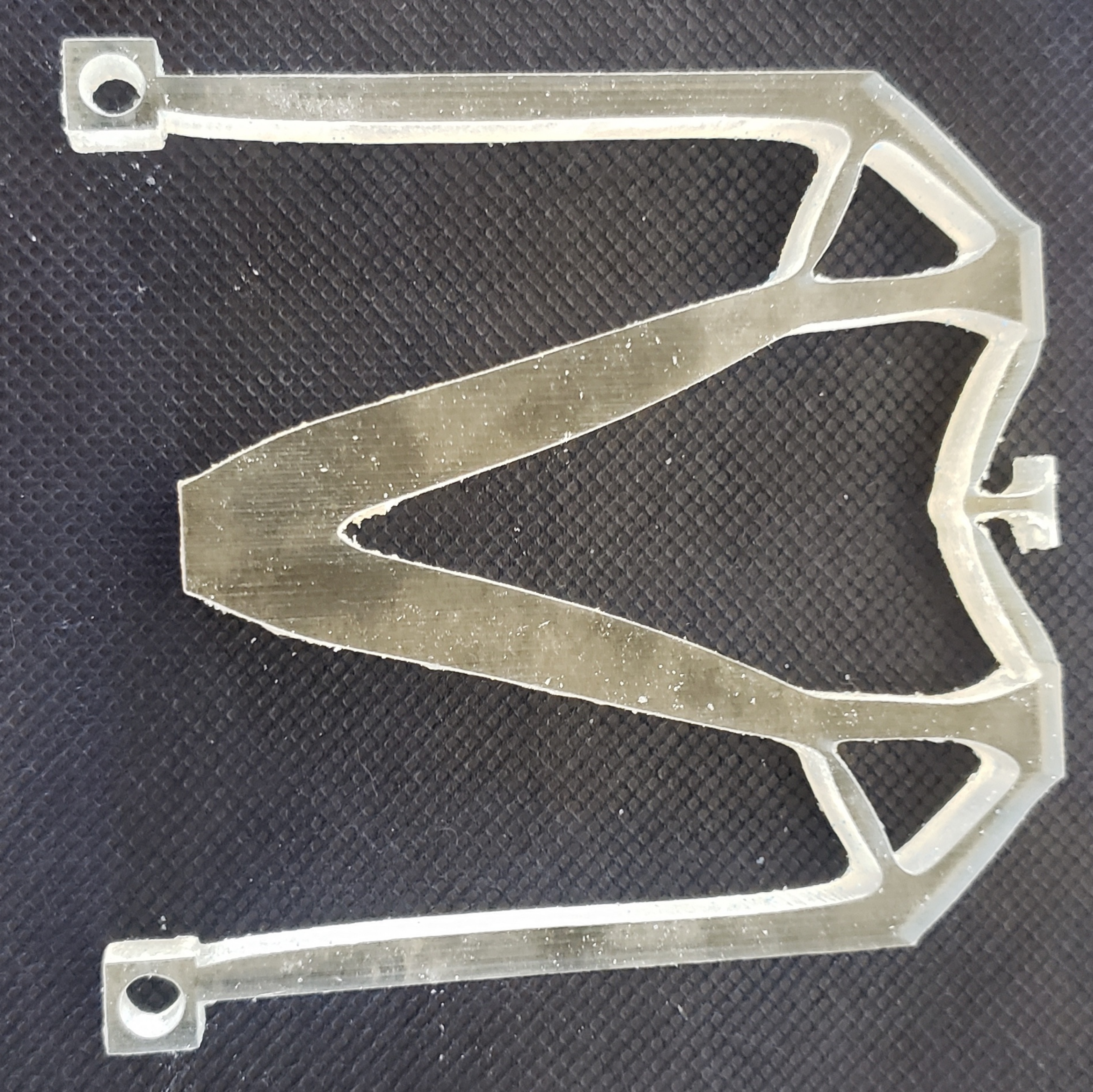}}\qquad
		\subfigure{\includegraphics[height=3.0cm]{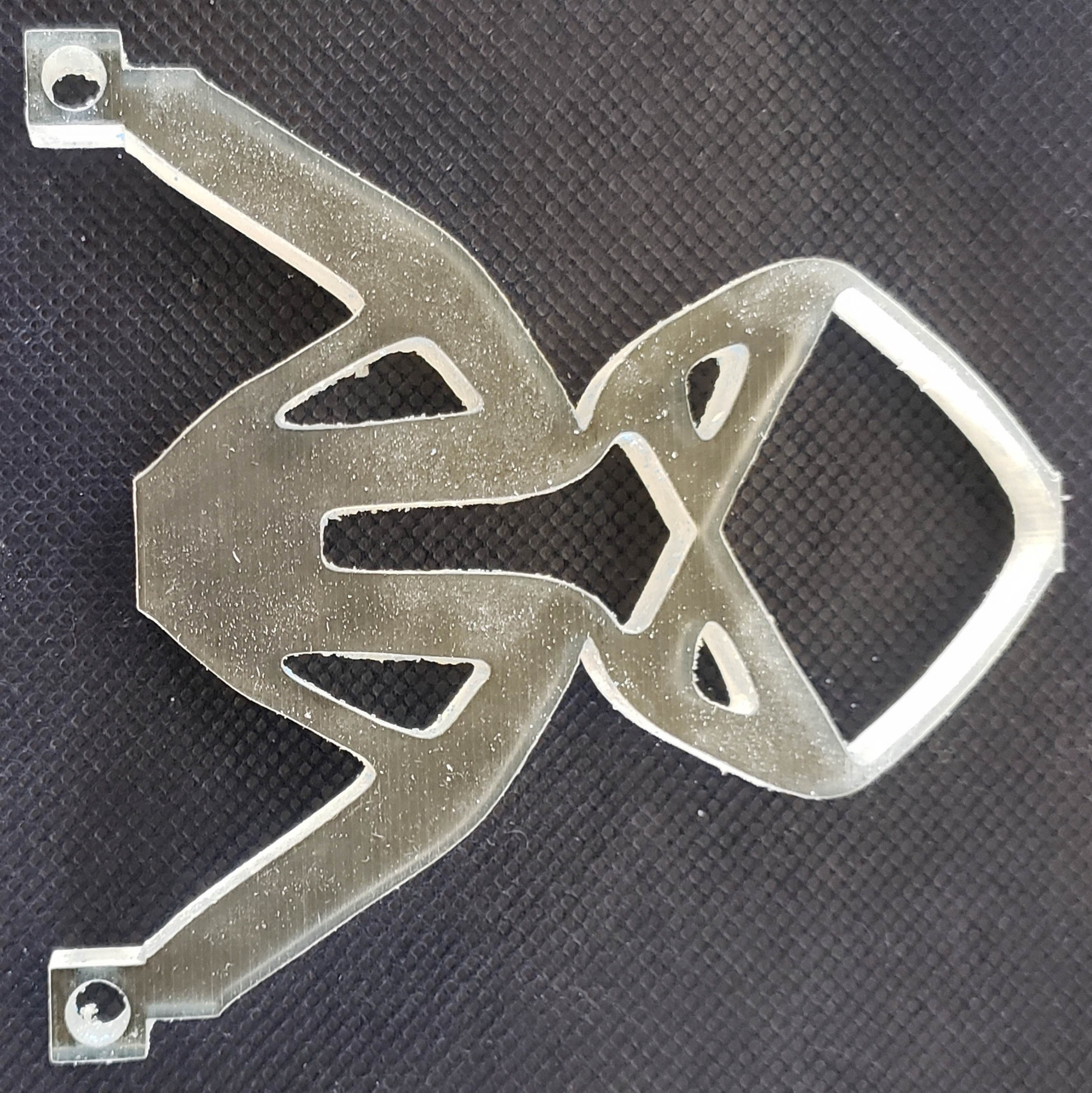}}
		\caption{Prototypes of displacement magnification mechanisms (left: $\mu=0$, right: $\mu=0.3$)}
		\label{fig:proto_magni}
	\end{center}
\end{figure*}

We prototyped the structure displayed in Figure \ref{fig:magni_tau} (d)-3 and (f)-3 using a 3D printer.
The size of the prototype was scaled down to L = 0.1 m.
First, the structure was extruded to a thickness of 0.1L to create a 3D structure.
A square of side 0.1 L was attached to the fixed boundary, and a hole of 0.066L in diameter was drilled in the rectangle and fixed with a M6 bolt.
The prototype was printed using STRATASYS Objet260 Connex3.
Figure \ref{fig:proto_magni} displays the printed prototypes.

\begin{figure*}[htb]
	\begin{center}
		\includegraphics[height=5.0cm]{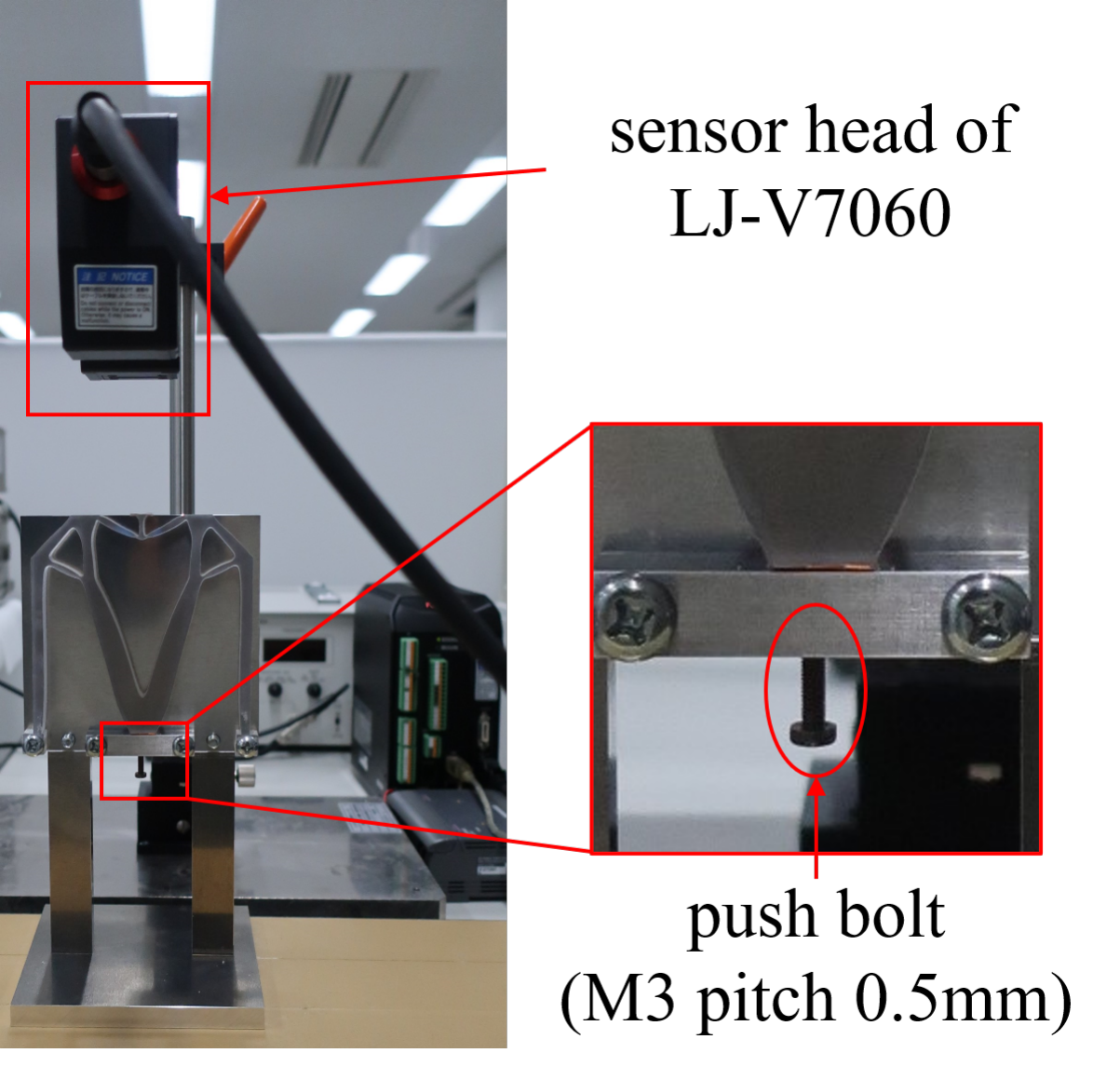}
		\caption{Measurement environment}
		\label{fig:measure}
	\end{center}
\end{figure*}

To measure the displacement magnification ratio, the prototype was fixed using a jig, and displacement was applied to the input port.
The displacement at the output port was measured using a Keyence LJ-V7060 sensor head.
Figure \ref{fig:measure} displays the measurement environment. 
The push bolt displayed in this figure had pitch of 0.5 mm.
Therefore, by turning the push bolt once, the input port was displaced by 0.5 mm.
The measurement procedure is described below.
\begin{itemize}
	\item[\bf{Step 1}] Set the mechanism on the jig and set the jig under the sensor head.
	\item[\bf{Step 2}] Turn the push bolt until the tip of the bolt touches the input port, and record the value of the sensor at that time. 
	\item[\bf{Step 3}] Turn the push bolt one full turn in the direction of pushing the input port using a torque wrench and record the value of the sensor and maximum torque at that time.
	\item[\bf{Step 4}] Turn the push bolt one more time in the direction of pushing the input port using a torque wrench and record the value of the sensor and maximum torque at that time.
	\item[\bf{Step 5}] Turn the push bolt two times in the direction away from the input port, and record the value of the sensor at that time.
	\item[\bf{Step 6}] Repeat steps 3--5 five times and record the data.
\end{itemize}

\begin{figure*}[htb]
	\begin{center}
		\subfigure{\includegraphics[height=3.0cm]{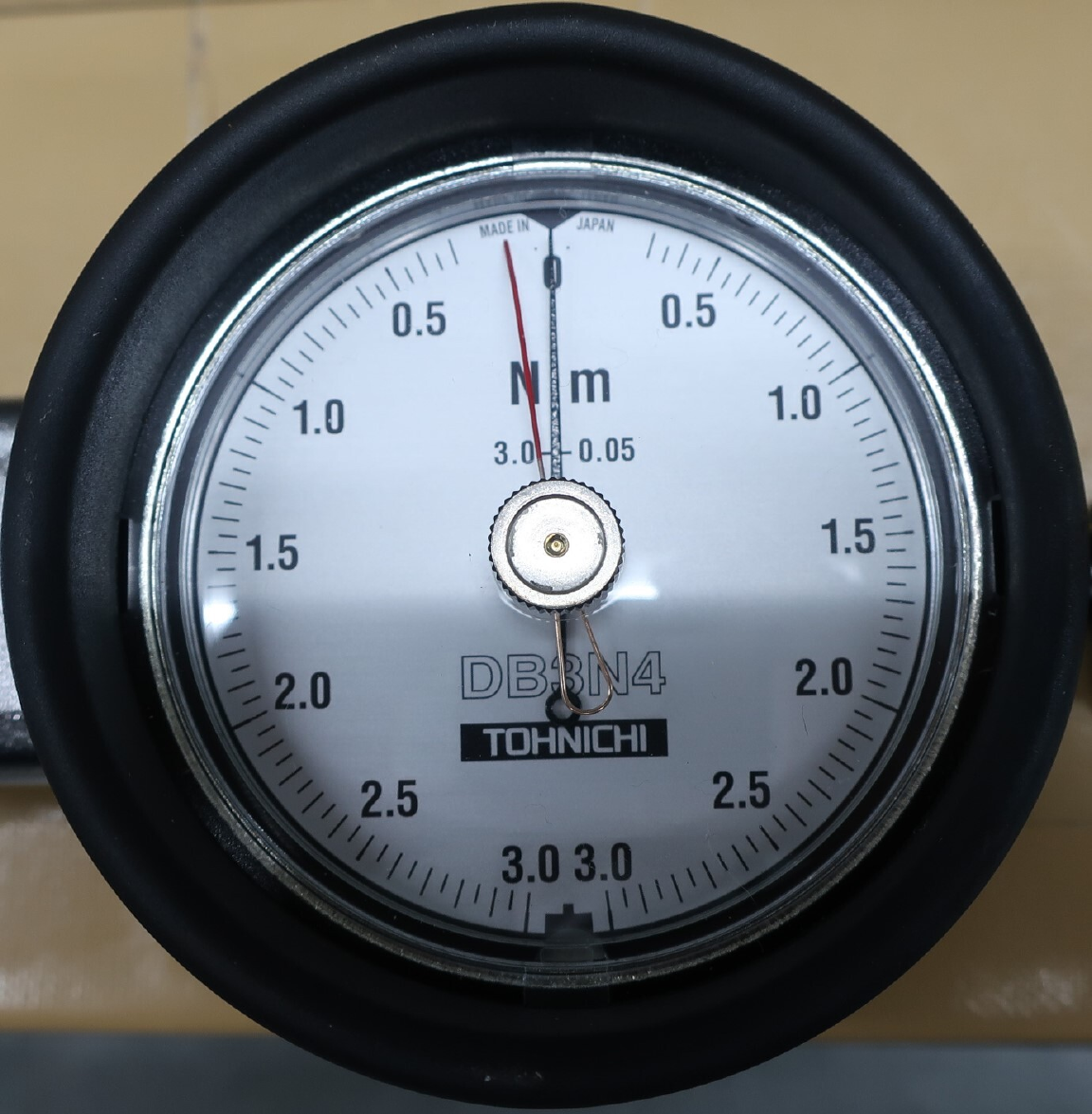}}\qquad
		\subfigure{\includegraphics[height=3.0cm]{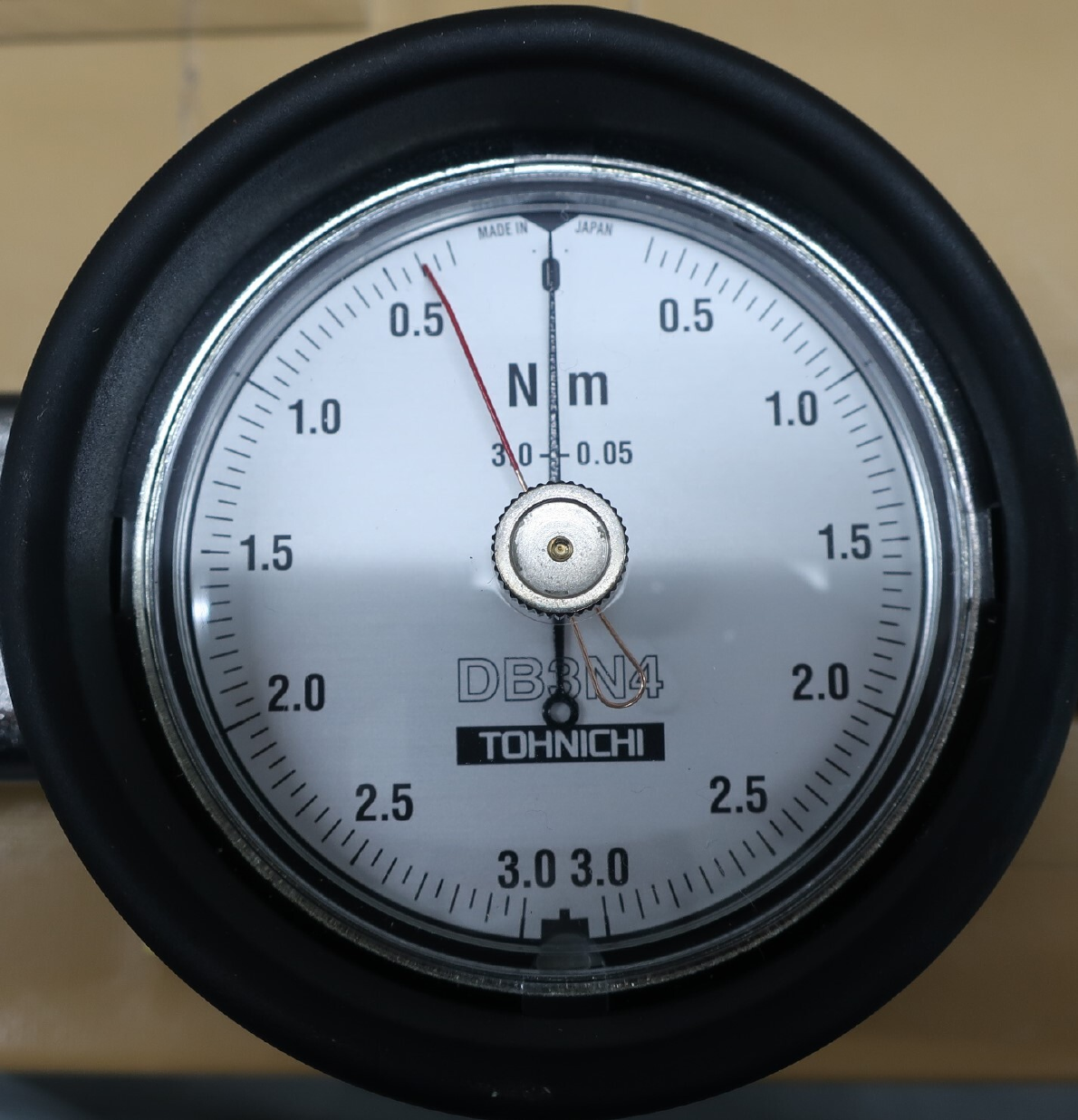}}
		\caption{Measured torque (left: $\mu=0$, right: $\mu=0.3$)}
		\label{fig:torque}
	\end{center}
\end{figure*}

In steps 3 and 4, the value of the sensor is the average distance from the sensor head to the output port in the thickness direction.
The displacement at the output port is defined as the difference between the sensor values in steps 3 and 4.
Table \ref{tab:disp} lists the displacement at the output port of each mechanism obtained in this measurement, while Figure \ref{fig:torque} presents the measured torque.

\begin{table}[htbp]
 \caption{Displacement at the output port}
 \label{tab:disp}
 \centering
 \begin{tabular}{ccc}
 \hline
  & \multicolumn{2}{c}{displacement [mm]}\\
 trial number & $\;\;\mu=0\;\;$ &$\mu=0.3$\\
 \hline
 1 & 1.60 & 0.95\\
 2 & 1.60 & 0.98\\
 3 & 1.57 & 0.97\\
 4 & 1.62 & 0.90\\
 5 & 1.63 & 0.97\\
 \hline
 average & 1.61 & 0.95\\
 \hline
 \end{tabular}
\end{table}

As illustrated in Table \ref{tab:disp}, the displacement of each mechanism was larger than the input displacement of 0.5 mm.
This signifies that the mechanisms magnified the input displacement as the output displacement.
For $\mu = 0.3$, the displacement at the output port was smaller than for $\mu = 0$.
However, as illustrated in Figure \ref{fig:torque}, the required torque to produce the displacement was larger when $\mu=0.3$ than when $\mu=0$.
This indicates that the stiffness against the input force of the mechanism was larger when $\mu=0.3$ instead of the output displacement being smaller.
These results are consistent with the calculation results obtained in Section \ref{sec:example}.

%--------------------------------
%   References
%--------------------------------
%\bibliographystyle{elsarticle-num-names} 
\bibliography{mybibfile}

\end{document}